# Temperature-dependent structure of methanol-water mixtures on cooling: X-ray and neutron diffraction and molecular dynamics simulations

Ildikó Pethes[1,a], László Pusztai[a,b], Koji Ohara[c], Shinji Kohara[d], Jacques Darpentigny[e], László Temleitner[a]

[a]Wigner Research Centre for Physics, Konkoly Thege út 29-33., H-1121 Budapest, Hungary

[b]International Research Organization for Advanced Science and Technology (IROAST), Kumamoto University, 2-39-1 Kurokami, Chuo-ku, Kumamoto 860-8555, Japan

[c]Diffraction and Scattering Division, Japan Synchrotron Radiation Research Institute (JASRI/SPring-8), 1-1-1 Kouto, Sayo-cho, Sayo-gun, Hyogo 679-5198, Japan

[d]Research Center for Advanced Measurement and Characterization, National Institute for Materials Science (NIMS), 1–1–1 Kouto, Sayo-cho, Sayo-gun, Hyogo 679–5148, Japan

[e]Laboratoire Léon Brillouin, CEA-Saclay 91191 Gif sur Yvette Cedex France

**Abstract**

Methanol-water liquid mixtures have been investigated by high-energy synchrotron X-ray and neutron diffraction at low temperatures. We are thus able to report the first complete sets of both X-ray and neutron weighted total scattering structure factors over the entire composition range (at 12 different methanol concentrations ($x_M$) from 10 to 100 mol%) and at temperatures from ambient down to the freezing points of the mixtures. The new diffraction data may later be used as reference in future theoretical and simulation studies. The measured data are interpreted by molecular dynamics simulations, in which the all atom OPLS/AA force field model for methanol is combined with both the SPC/E and TIP4P/2005 water potentials. Although the TIP4P/2005 water model was found to be somewhat more successful, both combinations provide at least semi-quantitative agreement with measured diffraction data. From the simulated particle configurations, partial radial distribution functions, as well as various distributions of the number of hydrogen bonds have been determined. As a general trend, the average number of hydrogen bonds increases upon cooling. However, the number of hydrogen bonds between methanol molecules slightly decreases with lowering temperatures in the concentration range between ca. 30 and 60 mol % alcohol content. The same is valid for water-water hydrogen bonds above 70 mol % of methanol content, from room temperature down to 193 K.

[1] Corresponding author: e-mail: pethes.ildiko@wigner.hu

1. Introduction

Alcohol-water mixtures have been, and still are, continuously in the center of attention in the physical and chemical sciences (for very recent works, see e.g. [1-15]). Apart from the theoretical and experimental scientific points interest, their technological importance is also notable.

Methanol, the simplest alcohol, is the closest analogue to water, in which one (hydrophilic) proton has been replaced by a (hydrophobic) methyl group. Despite their similar shape and size, this substitution leads to significant differences in terms of the structure of the two liquids, primarily due to the hydrophobic behavior of the methyl group. While the fourfold coordination of water molecules results in a three-dimensional hydrogen bonded (HB) network in liquid water, methanol molecules in pure methanol participate in at most two hydrogen bonds in practice, and form smaller chains [16-19]. In aqueous solutions of methanol, both methanol and water can act as H-donor and H-acceptor, and hydrogen bonded networks can be constructed via water-water, methanol-methanol and methanol-water hydrogen bonded pairs. Methanol molecules in such mixtures may form up to three hydrogen bonds (by one donor and two acceptor sites), whereas water molecules easily coordinate four H-bonded neighbors (via their two donor and two acceptor sites).

The structure of methanol-water mixtures has been the subject of numerous theoretical and simulation studies: the inverse Kirkwood-Buff integral method [20], Monte Carlo simulations [21-27], classical molecular dynamics simulations [2,8,28-40], empirical potential structure refinement [5,41-43], and first-principle density functional theory based (*ab initio*) molecular dynamics simulations [44-47] have been employed. Experimental works have been published in the past decades using a wide variety of techniques, such as neutron diffraction (ND) [5,41-43,48,49], X-ray diffraction (XRD) [14,50-52], mass-spectrometry [51,53], microwave dielectric relaxation analysis [54], Raman spectroscopy [55-57],

X-ray emission spectroscopy [58], X-ray absorption spectroscopy (XAS) [59], quasielastic neutron scattering [60] and nuclear magnetic resonance (NMR) [61-65].

According to the investigations listed above, a general picture seemed to have emerged as follows. At low methanol concentrations ($x_M < 0.3$, where $x_M$ is the mole fraction of methanol) methanol molecules are independently hydrated by water molecules. The 3D tetrahedral HB network of water molecules is hardly influenced by adding methanol to water; some authors even suggest an enhancement [9]. As the methanol concentration increases, the hydration spheres of methanol molecules overlap, and methanol molecules start to aggregate. According to some authors, the mixing of the two components is incomplete [36,49,58] and separate HB networks of water and methanol were also observed [42]. Around the equimolar composition, a change from the percolated to non-percolate regimes was reported [27]. At high methanol concentrations ($x_M > 0.7$), methanol molecules form (branched) chains (and rarely, rings) and 1D/2D hydrogen bonded clusters of methanol molecules dominate. Water molecules connect the terminals of methanol chains, or form small (water) clusters and these clusters bridge the neighboring methanol hydroxyl groups [49]. Water clusters at low water concentrations and methanol clusters at low methanol concentrations were also reported (see e. g. by NMR [63] or by XAS [59]).

Most of the publications listed above are limited to ambient temperature, only a much smaller number of studies deal with the temperature dependence of the structure [5,8,20,39,43,52,56,60,61,62,64,65]. The same is valid concerning diffraction: while several studies have considered neutron [41,42,48,49] and X-ray [14,50,51] measurements at ambient temperature, diffraction data at lower temperatures are insufficient. Up to now (and to our knowledge), only one set of X-ray diffraction data is available, for the water-rich solutions (at $x_M = 0.2, 0.3$ and $0.4$) [52], while two papers are available about neutron diffraction data [5,43] at two methanol concentrations ($x_M = 0.27$ and $0.54$), with 2 or 3 temperature points. Interpretation of the X-ray data suggested that the structure of dominant clusters formed in the mixtures at 298 K do not significantly change with lowering the temperatures. The neutron data mentioned above allowed speculations that the 'micro-segregation' between methanol and water clusters is enhanced by lowering the temperature.

The most complete investigation of the concentration and temperature dependences of methanol-water solutions was conducted in a series of NMR measurements [61,62,64,65]. Above a concentration dependent crossover temperature hydrophobicity was suggested to play a significant role while below this temperature the tetrahedral network of water molecules network appeared to determine the properties of the solutions.

In a recent publication molecular dynamics simulation data were compared with X-ray diffraction results of Takamuku et al. [8], where the H-bonding structure has been analyzed from simulated particle configurations. Necessarily, this work was limited to the water-rich regime. To extend the investigations to equimolar and methanol-rich mixtures would be desirable, but a complete diffraction data set over the whole range of methanol concentration at temperatures down to the freezing point was missing.

By the present study, we provide additional experimental data on methanol-water liquid mixtures for 12 methanol concentrations, from $x_M = 0.1$ to 1.0. Both X-ray and neutron diffraction total scattering structure factors have been determined at 1 bar and several temperatures, from 300 K down to the freezing points of the mixtures. For a primary interpretation of these, rather large amount of, data, molecular dynamics simulations have been performed using the SPC/E [66] and TIP4P/2005 [67] models for water and the all atom OPLS/AA model [68] for methanol molecules. Various distributions of the number of hydrogen bonds have also been calculated.

## 2. Experimental details

Both protonated ($CH_3OH$) and deuterated ($CD_3OD$) forms of methyl alcohol (methanol) were purchased from Sigma-Aldrich Co. 12 methanol-water samples were taken for the diffraction experiments, using protonated compounds for X-ray, and the fully deuterated form of methanol and water for neutron measurements. Mixtures with methanol concentrations from 10 mol% to 100 mol%, with 10 mol% equimolar steps, and additionally with 54.42 and 73.37 mol% (peritectic concentration and close to the eutectic concentration, respectively), were prepared. Nominal and exact compositions are shown in Table 1.

*2.1 X-ray diffraction experiments*
X-ray diffraction measurements were performed at the BL04B2 high energy X-ray diffraction beamline of the Japan Synchrotron Radiation Research Institute (SPring-8, Hyogo, Japan). The beam energy was 61.2 keV, which corresponds to a wavelength of 0.203 Å. Diffraction patterns were recorded in transmission mode, in the horizontal scattering plane, using an array of 6 solid state detectors. The available range of the scattering vector ($Q$) was between 0.16 and 16 Å$^{-1}$. Samples were contained in thin-walled capillaries, with an inner diameter of 2 mm and a wall thickness of 0.15 mm. The entire sample environment was under vacuum. To prevent evaporation, the capillaries were sealed with Torr-seal. After the seal has become sufficiently dry, samples were set in a triple sample holder made from copper and connected directly to the cold head of a closed circle refrigerator (CCR).

Temperature was controlled by a LakeShore 331 temperature controller and the measurement software of the beamline was modified to set and register temperatures. The temperature was measured on the surface of the triple sample holder. The possible bias of the temperature at the sample was checked by a calibrated thermocouple at room temperature (297.9 K, +0.1 K from the calibrated temperature) and the observation of monoclinic to fcc phase transition of $CCl_4$ at 223.8 K (225.35 K [69]). This way, the uncertainty of the temperature can be estimated as ±1.5 K. The cooling rate was 1 K/min. Measurements were performed after an equilibration time (15 to 30 min) at each temperature.

Measured intensities were normalized by the incoming beam intensity, and corrected for absorption, polarization and contributions from the empty capillary. The patterns over the entire $Q$-range were obtained by normalizing and merging each frame, then removing the Compton scattering contributions.

*2.2 Neutron diffraction experiments*

Series of neutron diffraction measurements performed at the 7C2 diffractometer of Laboratoire Léon-Brillouin [70] using the standard cryostat available at the beamline. Standard 6 mm vanadium cans were used for containing the liquid samples. The incoming wavelength was 0.72 Å, corresponding to a $Q$-range of 1.06 to 15.7 Å$^{-1}$. Due to some malfunctioning of a signal processing unit in the group of detectors in the $Q$-range between 12 and 13 Å$^{-1}$, datasets were taken only up to 11.8 Å$^{-1}$ for further investigations. Each measurement was controlled by software available at the beamline. The summarized datasets were corrected by efficiency, using a vanadium standard. Intensities from the empty container were also subtracted.

All temperatures and compositions visited by X-ray and neutron diffraction experiments are collected in Table 1, and shown superimposed on the phase-diagram of methanol-water mixtures in Fig. 1.

**3. Molecular dynamics simulations**

Details of the molecular dynamics simulations are provided in the Supplementary Material (SM), here only the main points are described briefly.

Classical molecular dynamics (MD) simulations have been performed by the GROMACS software package (version 2018.2) [73]. The OPLS/AA [68] all atom model was used for methanol, whereas both the SPC/E [66] and TIP4P/2005 [67] models were taken for water molecules. Cubic simulation boxes contained 2000 molecules. NPT (constant pressure 1 bar, and temperature) simulations were conducted in order to determine densities of the mixtures at a given temperature. Densities obtained are collected in Table S3 and S4 in the SM, and are shown in Figures 2 and S1.

101 particle configurations have been collected from long (20 ns) NVT (constant volume and temperature, where the volume is the average value from NPT runs) simulations, 200 ps apart. Neutron and X-ray weighted total scattering structure factors ($F^N(Q)$ and $F^X(Q)$, respectively) were calculated from the partial radial distribution functions (PRDF, $g_{ij}(r)$), obtained from the particle configurations.

Hydrogen bonds can be defined in various ways [74]: here a geometric definition was applied. Two molecules are identified as H-bonded if the intermolecular distance between an oxygen and a hydrogen atom is less than 2.5 Å, and the O...O-H angle is smaller than 30 degrees. Calculations concerning H-bonds and H-bonded networks were performed by an in-house programme, based on the HBTOPOLOGY code [75].

## 4. Results and discussion

*4.1 Total structure factors*

Measured XRD and ND structure factors, at three selected temperatures, are shown in Figures 3 and 4 for $x_M$ = 0.7. The complete set of measured curves are shown in the Supplementary Material, Figures S2-S30. At the investigated temperatures and concentrations presented here, Bragg peaks due to crystallization were not observed, not even at temperatures which are slightly below the solid-liquid coexistence curve.

Common features of the XRD structure factors over the entire concentration range are that the first (around $Q$ = 1.85 Å$^{-1}$) and second maxima (around $Q$ = 2.9 Å$^{-1}$) become more intense upon cooling. At room temperature the height of the 2nd peak is decreasing with increasing methanol concentration, and only a shoulder can be observed for the $x_M \geq 0.8$ mixtures. As temperature decreases a peak around $Q$ = 2.9 Å$^{-1}$ can be seen even in pure methanol.

The positions of the maxima also depend on temperature. At low methanol concentration ($x_M$ = 0.1) the position of the first peak moves to smaller $Q$ values as the temperature decreases. As the methanol content increases the 1st peak position becomes less sensitive to temperature variations -- even constant in the composition range $0.5 \leq x_M \leq 0.7$. Increasing methanol concentration further makes the position of the 1st peak move to slightly higher $Q$ values as the temperature decreases. The composition dependence of the position of the 1st peak is significant at room temperature, but becomes less pronounced upon cooling. The position of the second peak changes monotonously: it moves to higher $Q$ values on decreasing temperature over the entire composition range.

A similar behavior is observed for the first peak (around $Q = 1.8$ Å$^{-1}$) of the ND structure factors: its position moves slightly to lower $Q$ values with decreasing temperatures in mixtures with low methanol content, then stays nearly stable around the equimolar mixture, and later moves to higher $Q$ values in the methanol rich compositions.

The X-ray and neutron weighted total scattering structure factors ($F^X(Q)$ and $F^N(Q)$, respectively), obtained from MD simulations with the TIP4P/2005 water model (at the same temperatures and concentration as the experimental points), are also shown in Figures 3 and 4. The full set of simulated curves can be found in the Supplementary Material (Figures S2-S47), together with comparisons of the model curves obtained using the TIP4P/2005 and SPC/E water models. The simulated curves follow the same trend as the experimental ones. The temperature dependence of both the 1$^{st}$ and 2$^{nd}$ peaks of $F^X(Q)$ and that of the 1$^{st}$ peak of $F^N(Q)$, as judged by their heights and positions, are similar in the simulated and measured functions. A small difference concerning the position of the 1$^{st}$ peak of $F^X(Q)$ is that in the simulated curves the temperature independent position can be found at lower methanol concentrations ($x_M = 0.4$ and 0.5).

In general, the agreement between calculated and measured structure factors is satisfactory. The largest deviation can be seen in the XRD structure factors around $Q = 3$ Å$^{-1}$, where the height and position of the 2$^{nd}$ peaks are different: this is most pronounced for the $x_M = 0.4$ mixture. Only small discrepancies can be observed in the neutron weighted structure factors.

Simulations were carried out using the OPLS/AA model for methanol and both the TIP4P/2005 and SPC/E water models (thus only the water model is mentioned in the figure captions). All calculated curves are compared in Figures S2-S47 (in SM); here only the $F^N(Q)$ curves for $x_M = 0.2$ are shown (Fig. 5). Differences between the calculated curves are obviously larger in water-rich mixtures. Results obtained by the two water models were also compared quantitatively by calculating the goodness of fit ($R$-factor) values: an overall better agreement between experimental and calculated curves was found when using the TIP4P/2005 water model. For this reason, only structural properties obtained from simulations using the TIP4P/2005 water model will be discussed in the following; results for the SPC/E model are shown in the Supplementary Material.

*4.2 Partial radial distribution functions*

As simulated total structure factors reproduce the measured data satisfactorily; the simulations are thus in some sense validated and the PRDFs may be considered as representative. Note that since the number of atom types in the system is 6 (methanol: C, O, methyl H denoted as H, hydroxyl H denoted as $H_M$; water: $O_W$ and $H_W$), the number of partial radial distribution functions is 21. Thus it is impossible to separate the partials using only these two experimental datasets, without numerous further assumptions.

The temperature dependence of the PRDFs related to H-bonding is shown for $x_M = 0.7$ in Figure 6, where the TIP4P/2005 water model has been used. Qualitatively the same behavior can be observed in the simulations with the SPC/E model, shown in the SM (Fig. S48). The temperature dependency for other concentrations (Figs. S49-S59), and concentration dependencies at selected temperatures (Figs. S60-S62) are displayed as well. From these pictures the following observations can be made. The first peaks in the oxygen-oxygen and oxygen-hydrogen PRDFs (around 2.75 Å and 1.8 Å, respectively) correspond to H-bonding. They become sharper upon cooling: their heights increase while their widths decrease. Only slight shifts in terms of their positions occur: these distances are about 0.02-0.04 Å shorter at 163 K than at 300 K. The first minima following these peaks are more pronounced at lower temperatures. These observations are in agreement with what was found earlier at lower methanol content [8].

Focusing on the concentration dependency (see Figs. S60-S62) similar trends are found with increasing methanol concentrations for the first maxima at all examined temperatures : their heights increase and their positions shift slightly to shorter values. Although the position of some peaks (most notably, for $O_W$-$O_W$) depend on the water model, the general behavior observed with the SPC/E potential is the same. The number of H-bonded molecules that corresponds to the integral of the curves up to the first minima is discussed in the next section.

*4.3 H-bond analysis*

H-bonded pairs are identified through a geometric definition in which the O...H bond length and the O...O-H angle of the bonded molecules must be below a limiting value: 2.5 Å and 30 degrees, respectively (see Section 3).

The temperature dependence of the average number of H-bonds per molecule ($N_{Hb}$) is shown in Figure 7. For the calculation of $N_{Hb}$ all molecules (methanol and water) were considered. The average number of H-bonds increases monotonously with decreasing temperature and tends toward the value: $2x_M+4(1-x_M)$ (this is best observable for methanol concentrations $x_M \geq 0.5$).

The number of H-bonds between different kinds of pairs of molecules is presented in Figure 8. The number of H-bonded water-water pairs ($N_{WW}$, Fig. 8a) increases with decreasing temperature in the mixtures with water concentration $x_W > 0.3$ ($x_M < 0.7$), while no significant change could be detected at lower water concentrations. Moreover, in the temperature range between 300 and 193 K, even a very slight decrease is observable in some cases. The number of H-bonded water-methanol pairs ($N_{WM}$, Fig. 8b) increases in each mixture with decreasing temperature; the increment is significant in mixtures with higher methanol contents and most pronounced around $x_M = 0.7$. (Throughout this section, the term 'pairs' will refer to 'hydrogen bonded pairs'.)

A similar trend can be observed around methanol molecules. The number of methanol-water pairs ($N_{MW}$, Fig. 8d) is rising at high and low methanol concentrations as well whilst the number of methanol-methanol pairs ($N_{MM}$, Fig. 8e) increases with decreasing temperatures only in mixtures with high methanol content ($x_M \geq 0.7$). The number of hydrogen bonds between methanol molecules decreases visibly with lowering temperatures in the concentration range between ca. 30 and 60 mol % (between $x_M = 0.3$ and 0.7) alcohol content.

Based on the above findings, we conjecture that lowering the temperature forces enhanced mixing of the two species in certain concentration and, perhaps more importantly, temperature regions. It is important to point out that below 193 K the number of like-like (particularly of water-water) HB-s starts to rise again on further cooling. It may be speculated that this is connected with some precursor of water freezing out of the liquid mixtures in these regions. This may be an explanation for the fact that in some of the mixtures the number of HB-s between unlike molecules starts to decrease below ca. 193 K (cf. Fig. 8 parts b) and d) ). Note, however, that statistical uncertainties, as well as simulated annealing related issues in MD simulations, would necessitate at least one order of magnitude longer simulation times for confirming these speculations.

According to the tendencies mentioned above, by and large, three concentration regions may be identified with respect to changes of hydrogen bonding on cooling. At low methanol concentrations, roughly $x_M \leq 0.3$, the numbers of water-water and water-methanol pairs increase similarly with decreasing temperature, while the number of methanol-methanol pairs is nearly constant. (Similar trends were observed in Ref. [8].) In the $0.3 < x_M \leq 0.7$ region the number of water-water pairs is growing more slowly than that of water-methanol pairs, whereas the number of methanol-methanol pairs is constant (or slightly decreasing). In the methanol rich region, $x_M > 0.7$, the number of water-water pairs is nearly constant, the number of water-methanol pairs is growing like in the other regions, and the number of methanol-methanol pairs is also increasing with decreasing temperature.

The number of H-bonded molecules, water and methanol together, around both water and methanol is larger at lower temperatures ($N_W = N_{WW} + N_{WM}$ and $N_M = N_{MW} + N_{MM}$, Figs. 8c and 8f). However, the behavior of the two components at the lowest investigated temperatures is different: the number of H-bonded molecules around methanol is saturating for mixtures where $x_M > 0.4$, but the number of H-bonded molecules around water is increasing significantly even at the lowest investigated temperatures in each mixture. This may be explained by remembering that water molecules are easily capable of forming four hydrogen bonds, whereas methanol molecules, due to simple steric considerations, rarely form more than two (even though methanol molecules have three H-bonding sites).

The concentration dependence of the number of H-bonded pairs at some selected temperatures are shown in Figure 9. The number of water molecules around both water and methanol ($N_{WW}$ and $N_{MW}$, Figs. 9a and 9c) increases as the methanol concentration decreases. The shape of the curves is not linear: they both differ from linear upward, i.e. the number of hydrogen bonds to water molecules is higher than it would be if proportional to the concentration. At 300 K the number of H-bonded methanol molecules around water and methanol ($N_{WM}$ and $N_{MM}$, Figs. 9b and 9d) also show non-linear concentration dependence, but the deviation from linearity is negative for these curves. The same behavior was found at 300 K previously [14] by MD simulations using the SPC/E water model. As temperature decreases, the shape of the curves changes slightly as the number of H-bonds increases. The most significant change can be observed in terms of the number of methanol molecules around water ($N_{WM}$, Fig. 9b): as temperature is decreasing the deviation from linearity becomes at first less negative, and at 193 K it even switches to positive. These observations suggest that (at least at high methanol concentrations) preference of the H-bonded mixed pairs is increasing with decreasing temperature. This is in line with the observation made above, i.e., that decreasing temperature enhances mixing in certain concentration ranges.

## 5. Conclusions

Synchrotron X-ray and neutron diffraction experiments have been conducted, as a function of decreasing temperature, over the entire composition range of methanol-water liquid mixtures. Molecular dynamics simulations have been used for the interpretation of experimental data. The all atom OPLS/AA force field model for methanol has been combined with both the SPC/E and TIP4P/2005 water potentials. Based on results obtained from these investigations, the following statements can be made:

(i) The TIP4P/2005 water model is found to be somewhat more successful in reproducing measured X-ray and neutron diffraction data in the reciprocal space.
(ii) Hydrogen bond related partial radial distribution functions, as calculated from simulation trajectories, show a sharpening of the first (and, wherever well distinguishable, second) maxima and minima. The positions of the extrema do not change significantly.

(iii) As a general trend, the average number of hydrogen bonds increases upon cooling. This is strictly true for the 'water-all' and 'methanol-all' cases.

(iv) Interestingly, the number of hydrogen bonds between methanol molecules slightly decreases with lowering temperatures in the concentration range between ca. 30 and 60 mol % alcohol content. The same is valid for water-water hydrogen bonds above 70 mol % of methanol content, from room temperature down to 193 K.

(v) The effects of decreasing temperature is most significant for hydrogen bonding between unlike (water and methanol) molecules; this indicates that decreasing temperature enhances mixing between the constituents.

**Acknowledgments**

IP, LP and LT acknowledge financial support from the National Research, Development and Innovation Office (NKFIH), under grant no. KH 130425. IP, LP, SK and LT are grateful to the NKFIH of Hungary for making joint research possible under mobility grant No. TÉT_16-1-2016-0202. Synchrotron radiation experiments were performed on the BL04B2 of SPring-8, with the approval of the Japan Synchrotron Radiation Research Institute (JASRI) (Proposal Nos. 2017B1246 and 2018A1132). Neutron diffraction measurements were carried out on the 7C2 diffractometer at the Laboratoire Léon Brillouin (LLB), under Proposal id. 52/2017. LT is grateful for a János Bolyai Research Scholarship of the Hungarian Academy of Sciences. For the careful preparation of the mixtures Ms. A. Szuja (Centre for Energy Research, Hungary) is gratefully acknowledged.

**Tables**

Table 1. Investigated methanol-water mixtures: nominal and exact compositions, temperatures examined in X-ray and neutron diffraction measurements.

| Nominal methanol content, $x_M$ | Measured methanol content of samples investigated by XRD/ND [mol%] | Temperature points examined by XRD [K] | Temperature points examined by ND [K] |
|---|---|---|---|
| 0.1 | 10.02/10 | 300, 263 | 300, 263 |
| 0.2 | 20.037/20.01 | -- | 300, 268, 253, 243 |
| 0.3 | 30.013/30 | -- | 300, 268, 253, 243, 233, 223 |
| 0.4 | 39.986/39.96 | 300, 233, 213 | 300, 268, 243, 233, 223, 213 |
| 0.5 | 50.068/50.06 | 300, 233 | 300, 268, 243, 233, 213, 203, 193 |
| 0.5442 | 54.42 | 300, 233 | -- |
| 0.6 | 60.051/60.12 | 300, 233, 178 | 300, 268, 243, 213, 193, 178 |
| 0.7 | 70.011/70.04 | 300, 233, 178, 163 | 300, 268, 233, 203, 178, 163 |
| 0.7337 | 73.37 | 300, 233, 178 | -- |
| 0.8 | 80.102/79.95 | 300, 233, 178, 163 | 300, 268, 233, 203, 193, 178, 163 |
| 0.9 | 90.017 | 300, 233, 178, 163 | -- |
| 1.0 | 100 | 300, 233, 178, 163 | 300, 268, 233, 203, 193, 178, 163 |

**Figures**

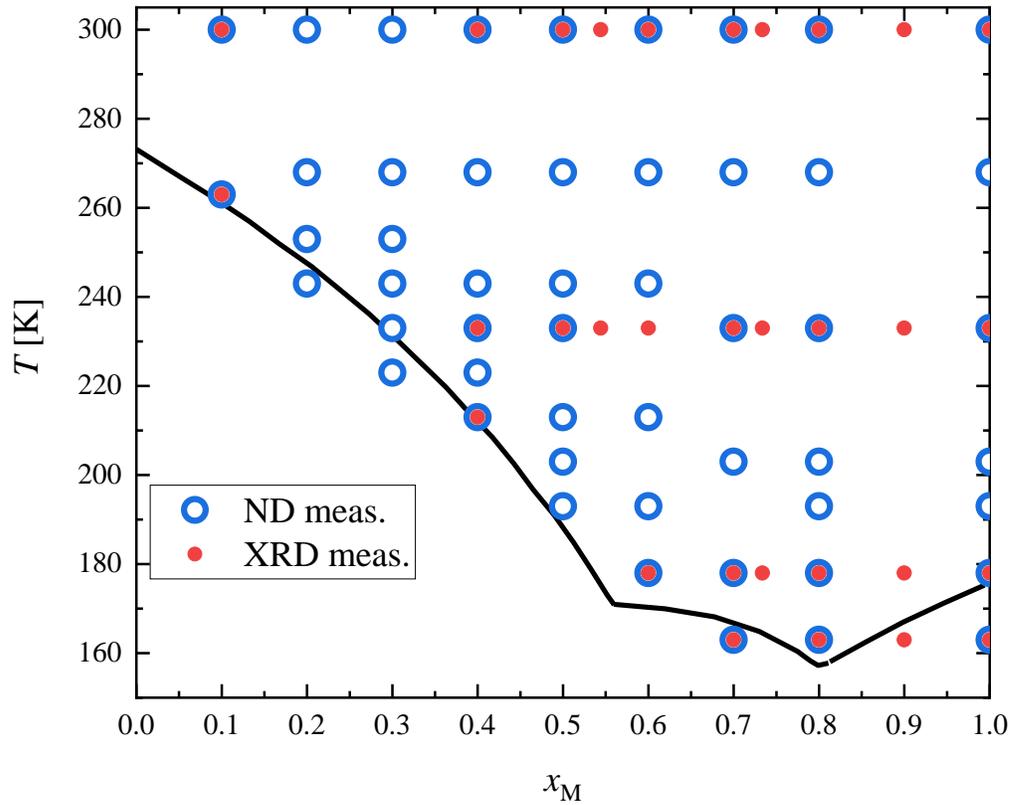

**Figure 1** Phase diagram for methanol-water mixtures at $p$ = 1 bar [71,72]. The thick curve is the solid-liquid coexistence curve. Temperatures examined by X-ray diffraction (solid red circles) and neutron diffraction (open blue circles) at different methanol concentrations of methanol-water mixtures are also shown.

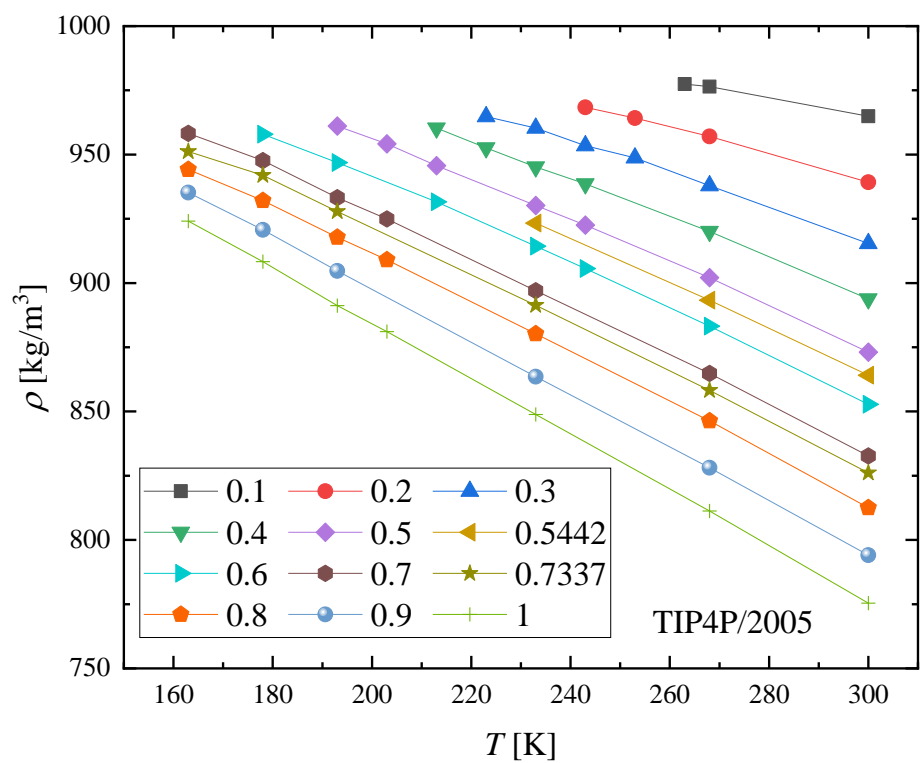

**Figure 2** Temperature dependence of densities of methanol-water mixtures at 1 bar obtained by MD simulations using the TIP4P/2005 water model.

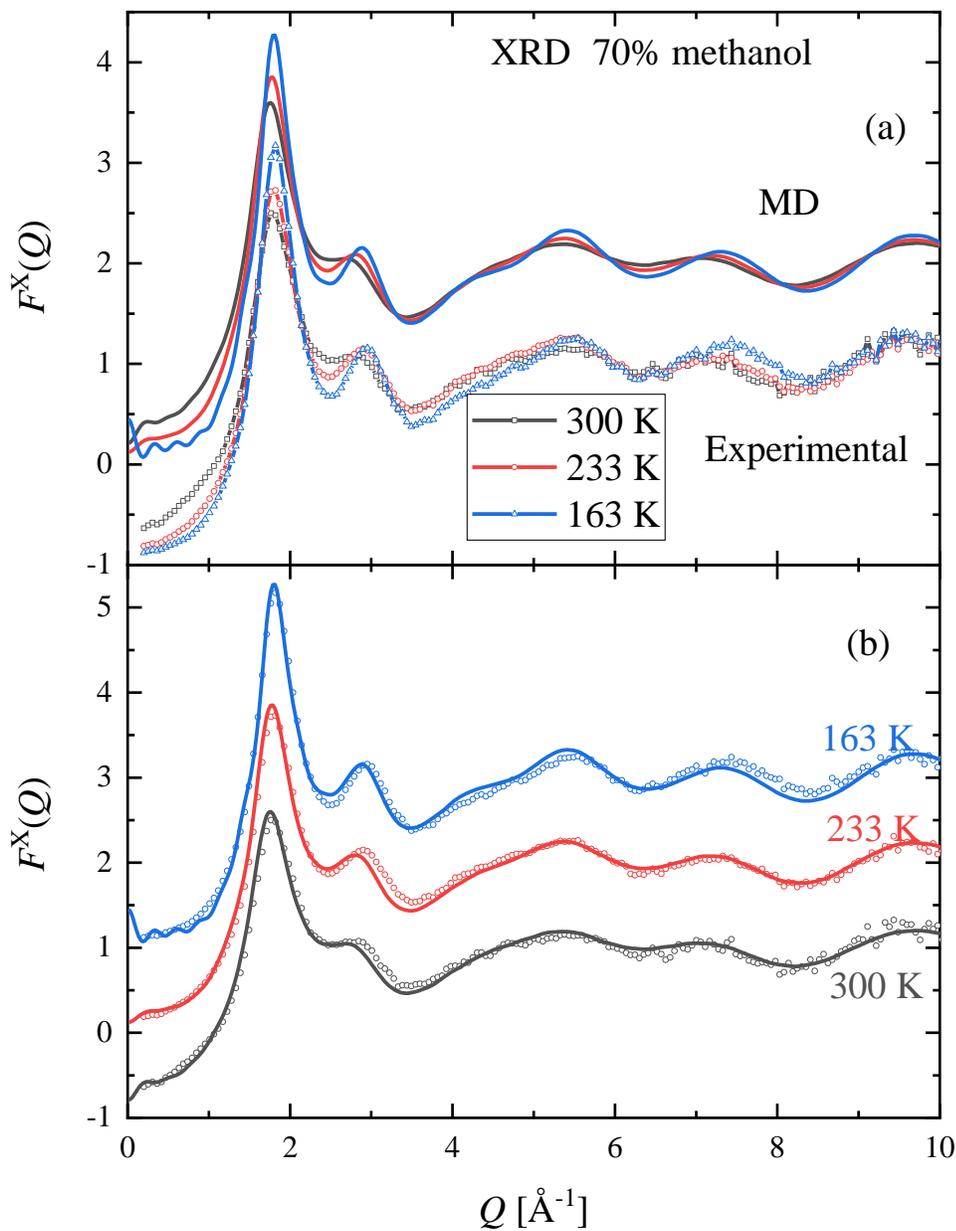

**Figure 3** Temperature dependence of measured (symbols) and simulated (lines) XRD structure factors for the methanol-water mixture with 70 mol % methanol. (a) Comparison of (a) trends observed upon cooling, (b) measured and simulated curves at three selected temperatures: 300 K (black lines and symbols), 233 K (red) and 163 K (blue). The simulated curves were obtained by using the TIP4P/2005 water model. (The curves are shifted by one unit for clarity.)

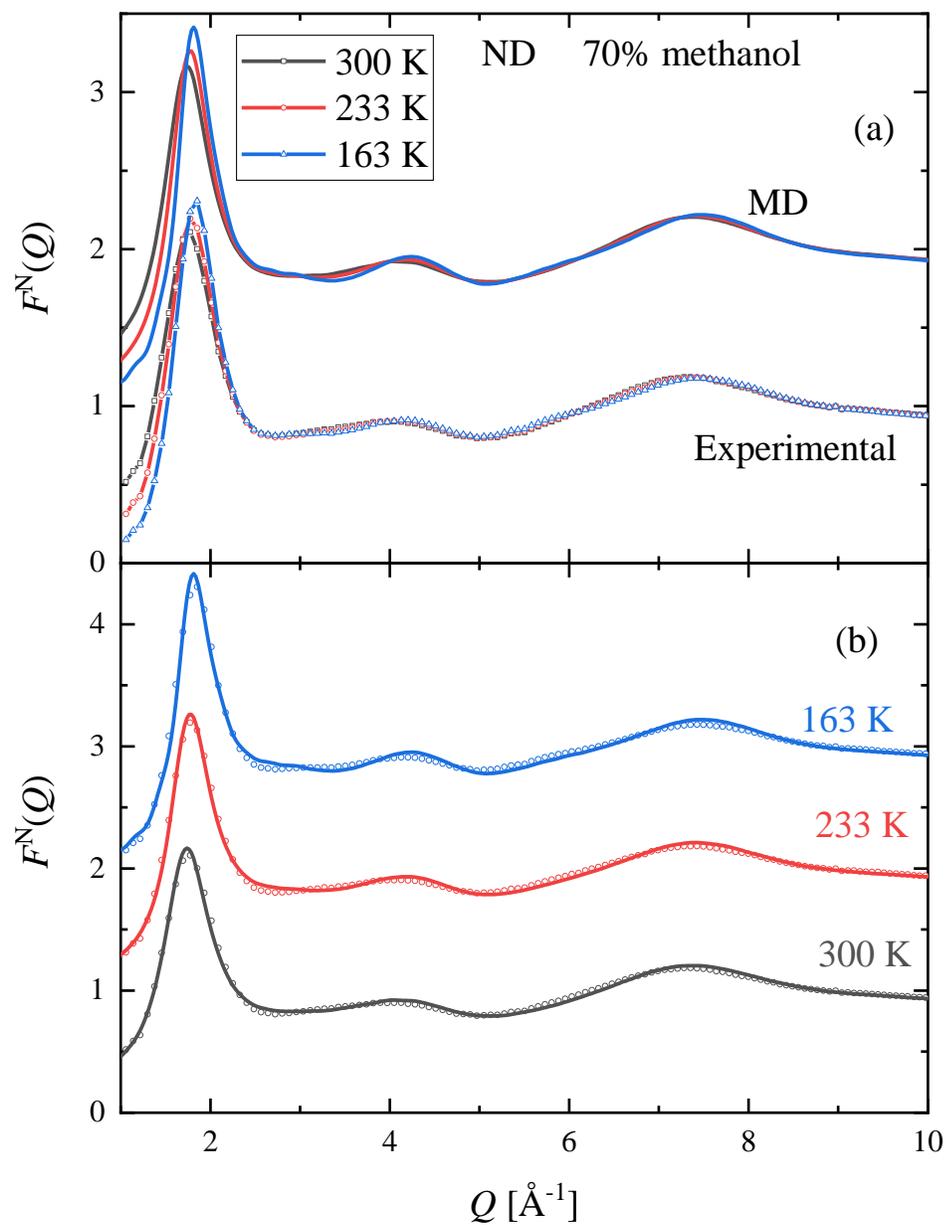

**Figure 4** Temperature dependence of measured (symbols) and simulated (lines) ND structure factors for the methanol-water mixture with 70 mol % methanol. Comparison of (a) trends observed upon cooling, (b) measured and simulated curves at three selected temperatures: 300 K (black lines and symbols), 233 K (red) and 163 K (blue). Simulated curves were obtained using the TIP4P/2005 water model. (The curves are shifted for clarity.)

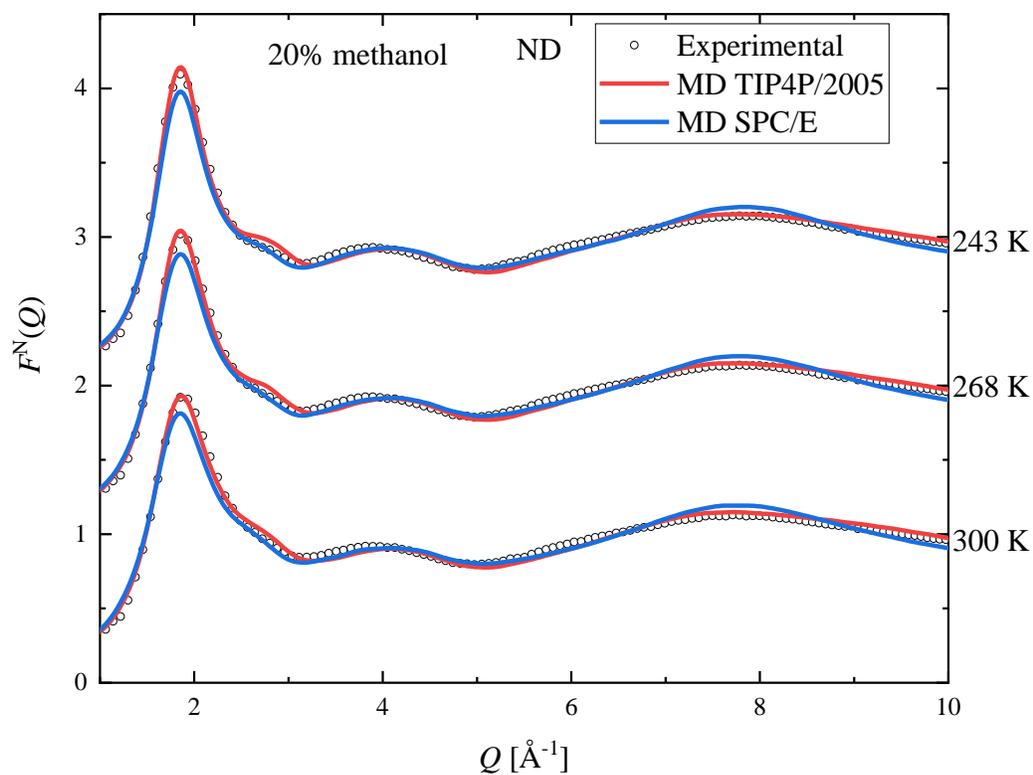

**Figure 5** Comparison of ND structure factors obtained from experiments (symbols) and simulations using TIP4P/2005 (red lines) and SPC/E (blue lines) water models for the methanol-water mixture with 20 mol% methanol, at three selected temperatures (300 K, 268 K and 243 K). (The curves are shifted for clarity.)

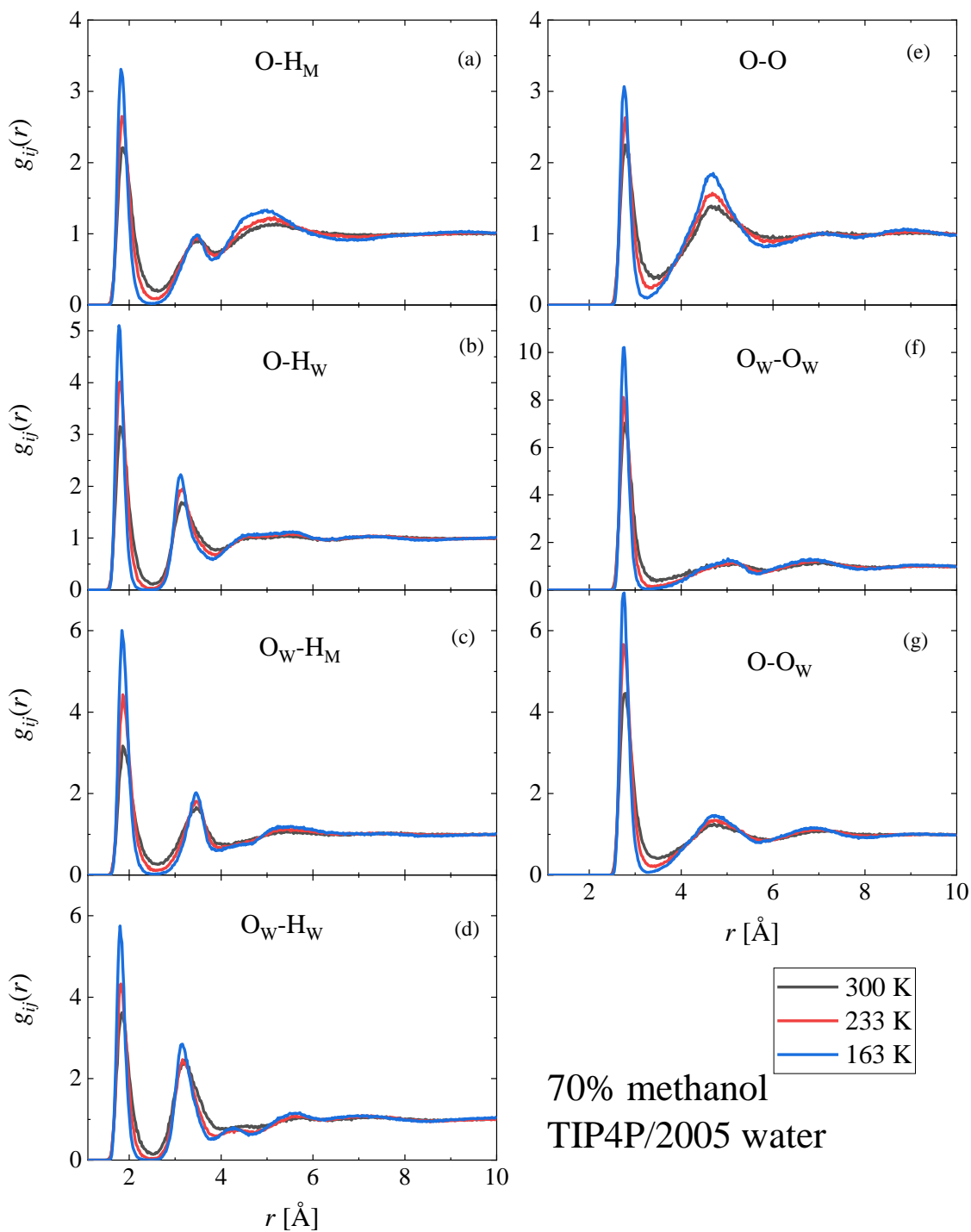

**Figure 6** Temperature dependence of simulated partial radial distribution functions of the methanol-water mixture with 70 mol % methanol. The H-bonding related partials are shown: (a) methanol O (denoted as O) – hydroxyl H of methanol (denoted as $H_M$), (b) methanol O – water H (denoted as $H_W$), (c) water O (denoted as $O_W$) – hydroxyl H of methanol, (d) water O – water H, (e) methanol O – methanol O, (f) water O – water O, (g) methanol O – water O. The curves were obtained using the TIP4P/2005 water model.

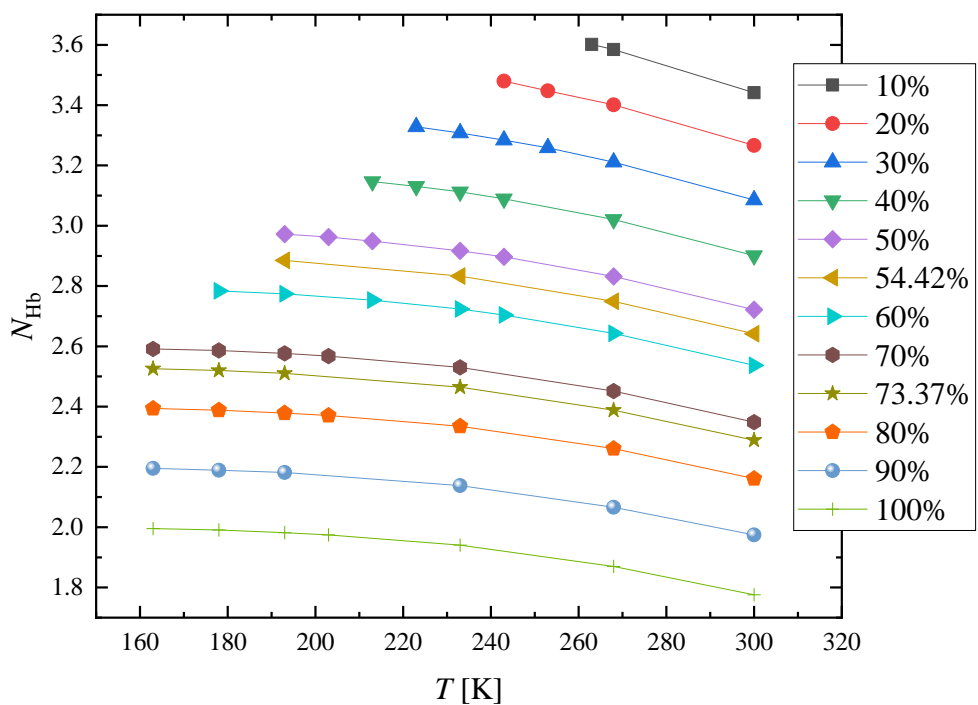

**Figure 7** Average number of hydrogen bonds per molecule (either water or methanol) in methanol-water mixtures as a function of temperature at different concentrations, obtained from MD simulations using the TIP4P/2005 water model.

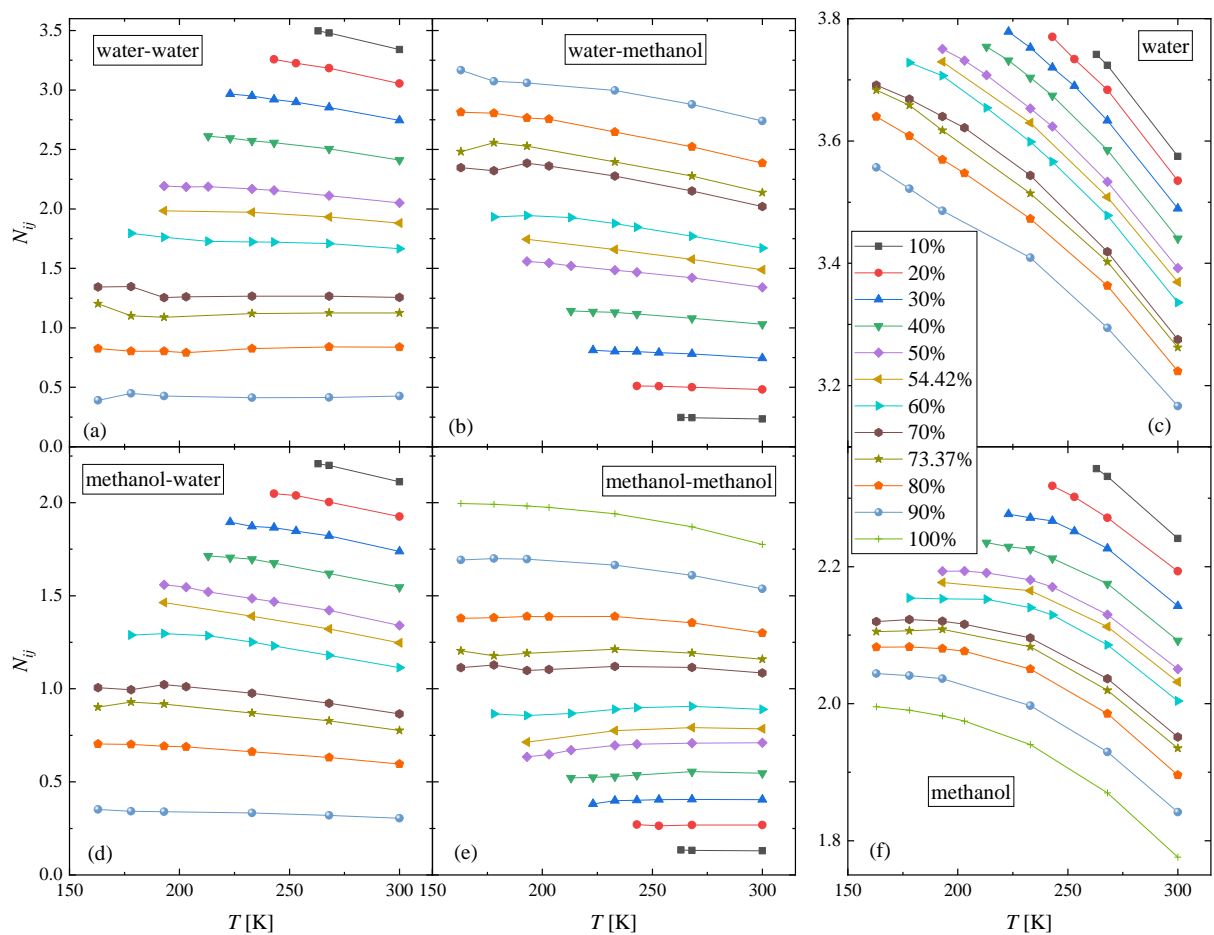

**Figure 8** Temperature dependence of the number of hydrogen bonds at different concentrations, as obtained from MD simulations using the TIP4P/2005 water model: (a) average number of H-bonded water molecules around water, (b) average number of H-bonded methanol molecules around water, (c) average number of H-bonded (water and methanol) molecules around water, (d) average number of H-bonded water molecules around methanol, (e) average number of H-bonded methanol molecules around methanol, (f) average number of H-bonded (water and methanol) molecules around methanol.

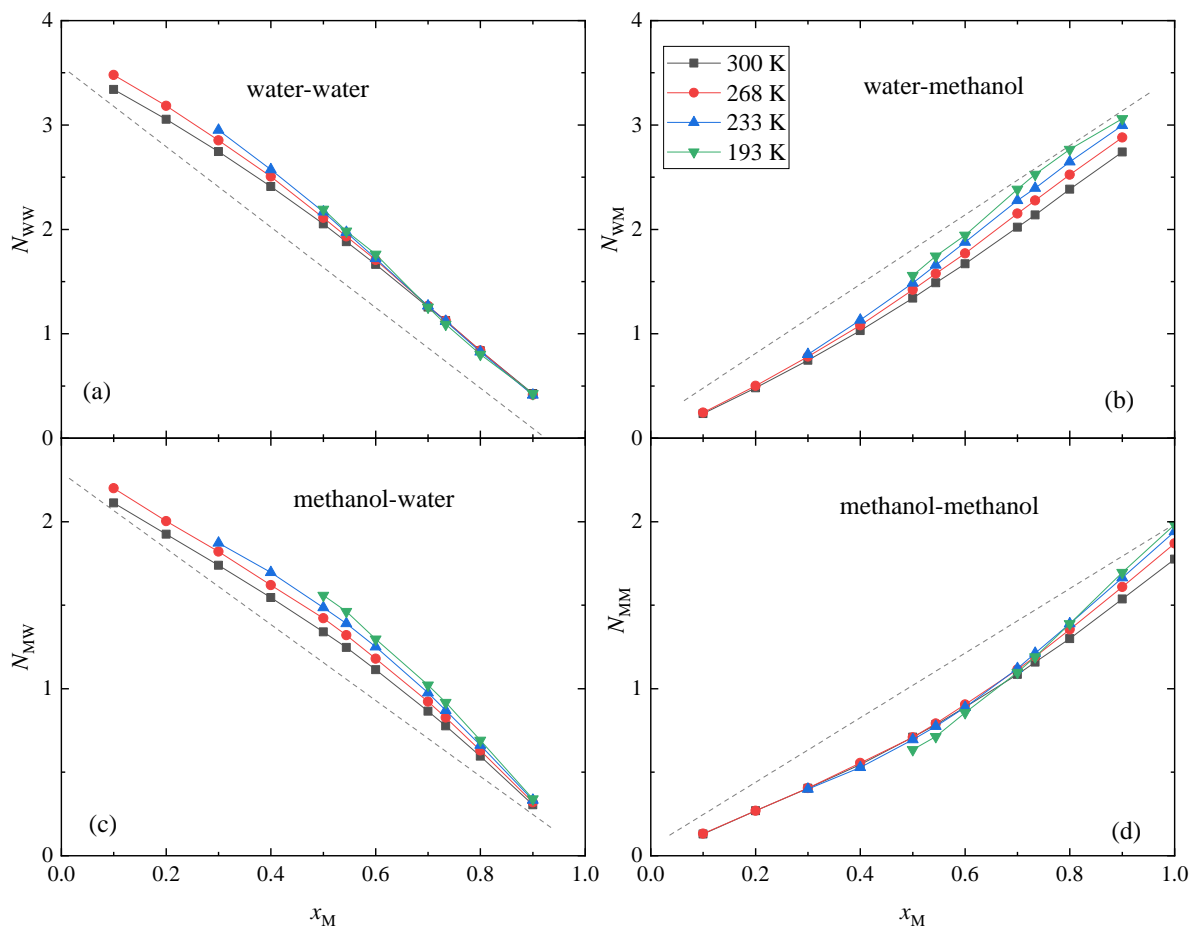

**Figure 9** Concentration dependence of the number of hydrogen bonds at different temperatures, as obtained from MD simulations using the TIP4P/2005 water model: (a) average number of H-bonded water molecules around water, (b) average number of H-bonded methanol molecules around water, (c) average number of H-bonded water molecules around methanol, (d) average number of H-bonded methanol molecules around methanol. The dashed linear lines are only guides to the eye.

# Supplementary Material

# Temperature-dependent structure of methanol-water mixtures on cooling: X-ray and neutron diffraction and molecular dynamics simulations

Ildikó Pethes[1,a], László Pusztai[a,b], Koji Ohara[c], Shinji Kohara[d], Jacques Darpentigny[e], László Temleitner[a]

[a]Wigner Research Centre for Physics, Konkoly Thege út 29-33., H-1121 Budapest, Hungary
[b]International Research Organization for Advanced Science and Technology (IROAST), Kumamoto University, 2-39-1 Kurokami, Chuo-ku, Kumamoto 860-8555, Japan
[c]Diffraction and Scattering Division, Japan Synchrotron Radiation Research Institute (JASRI/SPring-8), 1-1-1 Kouto, Sayo-cho, Sayo-gun, Hyogo 679-5198, Japan
[d]Research Center for Advanced Measurement and Characterization, National Institute for Materials Science (NIMS), 1–1–1 Kouto, Sayo-cho, Sayo-gun, Hyogo 679–5148, Japan
[e]Laboratoire Léon Brillouin, CEA-Saclay 91191 Gif sur Yvette Cedex France

---

1 Corresponding author: e-mail: pethes.ildiko@wigner.hu

**Molecular dynamics simulations**

Classical molecular dynamics (MD) simulations were performed with the GROMACS software package (version 2018.2) [1]. Simulations were conducted in cubic simulation boxes with periodic boundary conditions. The total number of methanol and water molecules in the simulation boxes was 2000. The initial box sizes were determined from the room temperature densities [2]. At first, calculations were run at constant pressure and temperature (NPT ensemble), at each experimentally examined temperature point (by decreasing from 300 K). Densities applied later were calculated from these NPT simulation runs. In the second set of the simulations the temperature and volume were kept at constant (NVT ensemble), using the box sizes calculated by the densities obtained from the NPT simulations.

The all atom OPLS-AA [3] force field was applied for methanol while for water two models, the SPC/E [4] and the TIP4P/2005 [5], were tested. Non-bonded interactions were described by the 12-6 Lennard-Jones interaction and the Coulomb potential (see Eq. 1):

$$V_{ij}^{NB}(r_{ij}) = \frac{1}{4\pi\varepsilon_0}\frac{q_i q_j}{r_{ij}} + 4\varepsilon_{ij}\left[\left(\frac{\sigma_{ij}}{r_{ij}}\right)^{12} - \left(\frac{\sigma_{ij}}{r_{ij}}\right)^{6}\right], \qquad (1)$$

where $r_{ij}$ is the distance between particles $i$ and $j$, $q_i$ and $q_j$ are the partial charges on these particles, $\varepsilon_0$ is the vacuum permittivity, and $\varepsilon_{ij}$ and $\sigma_{ij}$ represent the energy and distance parameters of the LJ potential. The LJ parameters ($\varepsilon_{ii}$ and $\sigma_{ii}$) and the partial charges applied ($q_i$) to the different atoms are collected in Table S1. The $\varepsilon_{ij}$ and $\sigma_{ij}$ parameters between unlike atoms are calculated as the geometric average of the homoatomic parameters (geometric combination rule, in accordance with the OPLS/AA force field).

Intramolecular non-bonded interactions between first and second neighbor atoms were neglected, whereas between third neighbors (atoms separated by 3 bonds, $H_C – H_O$ bonds) they were reduced by a factor of 2. The intramolecular (or bonded) forces considered here are the bond-stretching (2-body), angle bending (3-body) and the dihedral angle torsion (4-body) interactions. Bond lengths in methanol molecules were fixed using the LINCS [6] algorithm, while bond angles and torsional angles were flexible. The rigid water geometry was handled by the SETTLE algorithm [7]. Bond lengths, equilibrium angles and force constants are given in Table S2.

The smoothed particle-mesh Ewald (SPME) method [8,9] was used for treating the Coulomb interactions, using a 20 Å cutoff (15 Å for methanol concentrations 0.1 and 0.2 molar fraction, due to the smaller box sizes) in real space. Non-bonded LJ interactions were cut-off at 20 Å (15 Å for methanol concentration of 10 and 20 mol%), with added long-range corrections to energy and pressure [10].

Initial configurations for the NPT simulations at $T = 300$ K were obtained by placing the molecules into the simulation box randomly, following an energy minimization using the steepest-descent method (random configuration method). At lower temperatures the final configuration of the previous temperature point was used as initial configuration. The equations of motion were integrated via the leapfrog algorithm, the time step was 2 fs. At each temperature, at first a short (0.2 ns) NVT run was performed using the Berendsen thermostat [11], with $\tau_T = 0.1$, for relaxing the system to the target temperature. After that, the NPT ensemble was used. The temperature was kept constant by the Nose-Hoover thermostat [12,13], with $\tau_T = 2.0$, while the pressure was kept at $p = 10^5$ Pa, by the Parrinello-Rahman barostat [14,15], using a coupling constant of $\tau_p = 2.0$. After a 2 ns equilibration period, a 2 ns production run was completed, from which the densities were calculated.

Initial configurations for the NVT simulations were either obtained by the random configuration method or were adopted from the corresponding NPT simulations. (The latter method was used for low temperatures and high methanol content mixtures, in order to avoid artifacts resulting from close packing and low mobility of molecules). The leap-frog algorithm was used again, with the same time step as for the NPT runs (2 fs). Two equilibration runs were performed before the production run: during the first, short one (0.2 ns), the Berendsen thermostat was used; after that, and also for the production run, the Nose-Hoover thermostat was activated, with the same coupling constants as before.

Trajectories were saved in every 200 ps, for the duration of 20 ns: in this way, 101 configurations were used for further analyzes. Partial radial distribution functions (PRDF, $g_{ij}(r)$) were calculated from the collected configurations, by the 'gmx_rdf' programme of the GROMACS software. The model structure factor can be obtained from the PRDFs, according to the Faber-Ziman formalism [16], by the following equations:

$$S_{ij}(Q) - 1 = \frac{4\pi\rho_0}{Q} \int_0^\infty r\big(g_{ij}(r) - 1\big)\sin(Qr)\mathrm{d}r, \qquad (2)$$

where $Q$ is the amplitude of the scattering vector, and $\rho_0$ is the average number density.

The XRD ($F^X(Q)$) and ND ($F^N(Q)$) total structure factors can be composed from the partial structure factors $S_{ij}(Q)$ as:

$$F^X(Q) = \sum_{i \leq j} w_{ij}^X(Q) S_{ij}(Q) \qquad (3)$$

and

$$F^N(Q) = \sum_{i \leq j} w_{ij}^N S_{ij}(Q), \qquad (4)$$

where $w_{ij}^{X,N}$ denotes the X-ray and neutron scattering weights. For X-rays it is given by equation (5):

$$w_{ij}^X(Q) = (2 - \delta_{ij}) \frac{c_i c_j f_i(Q) f_j(Q)}{\sum_{ij} c_i c_j f_i(Q) f_j(Q)}. \qquad (5)$$

Here $\delta_{ij}$ is the Kronecker delta, $c_i$ denotes atomic concentrations, $f_i(Q)$ is the atomic form factor. The neutron weight factors are:

$$w_{ij}^N = (2 - \delta_{ij}) \frac{c_i c_j b_i b_j}{\sum_{ij} c_i c_j b_i b_j}, \qquad (6)$$

where $b_i$ is the coherent neutron scattering length.

The total structure factors obtained from simulations were compared with the measured curves by calculating the goodness-of-fit (R-factor) values:

$$R_X = \frac{\sqrt{\sum_i \left(F_{mod}^X(Q_i) - F_{exp}^X(Q_i)\right)^2}}{\sqrt{\sum_i \left(F_{exp}^X(Q_i)\right)^2}} \qquad (7)$$

$$R_N = \frac{\sqrt{\sum_i \left(F_{mod}^N(Q_i) - F_{exp}^N(Q_i)\right)^2}}{\sqrt{\sum_i \left(F_{exp}^N(Q_i)\right)^2}} \qquad (8)$$

where $Q_i$ denote the experimental points, 'mod' indicates the simulated and 'exp' the experimental curves.

**Tables**

**Table S1.** Non-bonded force field parameters. In the methanol molecule the H atoms of the hydroxyl and methyl groups are denoted as $H_M$ and H, respectively. In the TIP4P/2005 water model there is a fourth (virtual) site (M). It is situated along the bisector of the $H_W$-$O_W$-$H_W$ angle and coplanar with the oxygen and hydrogens. The negative charge is placed on site M.

| atom | $q$ [e] | $\sigma_{ii}$ [nm] | $\varepsilon_{ii}$ [kJ mol$^{-1}$] |
|---|---|---|---|
| OPLS/AA methanol [3] | | | |
| C | 0.145 | 0.35 | 0.276144 |
| O | -0.683 | 0.312 | 0.71128 |
| H | 0.04 | 0.25 | 0.12552 |
| $H_M$ | 0.418 | 0 | 0 |
| SPC/E water [4] | | | |
| $O_W$ | -0.8476 | 0.3166 | 0.6502 |
| $H_W$ | 0.4238 | 0 | 0 |
| TIP4P/2005 water [5] | | | |
| $O_W$ | 0 | 0.3159 | 0.7749 |
| $H_W$ | 0.5564 | 0 | 0 |
| M | -1.1128 | 0 | 0 |

**Table S2.** Equilibrium bond lengths, angle bending parameters, and dihedral angle torsion force constants. Bond angle vibrations are represented by harmonic potentials, $\theta^0_{ijk}$ is the equilibrium angle and $k^a_{ijk}$ is the force constant. Dihedral torsion angles in the OPLS/AA force field are given as the first three terms of a Fourier series: $V(\varphi_{ijkl})=1/2(F_1(1+\cos\varphi_{ijkl})+F_2(1-\cos2\varphi_{ijkl})+F_3(1+\cos3\varphi_{ijkl}))$, where $\varphi_{ijkl}$ is the angle between the *ijk* and *jkl* planes. $\varphi_{ijkl} = 0$ corresponds to the 'cis' conformation (*i* and *l* are on the same side).

| Bond type | Bond length [nm] | |
|---|---|---|
| C-H | 0.109 | |
| C-O | 0.141 | |
| O-$H_O$ | 0.0945 | |
| $O_W$-$H_W$ SPC/E | 0.1 | |
| $O_W$-$H_W$ TIP4P/2005 | 0.09572 | |
| $O_W$-M TIP4P/2005 | 0.01546 | |
| Angle type | $\theta^0_{ijk}$ [degree] | $k^a_{ijk}$ [kJ mol$^{-1}$ rad$^{-2}$] |
| H-C-H | 107.8 | 276.144 |
| C-O-$H_O$ | 108.5 | 460.24 |
| H-C-O | 109.5 | 292.88 |
| $H_W$-$O_W$-$H_W$ SPC/E | 109.47 | -- (rigid) |
| $H_W$-$O_W$-$H_W$ TIP4P/2005 | 104.52 | -- (rigid) |
| Dihedral type | $F_1$ [kJ mol$^{-1}$] | $F_2$ [kJ mol$^{-1}$] $F_3$ [kJ mol$^{-1}$] |
| H-C-O-$H_O$ | 0 | 0             1.8828 |

**Table S3.** Densities (in kg/m$^3$) of the methanol-water mixtures obtained in MD simulations using the TIP4P/2005 water model.

| | Temperature [K] | | | | | | | | | | | |
|---|---|---|---|---|---|---|---|---|---|---|---|---|
| $x_M$ | 300 | 268 | 263 | 253 | 243 | 233 | 223 | 213 | 203 | 193 | 178 | 163 |
| 0.1 | 964.9 | 976.4 | 977.4 | | | | | | | | | |
| 0.2 | 939.2 | 957.1 | | 964.2 | 968.4 | | | | | | | |
| 0.3 | 915.4 | 937.9 | | 948.7 | 953.5 | 960.3 | 964.8 | | | | | |
| 0.4 | 893.8 | 920.1 | | | 938.6 | 945.2 | 952.6 | 960.3 | | | | |
| 0.5 | 873.1 | 902.1 | | | 922.5 | 930.2 | | 945.7 | 954.2 | 961.1 | | |
| 0.5442 | 864.1 | 893.3 | | | | 923.3 | | | | 955.6 | | |
| 0.6 | 852.8 | 883.2 | | | 905.6 | 914.4 | | 931.6 | | 946.9 | 957.9 | |
| 0.7 | 832.7 | 864.8 | | | | 897.0 | | | 924.9 | 933.2 | 947.7 | 958.3 |
| 0.7337 | 826.1 | 858.2 | | | | 891.4 | | | | 927.8 | 941.9 | 951.2 |
| 0.8 | 812.5 | 846.3 | | | | 880.3 | | | 909.0 | 917.8 | 932.1 | 944.2 |
| 0.9 | 794.1 | 828.1 | | | | 863.5 | | | | 904.7 | 920.7 | 935.2 |
| 1 | 775.4 | 811.3 | | | | 848.9 | | | 881.1 | 891.2 | 908.3 | 924.1 |

**Table S4.** Densities (in kg/m$^3$) of the methanol-water mixtures obtained in MD simulations using the SPC/E water model.

| $x_M$ | Temperature [K] | | | | | | | | | | | |
|---|---|---|---|---|---|---|---|---|---|---|---|---|
| | 300 | 268 | 263 | 253 | 243 | 233 | 223 | 213 | 203 | 193 | 178 | 163 |
| 0.1 | 966.7 | 985.4 | 987.6 | | | | | | | | | |
| 0.2 | 939.6 | 962.9 | | 971.9 | 978.0 | | | | | | | |
| 0.3 | 914.1 | 941.2 | | 952.1 | 959.2 | 965.9 | 973.1 | | | | | |
| 0.4 | 890.9 | 919.8 | | | 941.0 | 949.0 | 956.1 | 963.7 | | | | |
| 0.5 | 869.2 | 900.2 | | | 922.1 | 931.4 | | 947.5 | 956.1 | 964.7 | | |
| 0.5442 | 859.8 | 891.6 | | | | 923.3 | | | | 958.2 | | |
| 0.6 | 850.8 | 881.5 | | | 904.6 | 913.6 | | 932.0 | | 949.1 | 962.1 | |
| 0.7 | 829.0 | 862.2 | | | | 896.4 | | | 924.1 | 933.7 | 947.5 | 960.5 |
| 0.7337 | 822.2 | 855.9 | | | | 890.4 | | | | 929.3 | 942.3 | 953.7 |
| 0.8 | 810.2 | 844.7 | | | | 879.2 | | | 909.0 | 919.0 | 933.5 | 946.9 |
| 0.9 | 792.1 | 827.3 | | | | 863.6 | | | | 904.8 | 919.9 | 933.6 |
| 1 | 775.4 | 811.3 | | | | 848.9 | | | 881.1 | 891.2 | 908.3 | 924.1 |

**Figures**

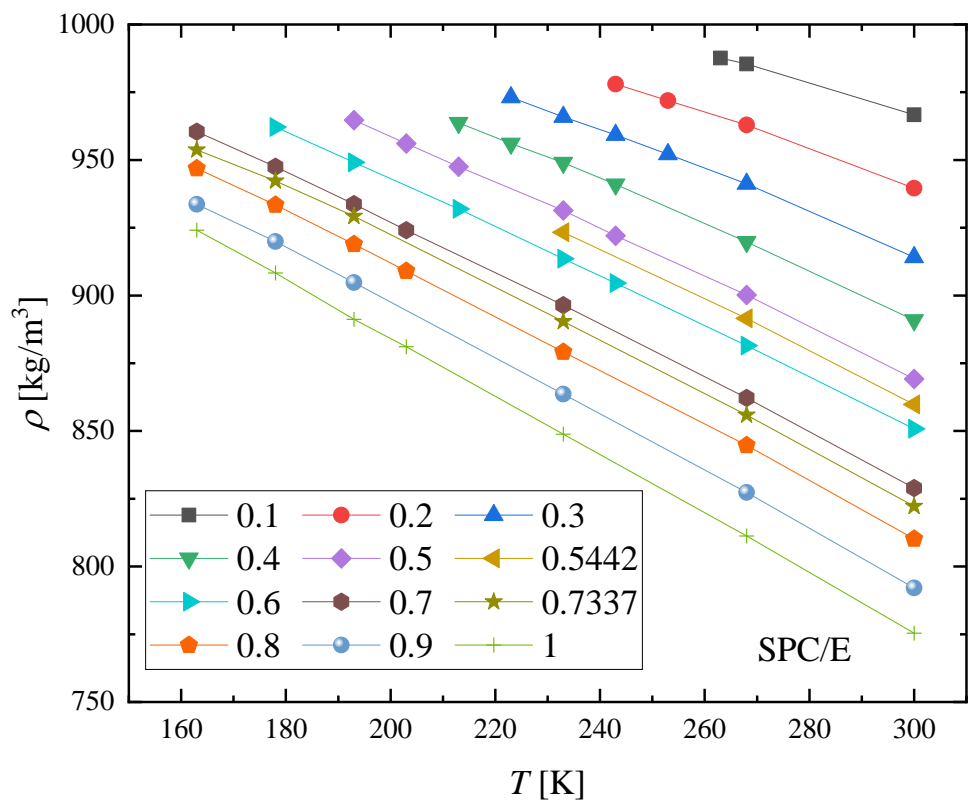

**Figure S1** Temperature dependence of densities of methanol-water mixtures at $p$ = 1 bar obtained in MD simulations using the SPC/E water model.

**Temperature dependence of the XRD total structure factors (at selected temperatures)**

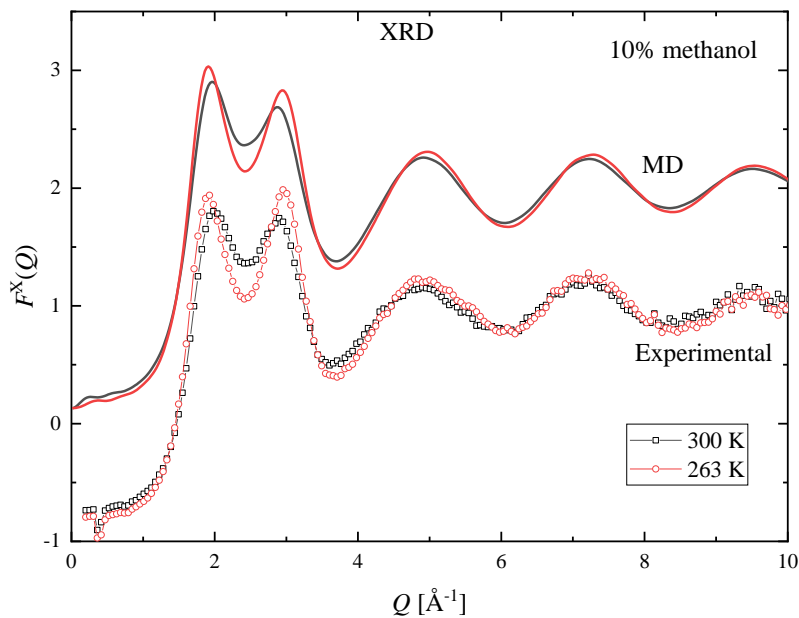

**Figure S2** Temperature dependence of measured (symbols) and simulated (lines) XRD structure factors for the methanol-water mixture with 10 mol % methanol. The simulated curves were obtained by using the TIP4P/2005 water model. (The MD curves are shifted for clarity.)

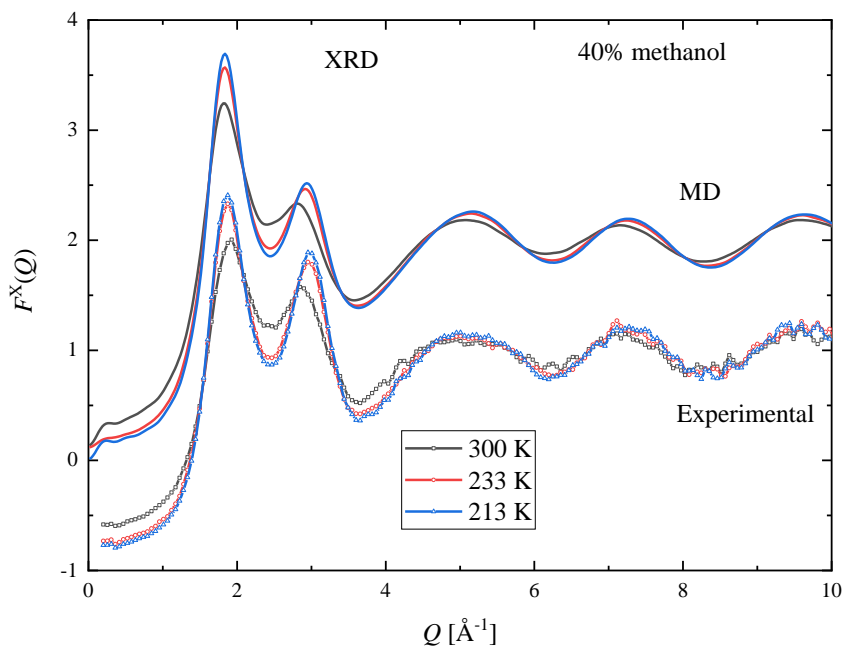

**Figure S3** Temperature dependence of measured (symbols) and simulated (lines) XRD structure factors for the methanol-water mixture with 40 mol % methanol. The simulated curves were obtained by using the TIP4P/2005 water model. (The MD curves are shifted for clarity.)

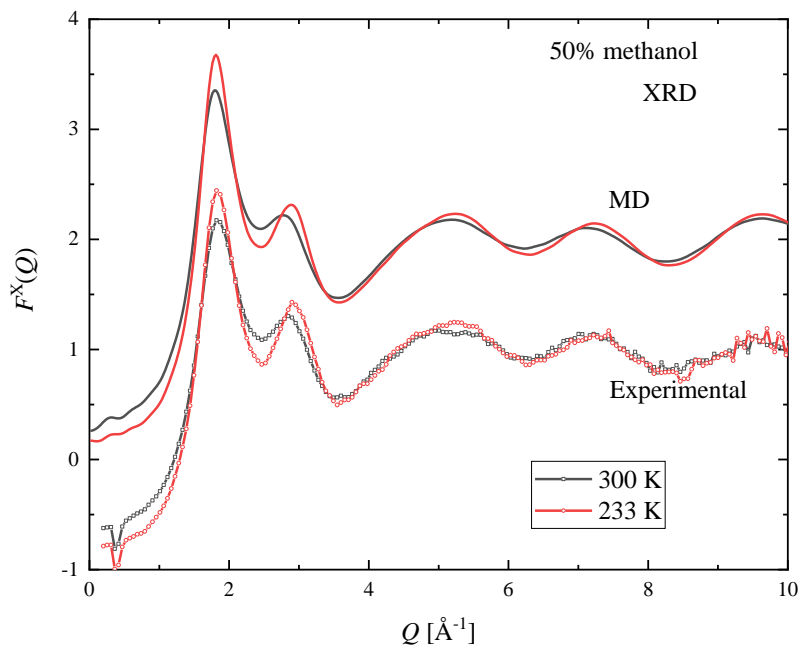

**Figure S4** Temperature dependence of measured (symbols) and simulated (lines) XRD structure factors for the methanol-water mixture with 50 mol % methanol. The simulated curves were obtained by using the TIP4P/2005 water model. (The MD curves are shifted for clarity.)

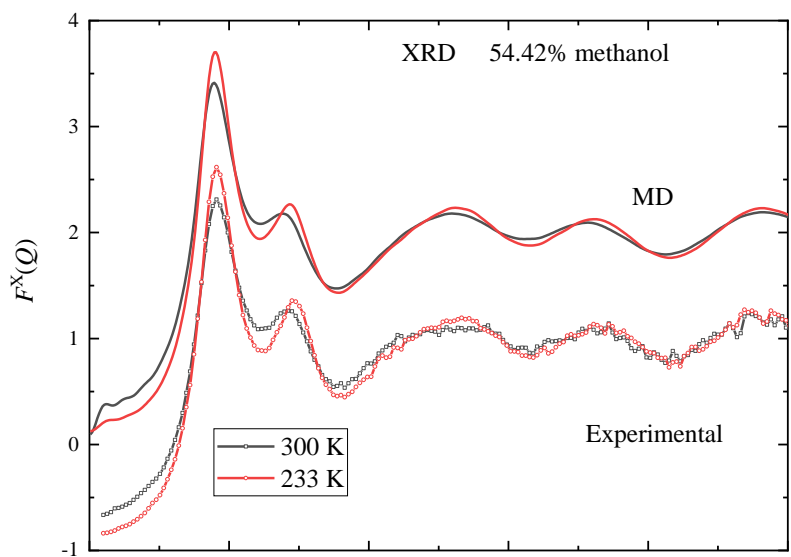

**Figure S5** Temperature dependence of measured (symbols) and simulated (lines) XRD structure factors for the methanol-water mixture with 54.42 mol % methanol. The simulated curves were obtained by using the TIP4P/2005 water model. (The MD curves are shifted for clarity.)

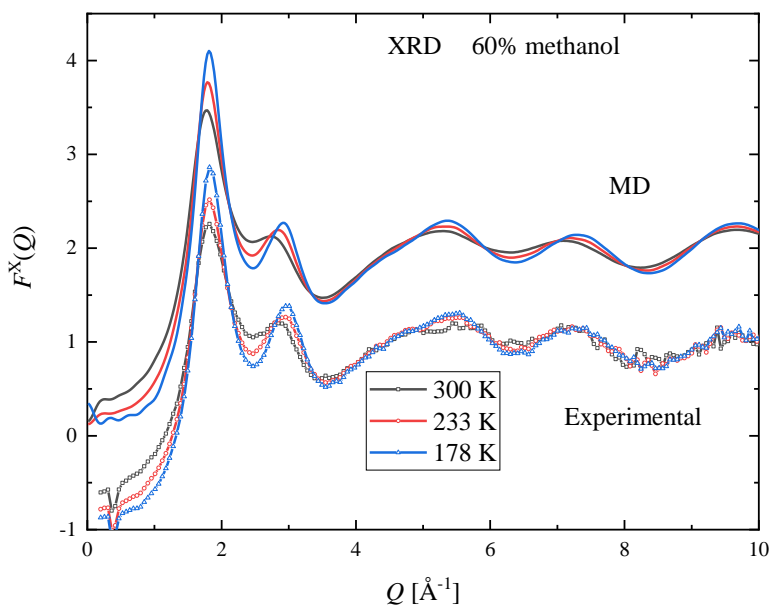

**Figure S6** Temperature dependence of measured (symbols) and simulated (lines) XRD structure factors for the methanol-water mixture with 60 mol % methanol. The simulated curves were obtained by using the TIP4P/2005 water model. (The MD curves are shifted for clarity.)

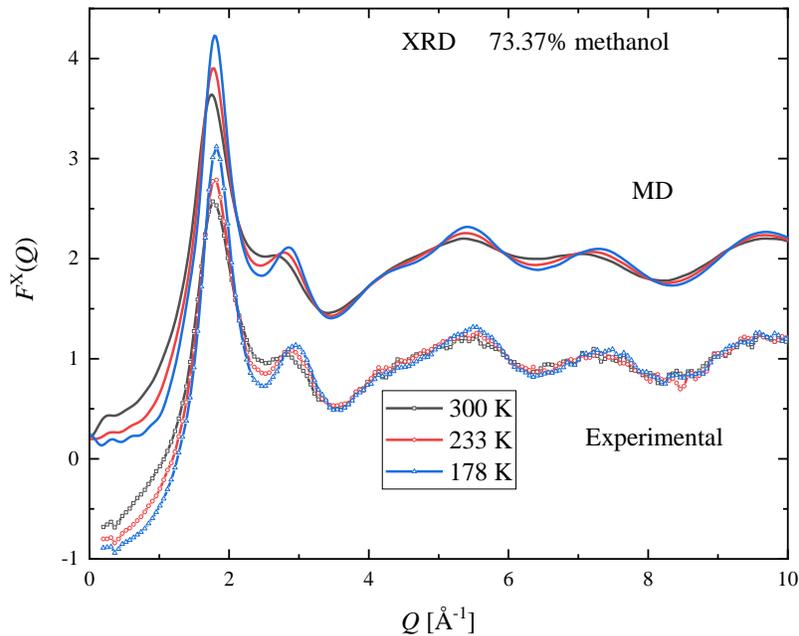

**Figure S7** Temperature dependence of measured (symbols) and simulated (lines) XRD structure factors for the methanol-water mixture with 73.37 mol % methanol. The simulated curves were obtained by using the TIP4P/2005 water model. (The MD curves are shifted for clarity.)

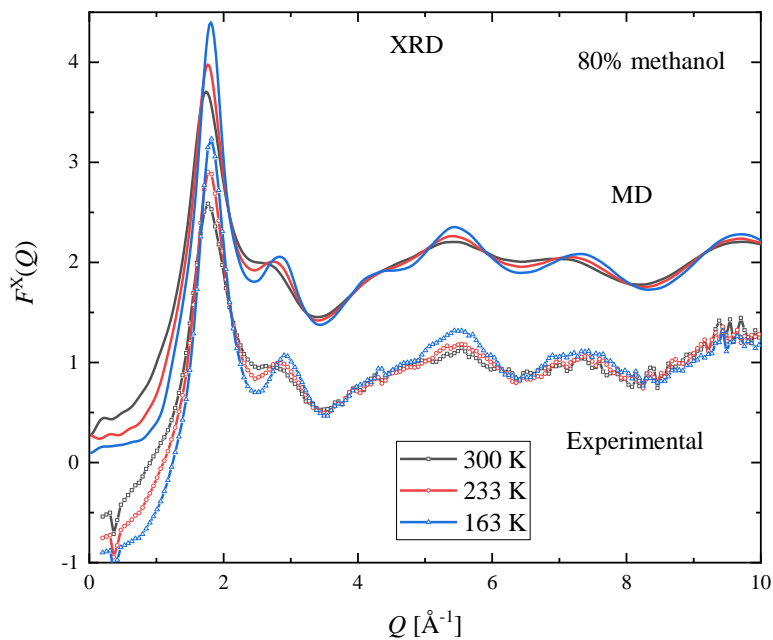

**Figure S8** Temperature dependence of measured (symbols) and simulated (lines) XRD structure factors for the methanol-water mixture with 80 mol % methanol. The simulated curves were obtained by using the TIP4P/2005 water model. (The MD curves are shifted for clarity.)

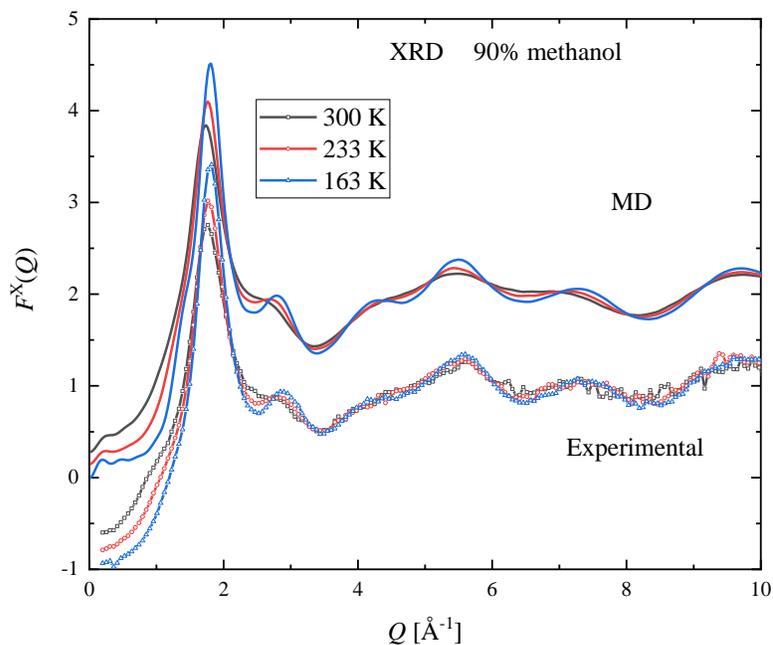

**Figure S9** Temperature dependence of measured (symbols) and simulated (lines) XRD structure factors for the methanol-water mixture with 90 mol % methanol. The simulated curves were obtained by using the TIP4P/2005 water model. (The MD curves are shifted for clarity.)

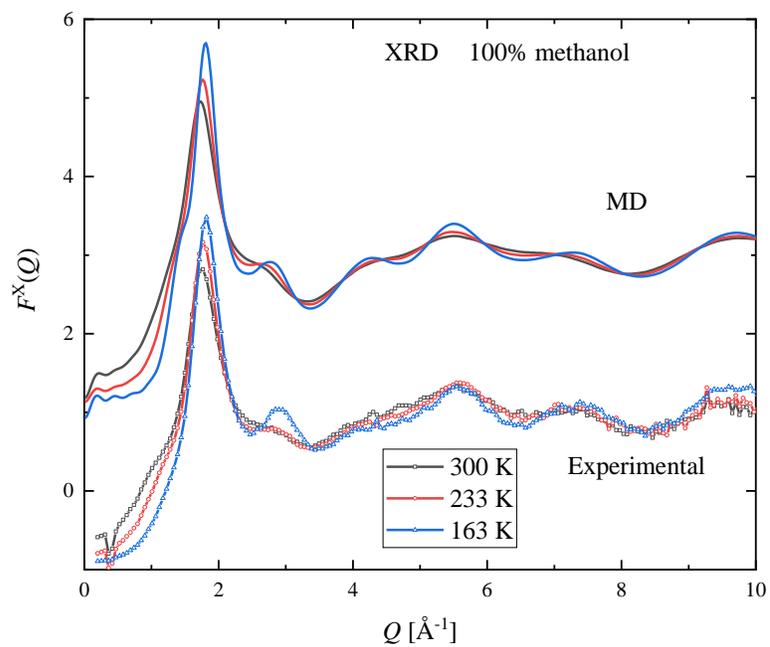

**Figure S10** Temperature dependence of measured (symbols) and simulated (lines) XRD structure factors for pure methanol. (The MD curves are shifted for clarity.)

**Temperature dependence of the ND structure factors (at selected temperatures)**

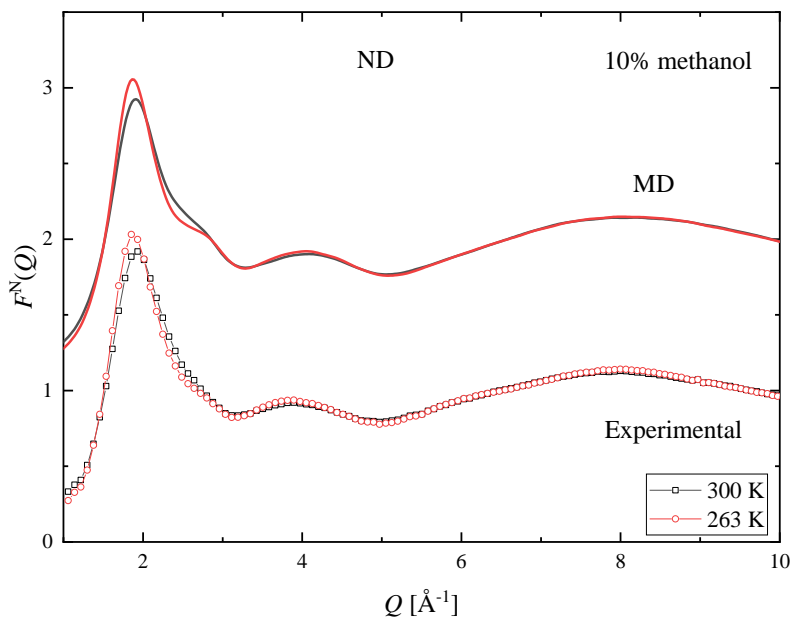

**Figure S11** Temperature dependence of measured (symbols) and simulated (lines) ND structure factors for the methanol-water mixture with 10 mol % methanol. The simulated curves were obtained by using the TIP4P/2005 water model. (The MD curves are shifted for clarity.)

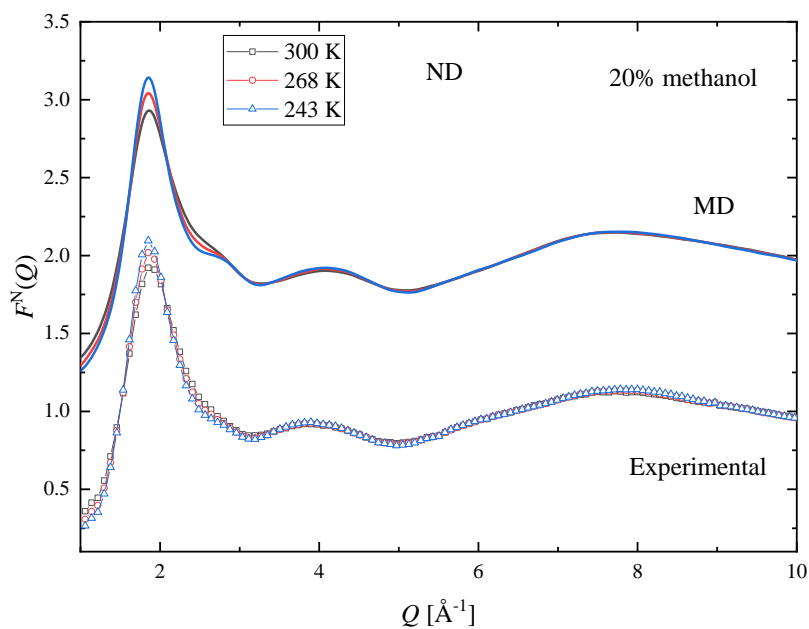

**Figure S12** Temperature dependence of measured (symbols) and simulated (lines) ND structure factors for the methanol-water mixture with 20 mol % methanol. The simulated curves were obtained by using the TIP4P/2005 water model. (The MD curves are shifted for clarity.)

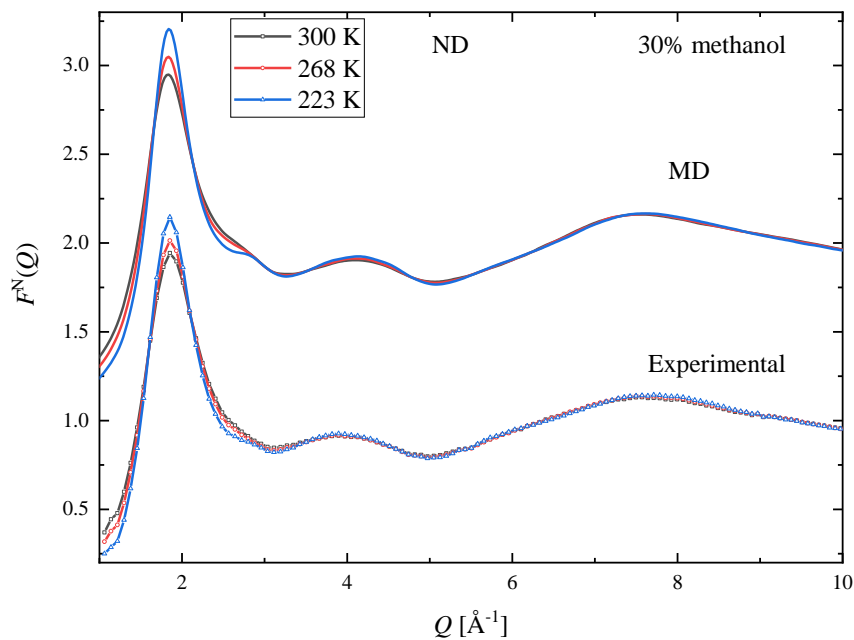

**Figure S13** Temperature dependence of measured (symbols) and simulated (lines) ND structure factors for the methanol-water mixture with 30 mol % methanol. The simulated curves were obtained by using the TIP4P/2005 water model. (The MD curves are shifted for clarity.)

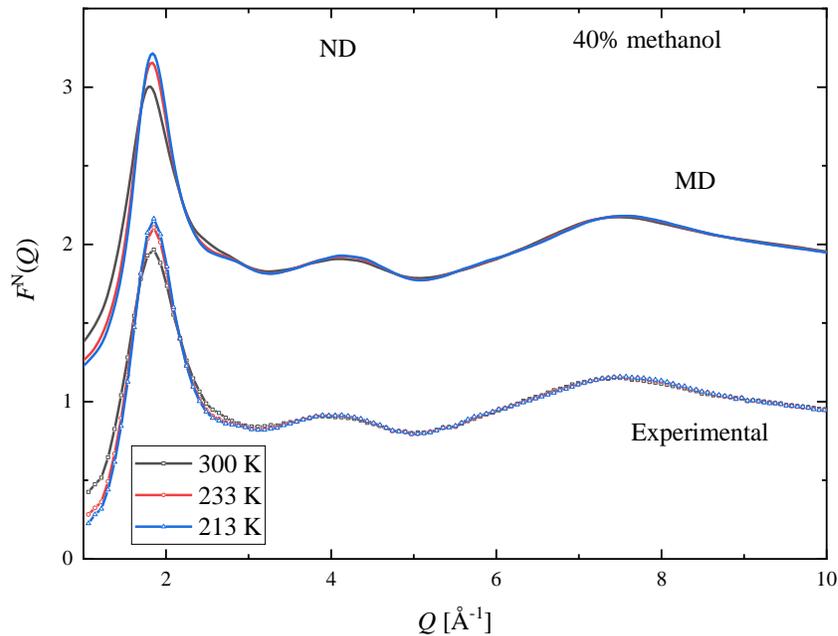

**Figure S14** Temperature dependence of measured (symbols) and simulated (lines) ND structure factors for the methanol-water mixture with 40 mol % methanol. The simulated curves were obtained by using the TIP4P/2005 water model. (The MD curves are shifted for clarity.)

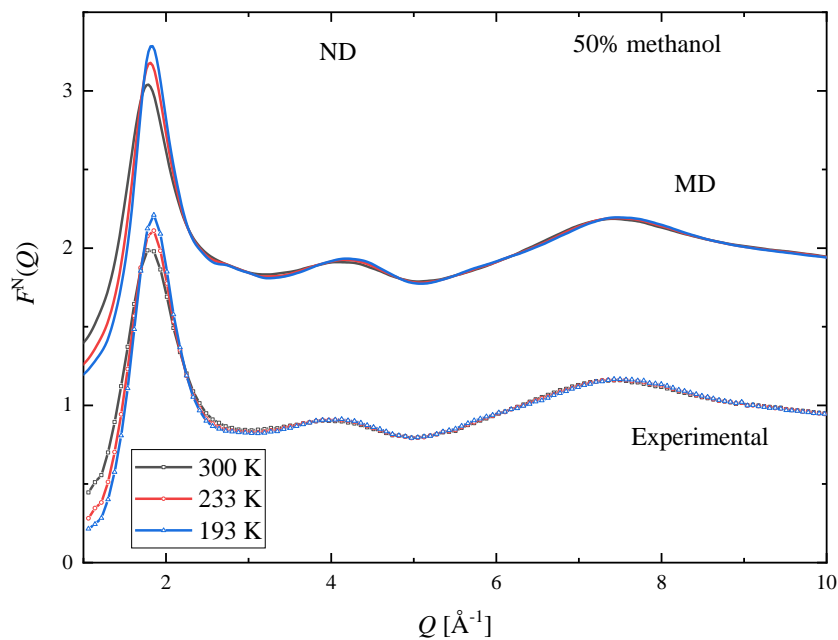

**Figure S15** Temperature dependence of measured (symbols) and simulated (lines) ND structure factors for the methanol-water mixture with 50 mol % methanol. The simulated curves were obtained by using the TIP4P/2005 water model. (The MD curves are shifted for clarity.)

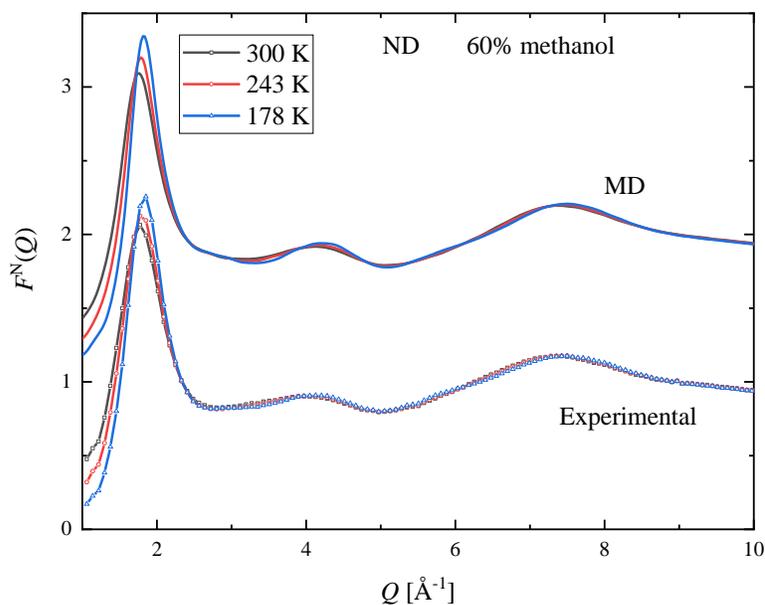

**Figure S16** Temperature dependence of measured (symbols) and simulated (lines) ND structure factors for the methanol-water mixture with 60 mol % methanol. The simulated curves were obtained by using the TIP4P/2005 water model. (The MD curves are shifted for clarity.)

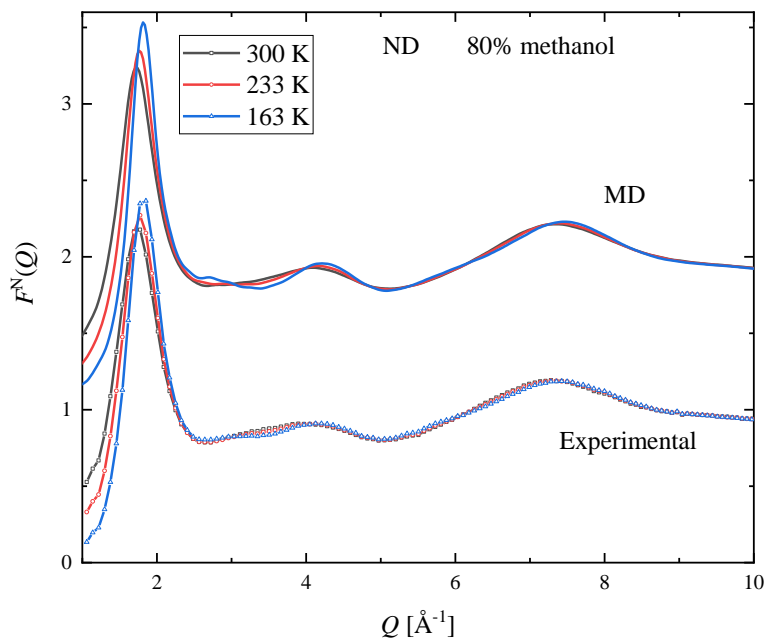

**Figure S17** Temperature dependence of measured (symbols) and simulated (lines) ND structure factors for the methanol-water mixture with 80 mol % methanol. The simulated curves were obtained by using the TIP4P/2005 water model. (The MD curves are shifted for clarity.)

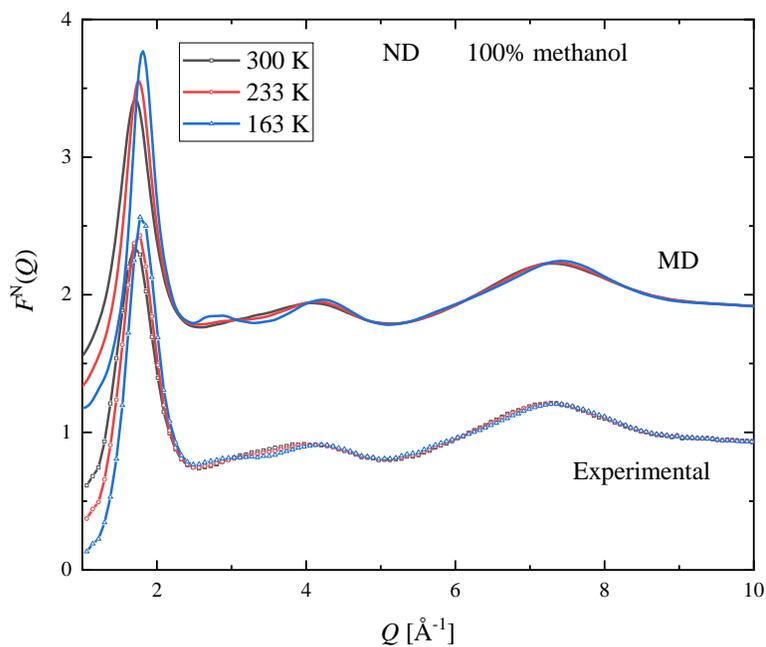

**Figure S18** Temperature dependence of measured (symbols) and simulated (lines) ND structure factors for pure methanol. (The MD curves are shifted for clarity.)

**XRD structure factors at temperatures not shown in the previous figures**

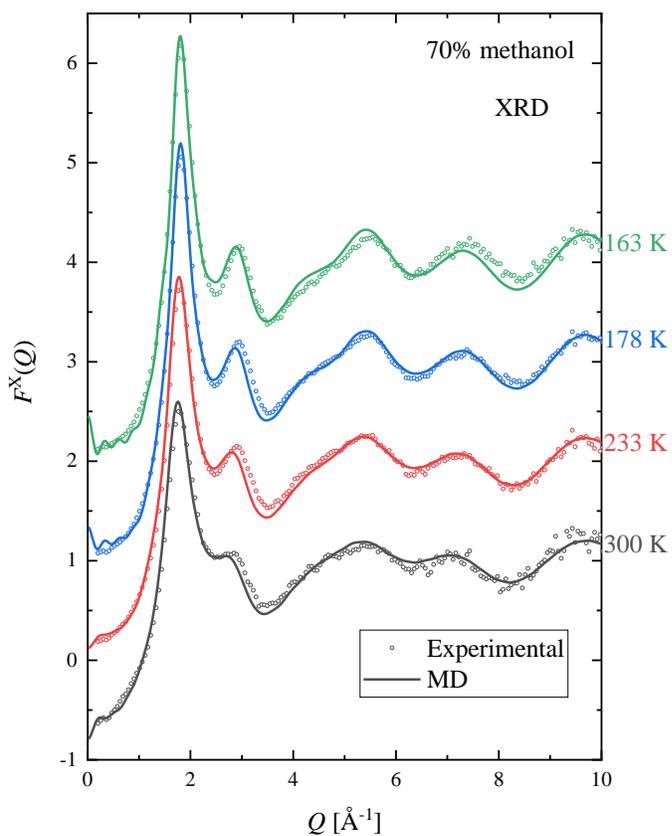

**Figure S19** Comparison of measured (symbols) and simulated (lines) XRD structure factors for the methanol-water mixture with 70 mol % methanol. Simulated curves were obtained by using the TIP4P/2005 water model. (The curves are shifted for clarity.)

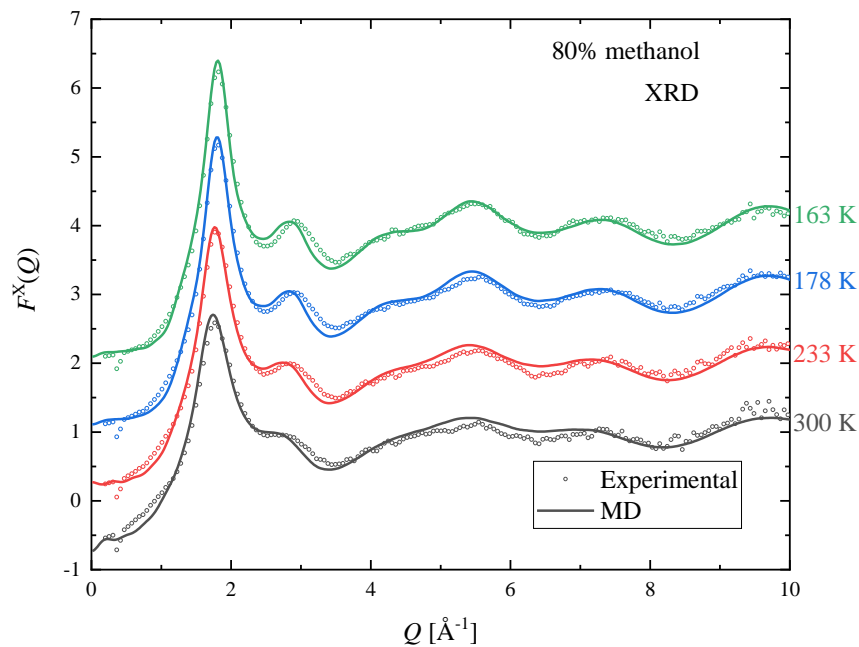

**Figure S20** Comparison of measured (symbols) and simulated (lines) XRD structure factors for the methanol-water mixture with 80 mol % methanol. Simulated curves were obtained by using the TIP4P/2005 water model. (The cuurves are shifted for clarity.)

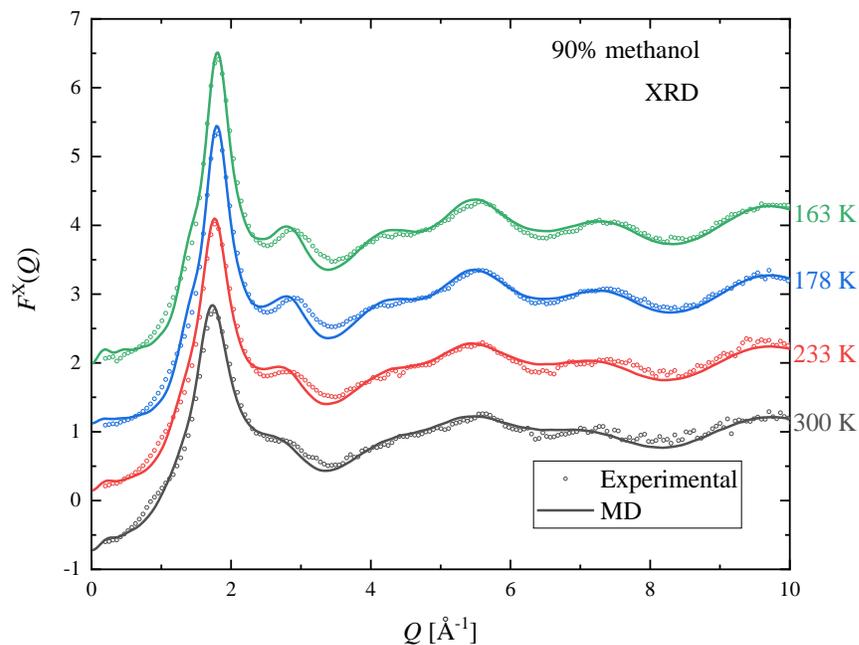

**Figure S21** Comparison of measured (symbols) and simulated (lines) XRD structure factors for the methanol-water mixture with 90 mol % methanol. Simulated curves were obtained by using the TIP4P/2005 water model. (The curves are shifted for clarity.)

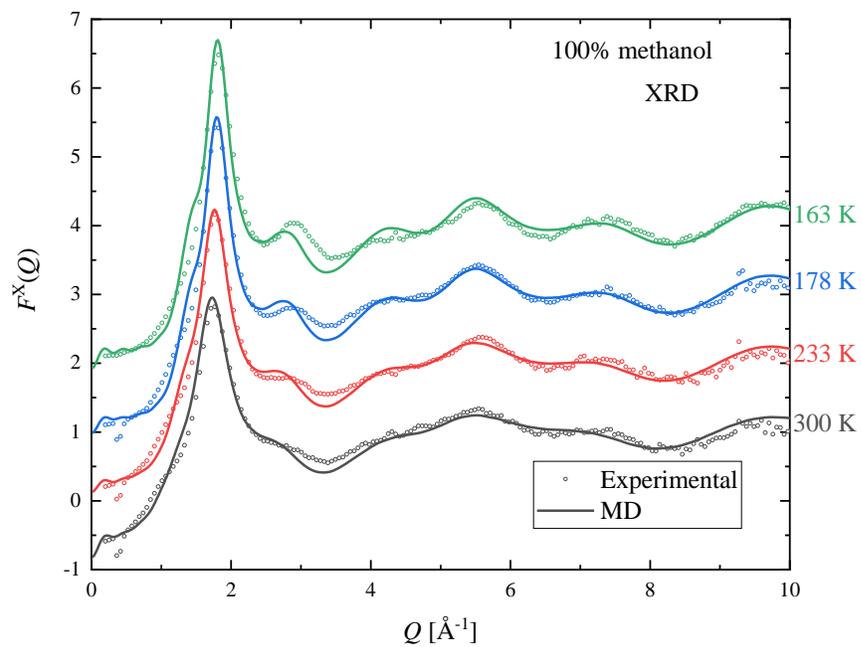

**Figure S22** Comparison of measured (symbols) and simulated (lines) XRD structure factors for pure methanol. (The curves are shifted for clarity.)

**ND structure factors at temperatures not shown in the previous figures**

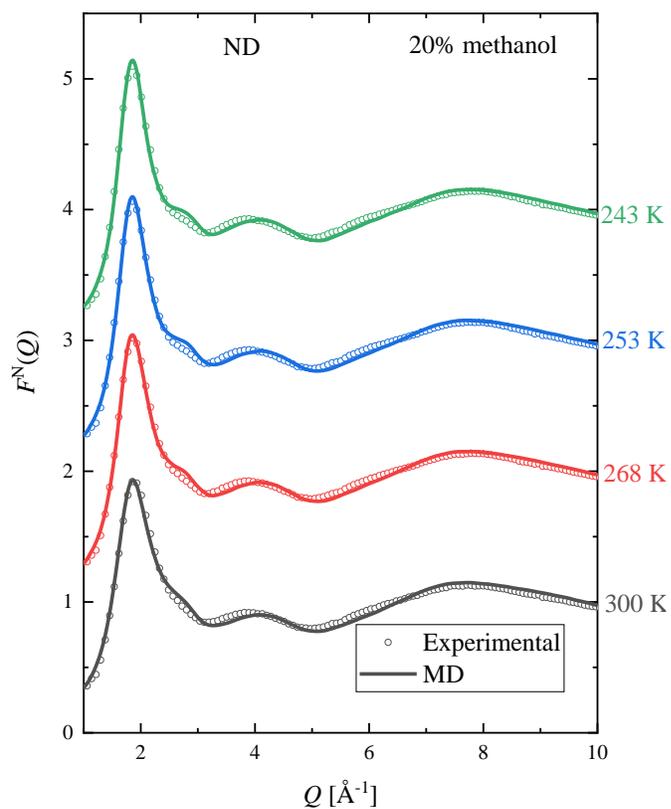

**Figure S23** Comparison of measured (symbols) and simulated (lines) ND structure factors for the methanol-water mixture with 20 mol % methanol. Simulated curves were obtained by using the TIP4P/2005 water model. (The curves are shifted for clarity.)

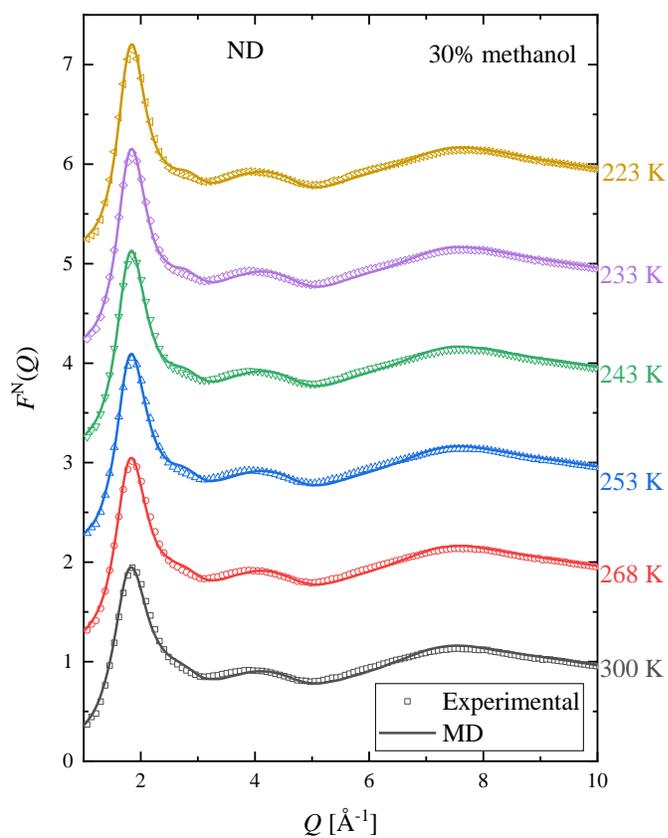

**Figure S24** Comparison of measured (symbols) and simulated (lines) ND structure factors for the methanol-water mixture with 30 mol % methanol. Simulated curves were obtained by using the TIP4P/2005 water model. (The curves are shifted for clarity.)

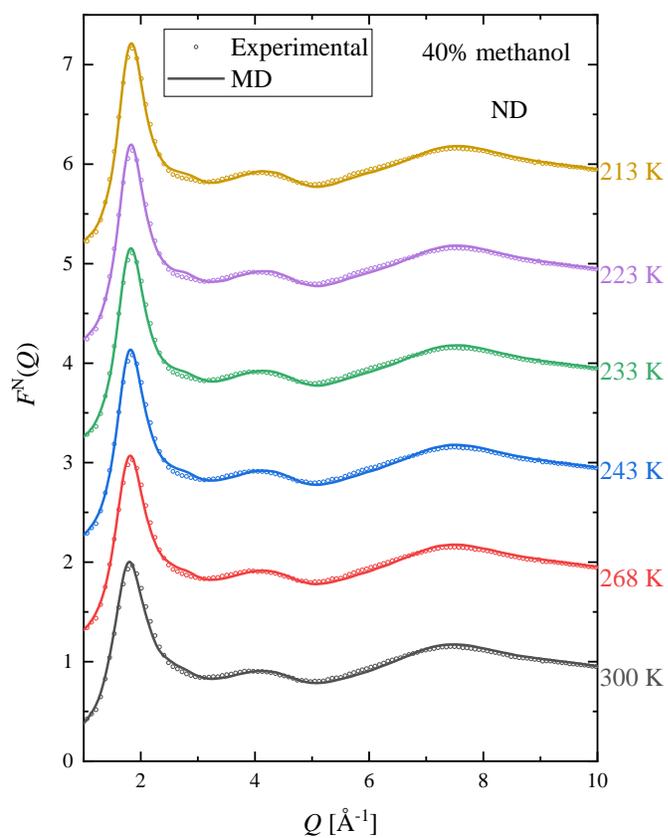

**Figure S25** Comparison of measured (symbols) and simulated (lines) ND structure factors for the methanol-water mixture with 40 mol % methanol. Simulated curves were obtained by using the TIP4P/2005 water model. (The curves are shifted for clarity.)

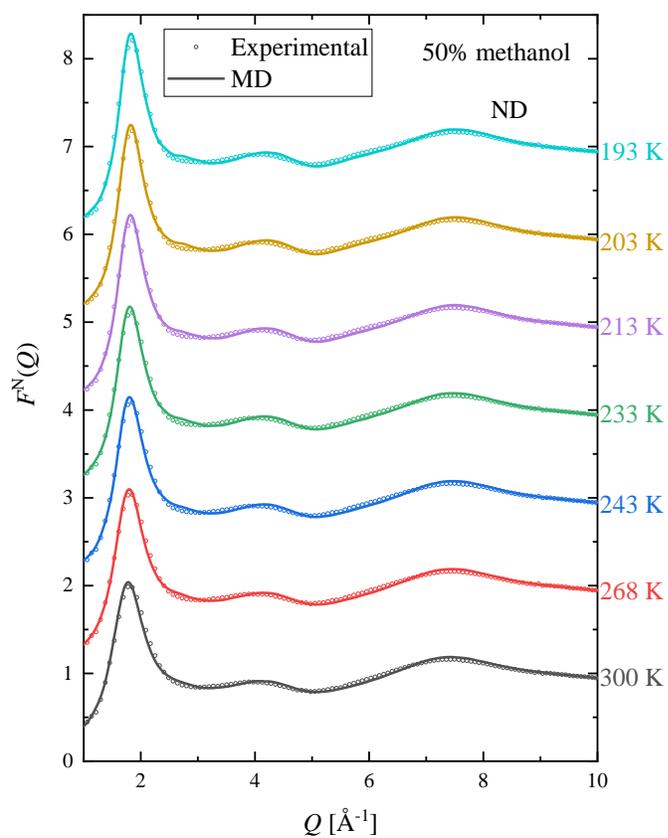

**Figure S26** Comparison of measured (symbols) and simulated (lines) ND structure factors for the methanol-water mixture with 50 mol % methanol. Simulated curves were obtained by using the TIP4P/2005 water model. (The curves are shifted for clarity.)

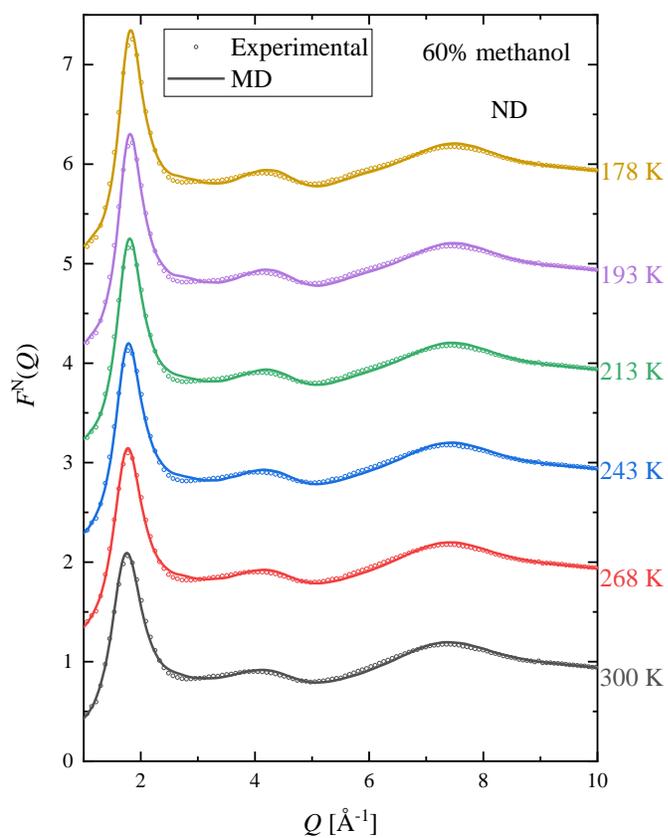

**Figure S27** Comparison of measured (symbols) and simulated (lines) ND structure factors for the methanol-water mixture with 60 mol % methanol. Simulated curves were obtained by using the TIP4P/2005 water model. (The curves are shifted for clarity.)

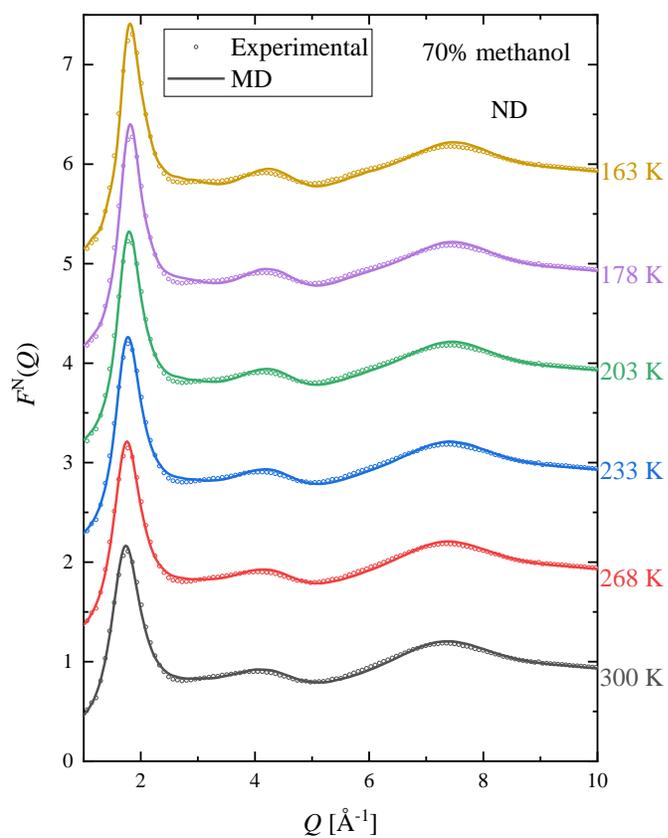

**Figure S28** Comparison of measured (symbols) and simulated (lines) ND structure factors for the methanol-water mixture with 70 mol % methanol. Simulated curves were obtained by using the TIP4P/2005 water model. (The curves are shifted for clarity.)

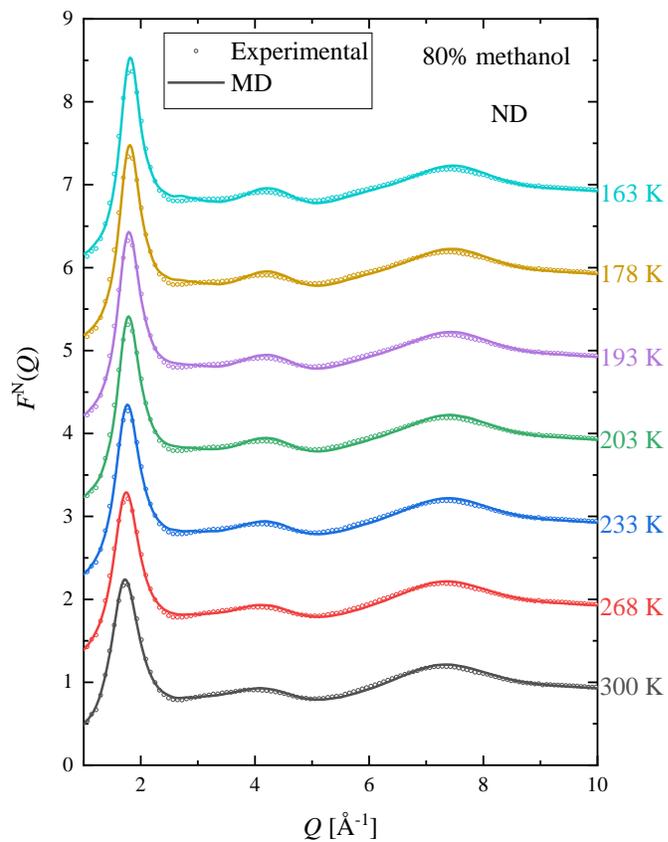

**Figure S29** Comparison of measured (symbols) and simulated (lines) ND structure factors for the methanol-water mixture with 80 mol % methanol. Simulated curves were obtained by using the TIP4P/2005 water model. (The curves are shifted for clarity.)

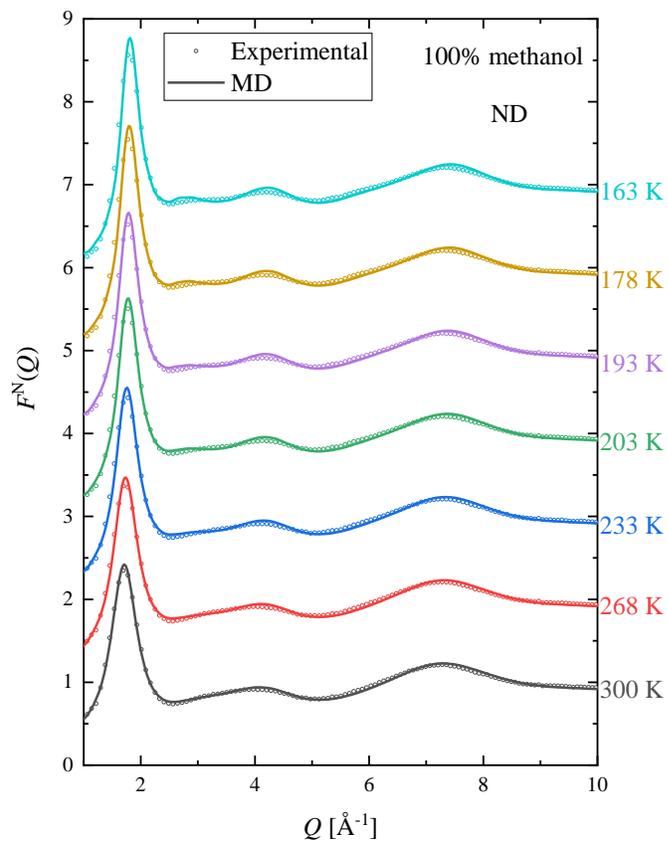

**Figure S30** Comparison of measured (symbols) and simulated (lines) ND structure factors for pure methanol. (The curves are shifted for clarity.)

**Comparison of the XRD structure factors obtained from experiments and simulations**

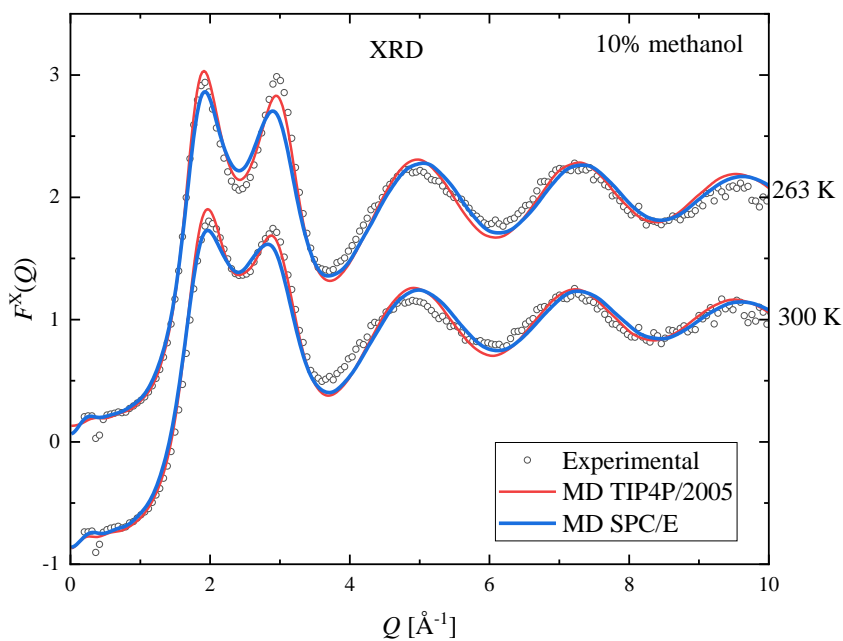

**Figure S31** Comparison of XRD structure factors obtained from experiments (symbols) and simulations using TIP4P/2005 (red lines) and SPC/E (blue lines) water models for the methanol-water mixture with 10% methanol, at 300 K and 263 K. (The curves are shifted for clarity.)

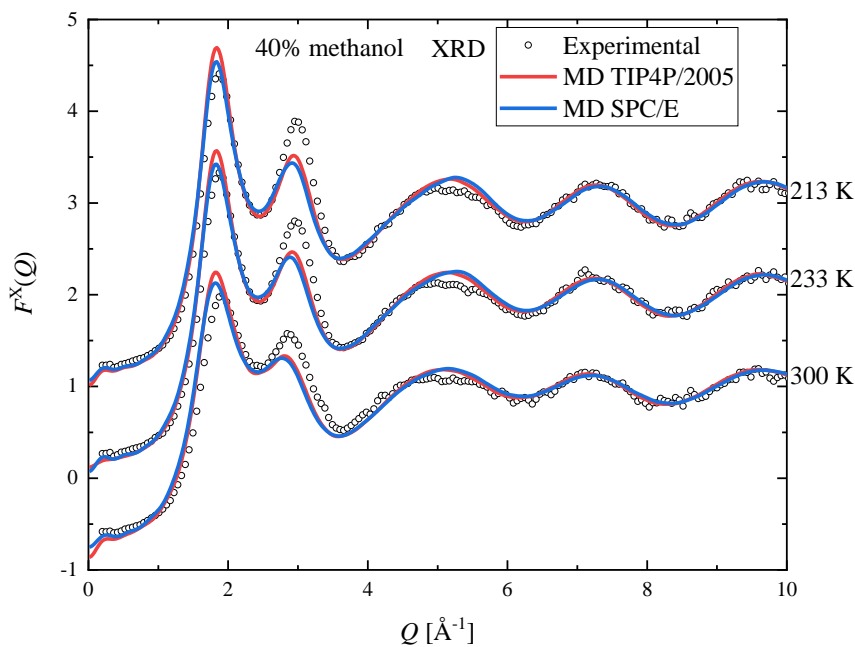

**Figure S32** Comparison of XRD structure factors obtained from experiments (symbols) and simulations using TIP4P/2005 (red lines) and SPC/E (blue lines) water models for the methanol-water mixture with 40% methanol, at three temperatures (300 K, 233 K and 213 K). (The curves are shifted for clarity.)

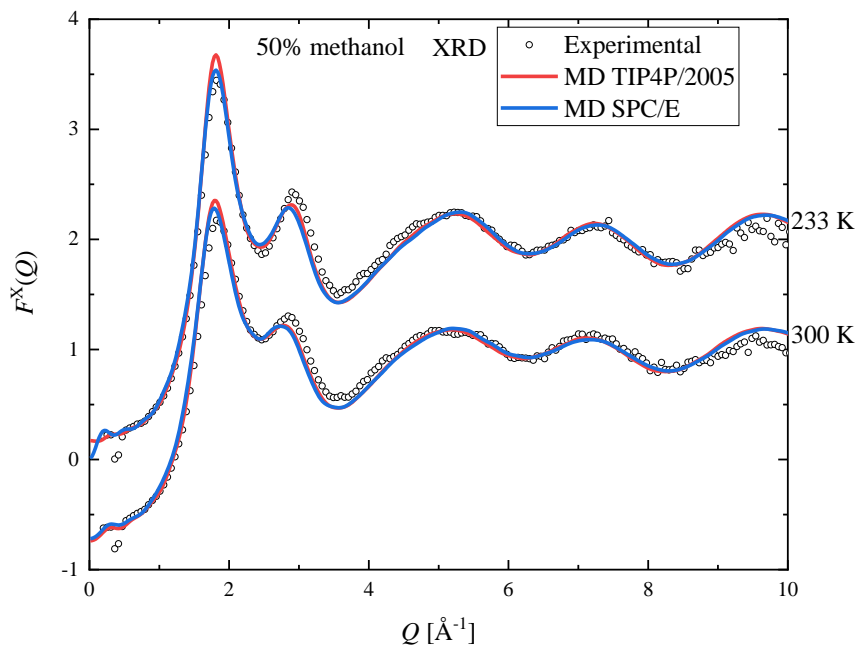

**Figure S33** Comparison of XRD structure factors obtained from experiments (symbols) and simulations using TIP4P/2005 (red lines) and SPC/E (blue lines) water models for the methanol-water mixture with 50% methanol, at 300 K and 233 K. (The curves are shifted for clarity.)

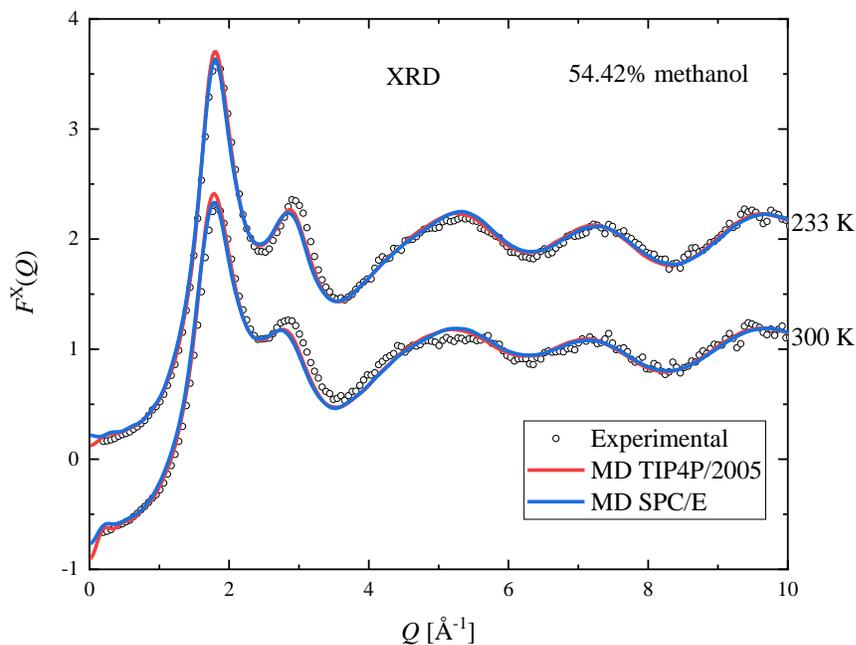

**Figure S34** Comparison of XRD structure factors obtained from experiments (symbols) and simulations using TIP4P/2005 (red lines) and SPC/E (blue lines) water models for the methanol-water mixture with 54.42% methanol, at 300 K and 233 K. (The curves are shifted for clarity.)

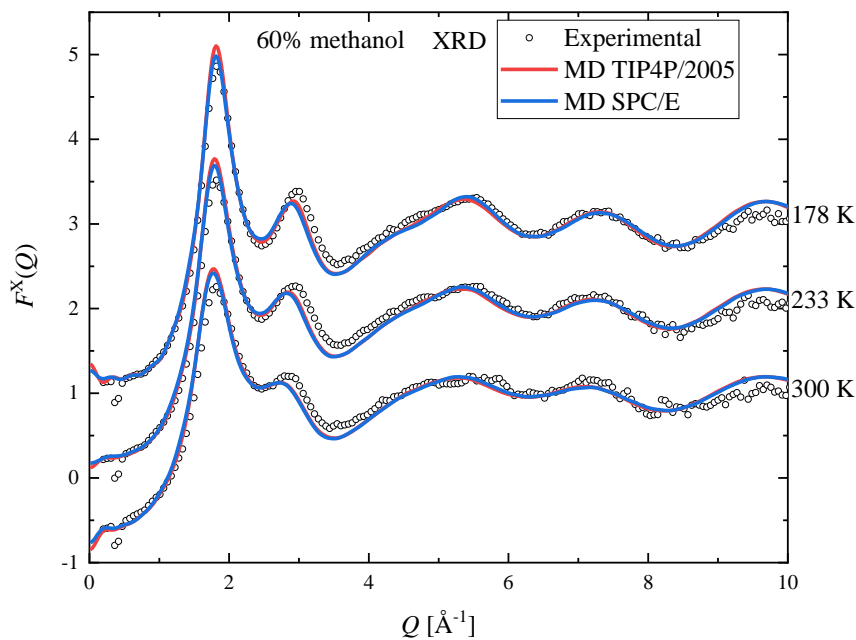

**Figure S35** Comparison of XRD structure factors obtained from experiments (symbols) and simulations using TIP4P/2005 (red lines) and SPC/E (blue lines) water models for the methanol-water mixture with 60% methanol, at three temperatures (300 K, 233 K and 178 K). (The curves are shifted for clarity.)

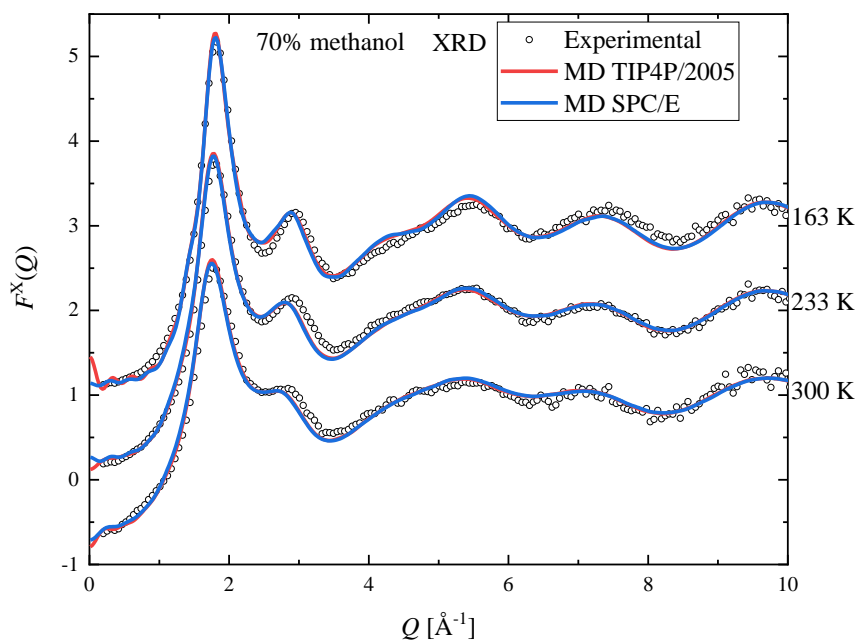

**Figure S36** Comparison of XRD structure factors obtained from experiments (symbols) and simulations using TIP4P/2005 (red lines) and SPC/E (blue lines) water models for the methanol-water mixture with 70% methanol, at three temperatures (300 K, 233 K and 163 K). (The curves are shifted for clarity.)

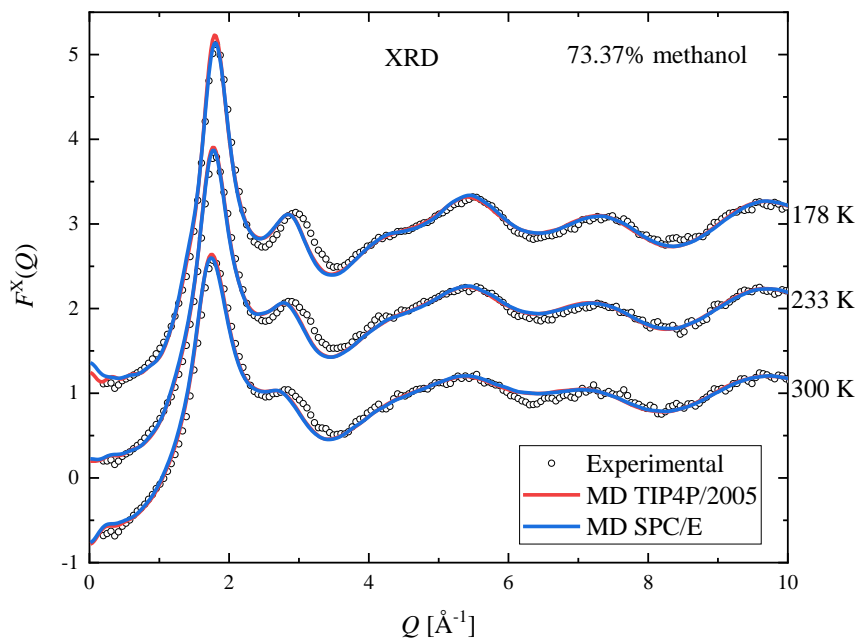

**Figure S37** Comparison of XRD structure factors obtained from experiments (symbols) and simulations using TIP4P/2005 (red lines) and SPC/E (blue lines) water models for the methanol-water mixture with 73.37% methanol, at three temperatures (300 K, 233 K and 178 K). (The curves are shifted for clarity.)

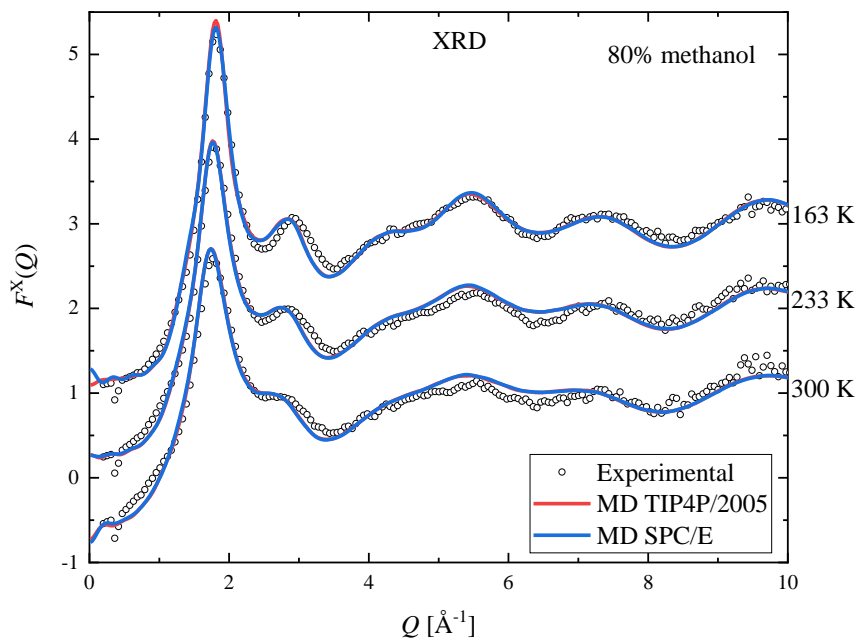

**Figure S38** Comparison of XRD structure factors obtained from experiments (symbols) and simulations using TIP4P/2005 (red lines) and SPC/E (blue lines) water models for the methanol-water mixture with 80% methanol, at three temperatures (300 K, 233 K and 163 K). (The curves are shifted for clarity.)

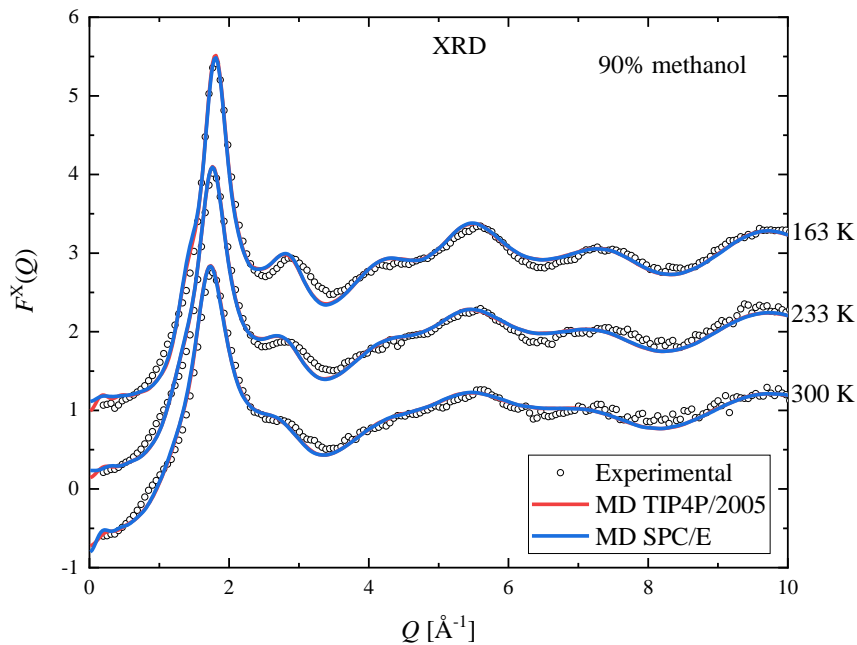

**Figure S39** Comparison of XRD structure factors obtained from experiments (symbols) and simulations using TIP4P/2005 (red lines) and SPC/E (blue lines) water models for the methanol-water mixture with 90% methanol, at three temperatures (300 K, 233 K and 163 K). (Th curves are shifted for clarity.)

**Comparison of the ND structure factors obtained from experiments and simulations**

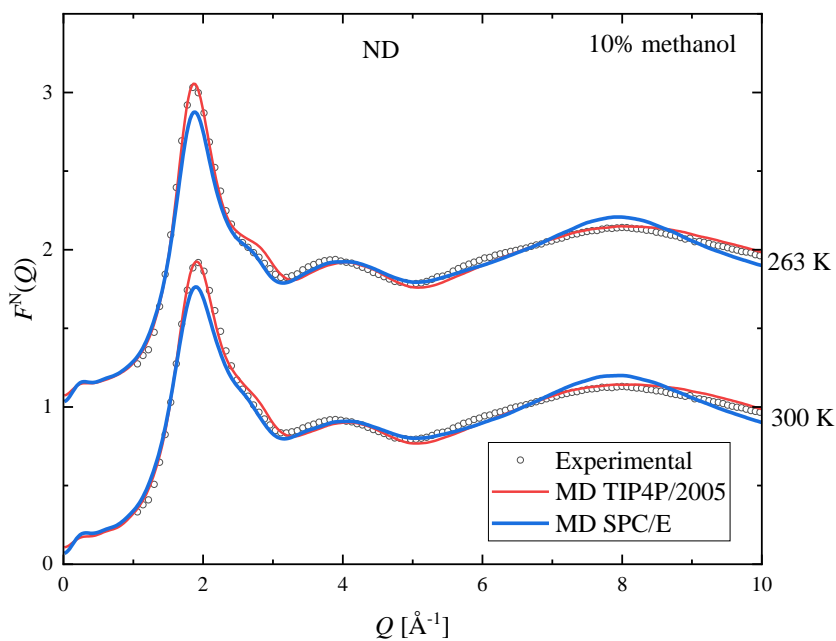

**Figure S40** Comparison of ND structure factors obtained from experiments (symbols) and simulations using TIP4P/2005 (red lines) and SPC/E (blue lines) water models for the methanol-water mixture with 10% methanol, at 300 K and 263 K. (The curves are shifted for clarity.)

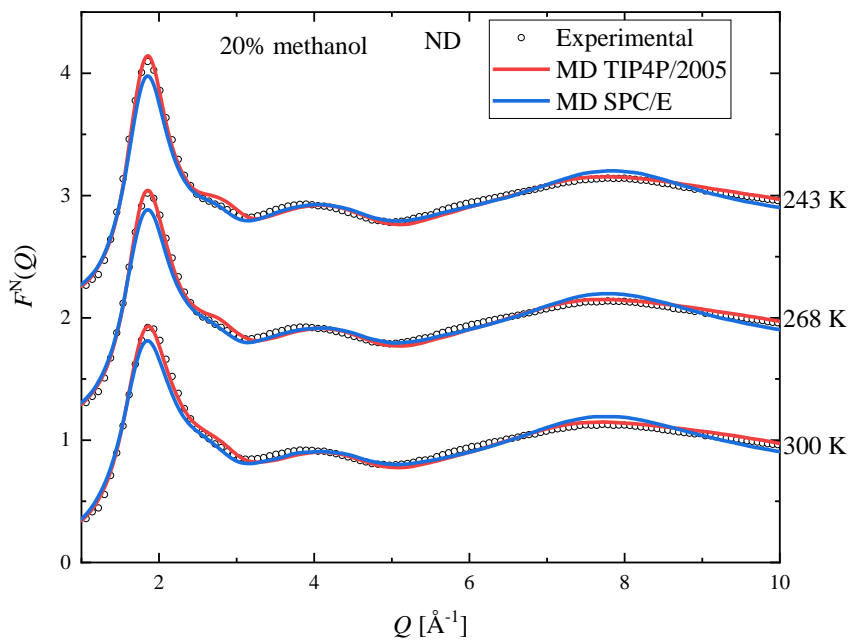

**Figure S41** Comparison of ND structure factors obtained from experiments (symbols) and simulations using TIP4P/2005 (red lines) and SPC/E (blue lines) water models for the methanol-water mixture with 20% methanol, at three selected temperatures (300 K, 268 K and 243 K). (The curves are shifted for clarity.)

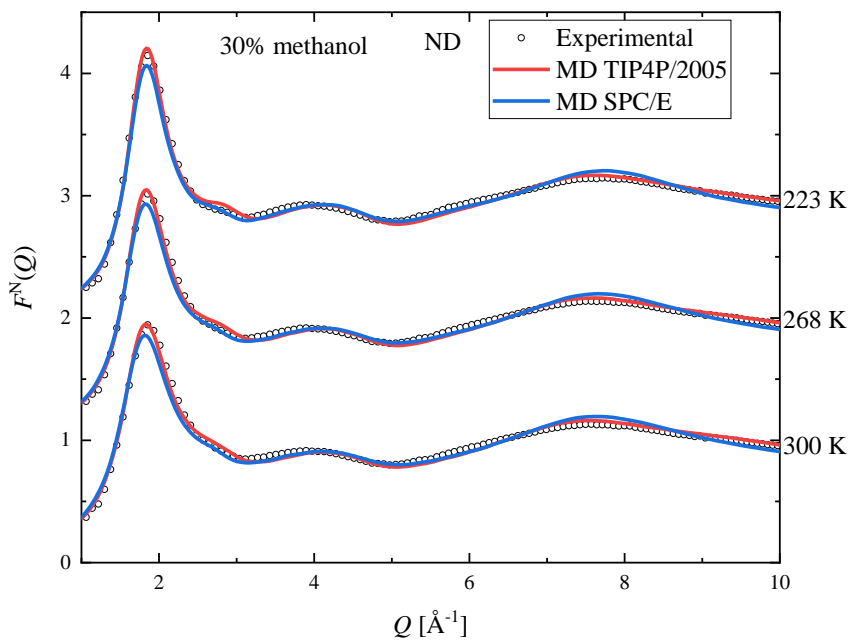

**Figure S42** Comparison of ND structure factors obtained from experiments (symbols) and simulations using TIP4P/2005 (red lines) and SPC/E (blue lines) water models for the methanol-water mixture with 30% methanol, at three selected temperatures (300 K, 268 K and 223 K). (The curves are shifted for clarity.)

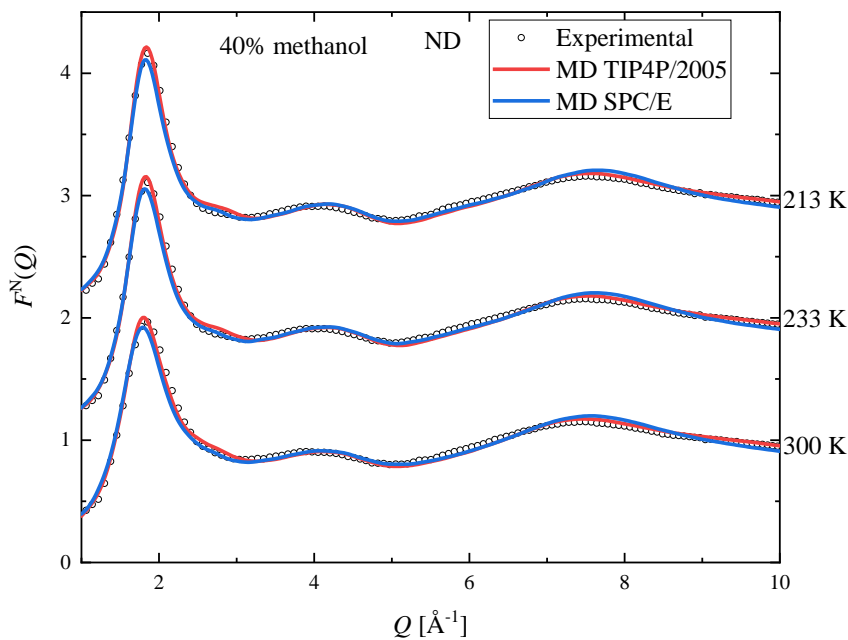

**Figure S43** Comparison of ND structure factors obtained from experiments (symbols) and simulations using TIP4P/2005 (red lines) and SPC/E (blue lines) water models for the methanol-water mixture with 40% methanol, at three selected temperatures (300 K, 233 K and 213 K). (The curves are shifted for clarity.)

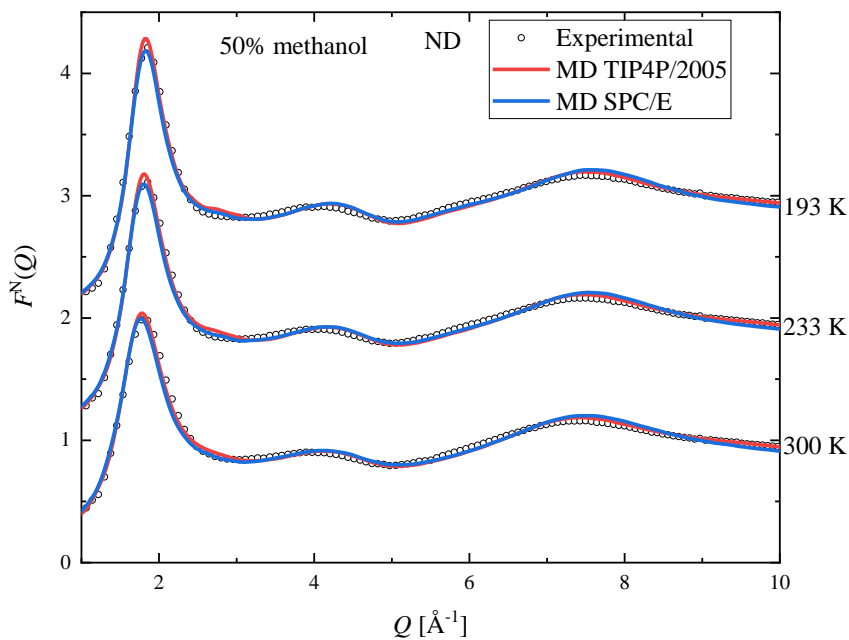

**Figure S44** Comparison of ND structure factors obtained from experiments (symbols) and simulations using TIP4P/2005 (red lines) and SPC/E (blue lines) water models for the methanol-water mixture with 50% methanol, at three selected temperatures (300 K, 233 K and 193 K). (The curves are shifted for clarity.)

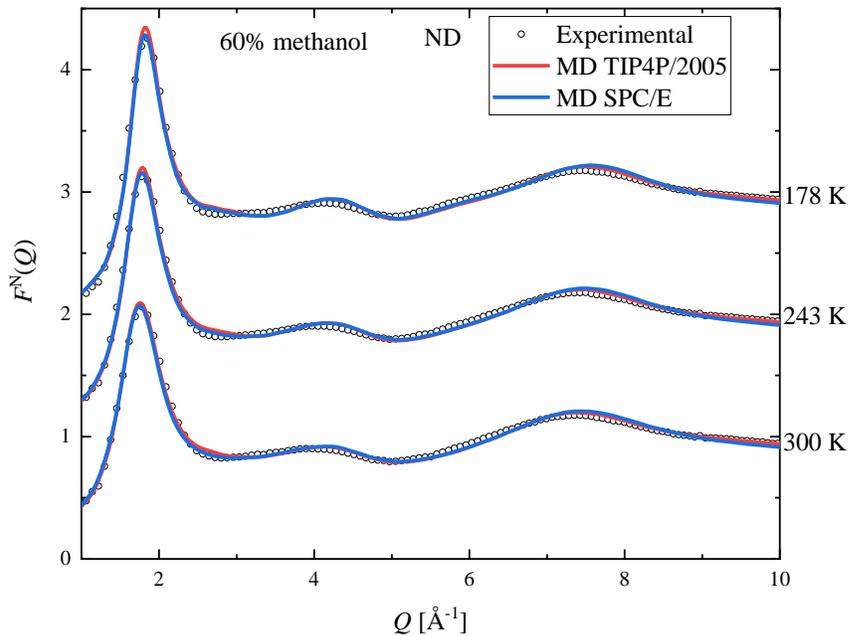

**Figure S45** Comparison of ND structure factors obtained from experiments (symbols) and simulations using TIP4P/2005 (red lines) and SPC/E (blue lines) water models for the methanol-water mixture with 60% methanol, at three selected temperatures (300 K, 243 K and 178 K). (The curves are shifted for clarity.)

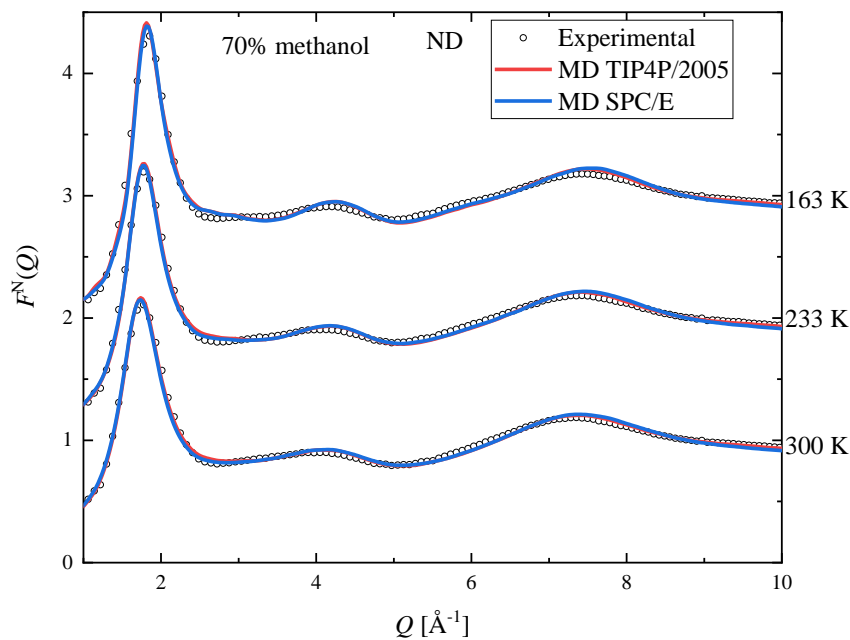

**Figure S46** Comparison of ND structure factors obtained from experiments (symbols) and simulations using TIP4P/2005 (red lines) and SPC/E (blue lines) water models for the methanol-water mixture with 70% methanol, at three selected temperatures (300 K, 233 K and 163 K). (The curves are shifted for clarity.)

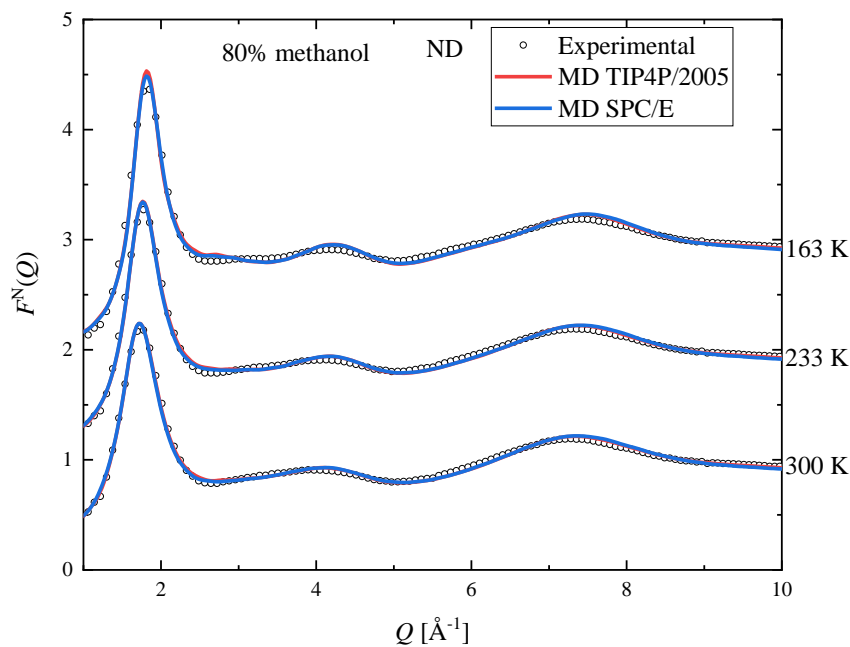

**Figure S47** Comparison of ND structure factors obtained from experiments (symbols) and simulations using TIP4P/2005 (red lines) and SPC/E (blue lines) water models for the methanol-water mixture with 80% methanol, at three selected temperatures (300 K, 233 K and 163 K). (The curves are shifted for clarity.)

**Partial radial distribution functions obtained from molecular dynamics simulations**

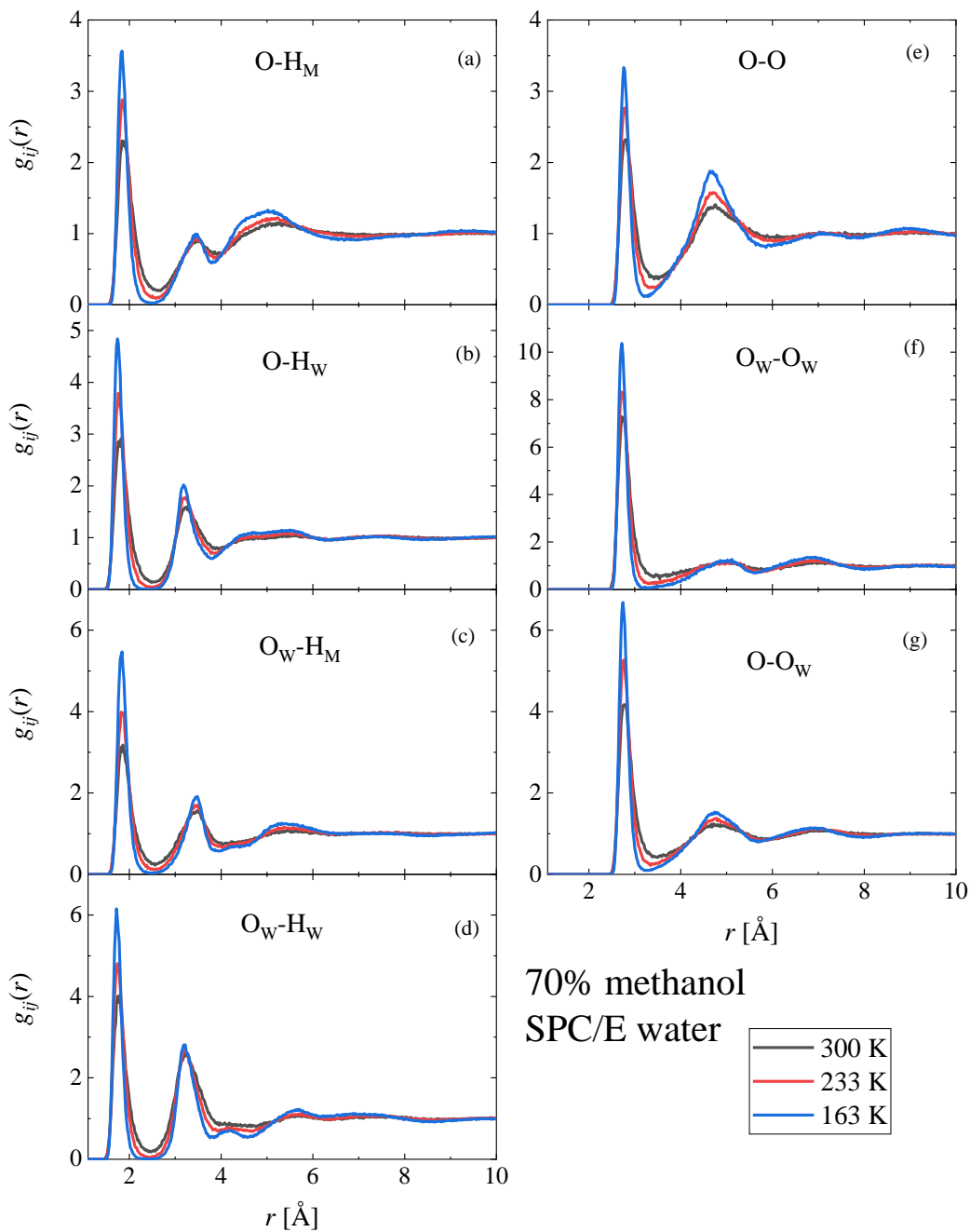

**Figure S48** Temperature dependence of simulated partial radial distribution functions of the methanol-water mixture with 70 mol % methanol. The H-bonding related partials are shown: (a) methanol O (denoted as O) – hydroxyl H of methanol (denoted as $H_M$), (b) methanol O – water H (denoted as $H_W$), (c) water O (denoted as $O_W$) – hydroxyl H of methanol, (d) water O – water H, (e) methanol O – methanol O, (f) water O – water O, (g) methanol O – water O. The curves were obtained using the SPC/E water model.

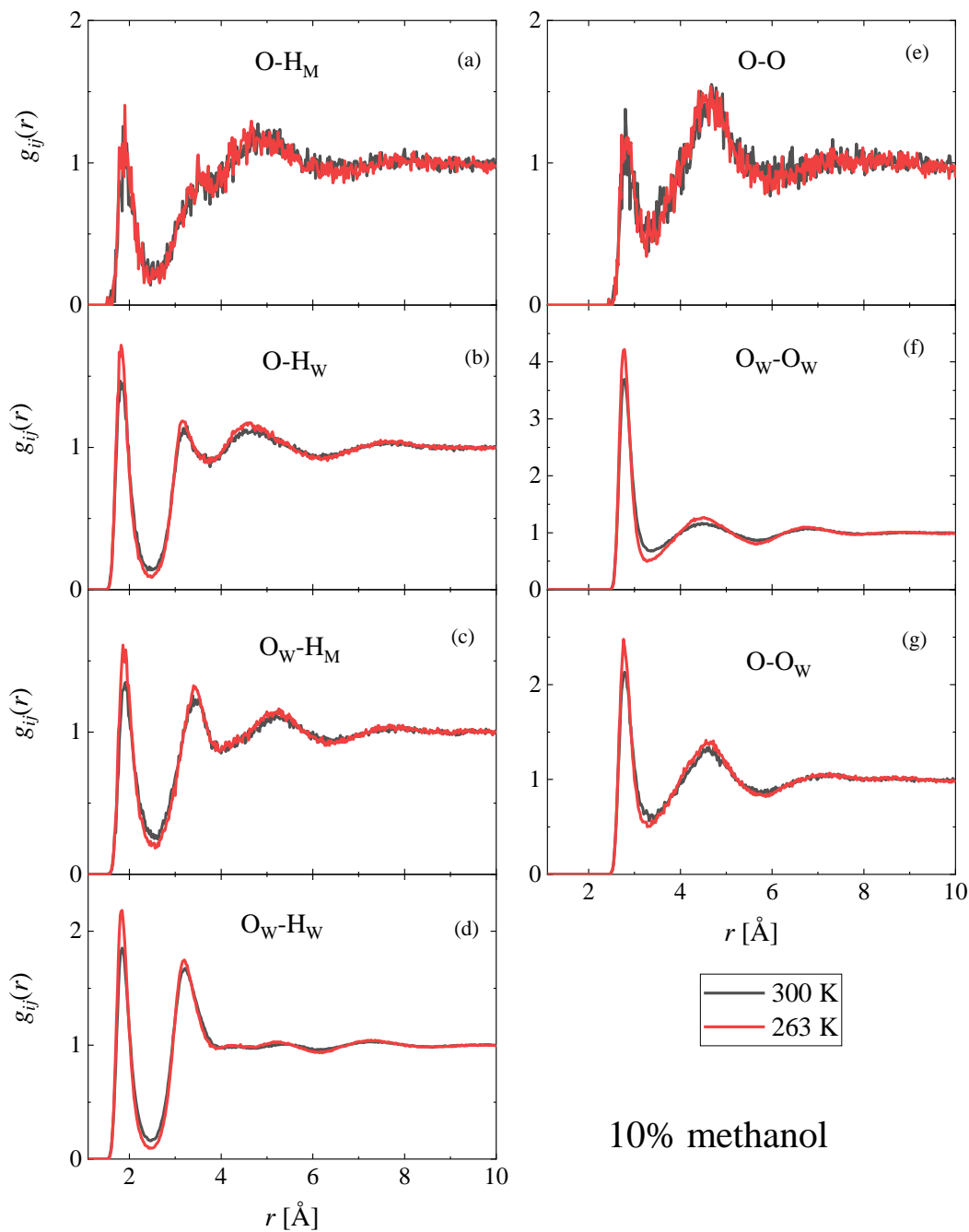

**Figure S49** Temperature dependence of simulated partial radial distribution functions of the methanol-water mixture with 10 mol % methanol. The H-bonding related partials are shown: (a) methanol O (denoted as O) – hydroxyl H of methanol (denoted as $H_M$), (b) methanol O – water H (denoted as $H_W$), (c) water O (denoted as $O_W$) – hydroxyl H of methanol, (d) water O – water H, (e) methanol O – methanol O, (f) water O – water O, (g) methanol O – water O. The curves were obtained using the TIP4P/2005 water model.

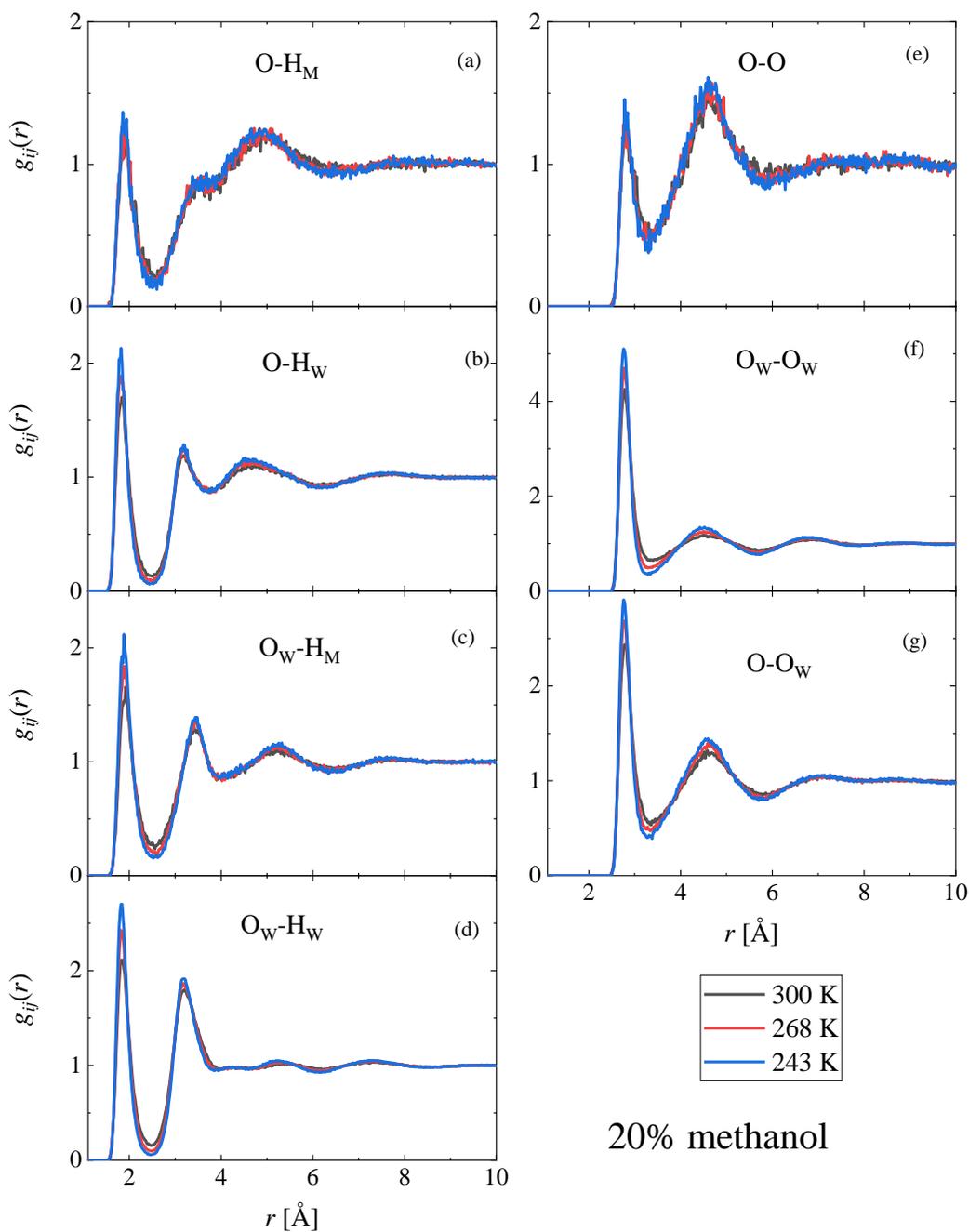

**Figure S50** Temperature dependence of simulated partial radial distribution functions of the methanol-water mixture with 20 mol % methanol. The H-bonding related partials are shown: (a) methanol O (denoted as O) – hydroxyl H of methanol (denoted as $H_M$), (b) methanol O – water H (denoted as $H_W$), (c) water O (denoted as $O_W$) – hydroxyl H of methanol, (d) water O – water H, (e) methanol O – methanol O, (f) water O – water O, (g) methanol O – water O. The curves were obtained using the TIP4P/2005 water model.

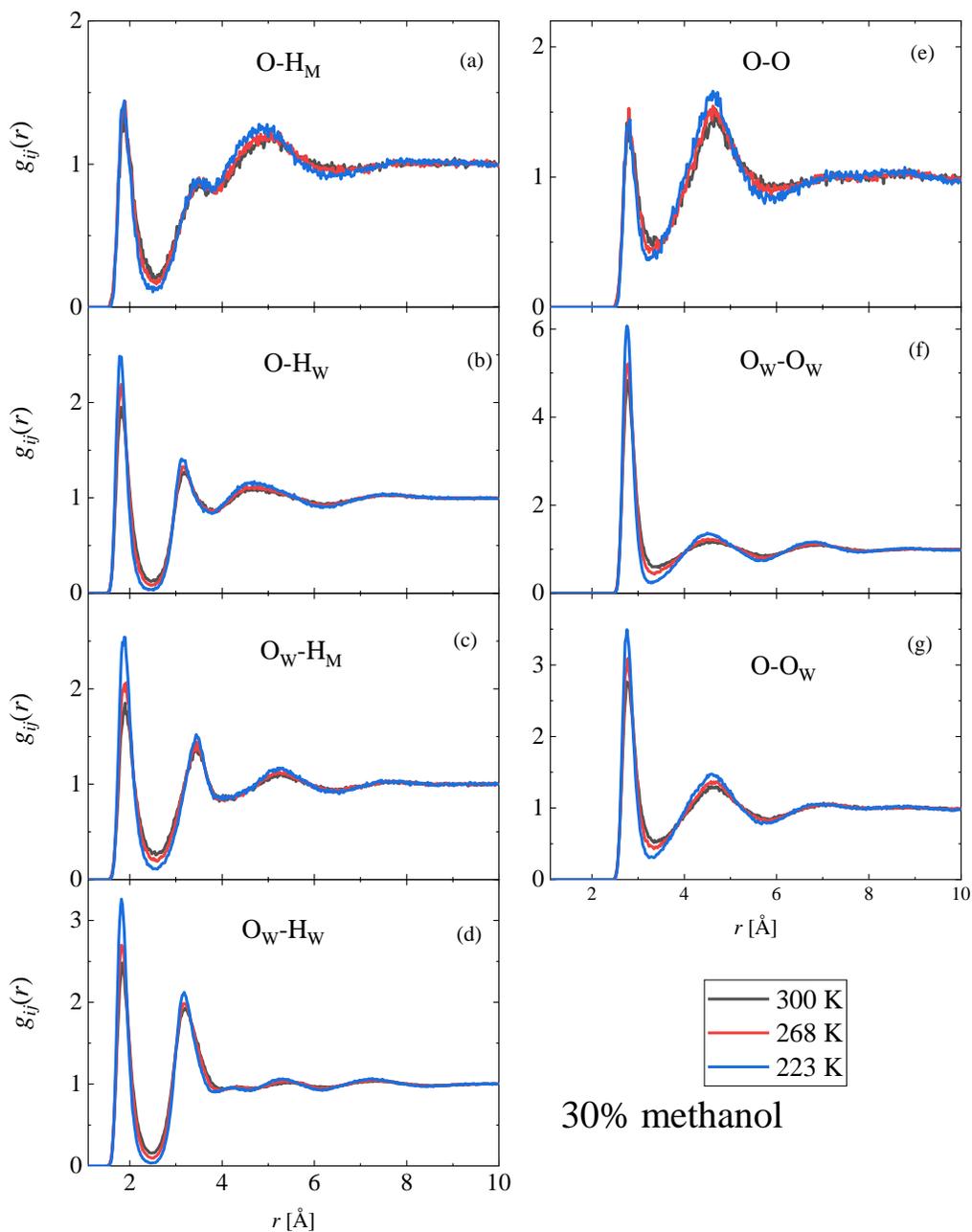

**Figure S51** Temperature dependence of simulated partial radial distribution functions of the methanol-water mixture with 30 mol % methanol. The H-bonding related partials are shown: (a) methanol O (denoted as O) – hydroxyl H of methanol (denoted as $H_M$), (b) methanol O – water H (denoted as $H_W$), (c) water O (denoted as $O_W$) – hydroxyl H of methanol, (d) water O – water H, (e) methanol O – methanol O, (f) water O – water O, (g) methanol O – water O. The curves were obtained using the TIP4P/2005 water model.

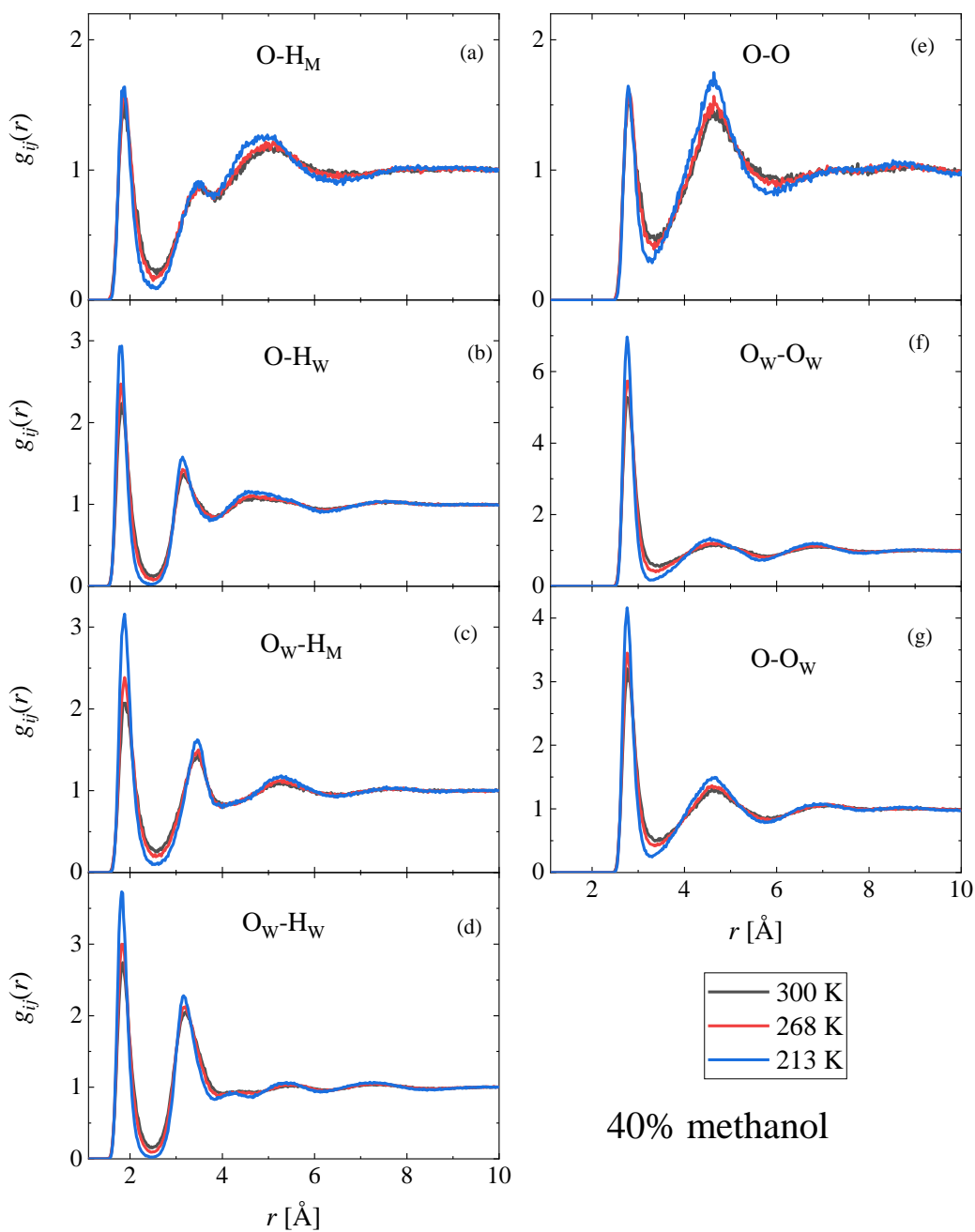

**Figure S52** Temperature dependence of simulated partial radial distribution functions of the methanol-water mixture with 40 mol % methanol. The H-bonding related partials are shown: (a) methanol O (denoted as O) – hydroxyl H of methanol (denoted as $H_M$), (b) methanol O – water H (denoted as $H_W$), (c) water O (denoted as $O_W$) – hydroxyl H of methanol, (d) water O – water H, (e) methanol O – methanol O, (f) water O – water O, (g) methanol O – water O. The curves were obtained using the TIP4P/2005 water model.

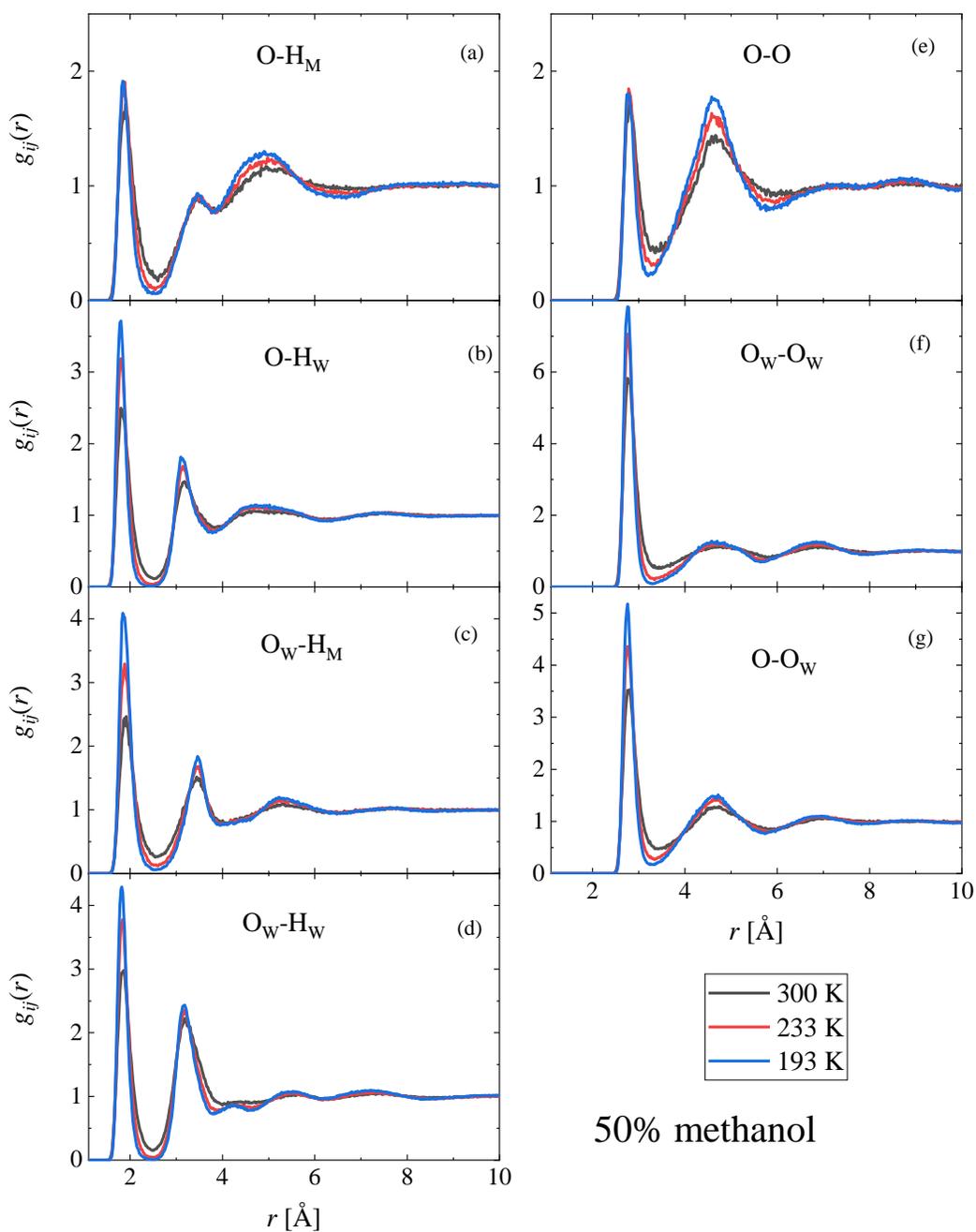

**Figure S53** Temperature dependence of simulated partial radial distribution functions of the methanol-water mixture with 50 mol % methanol. The H-bonding related partials are shown: (a) methanol O (denoted as O) – hydroxyl H of methanol (denoted as $H_M$), (b) methanol O – water H (denoted as $H_W$), (c) water O (denoted as $O_W$) – hydroxyl H of methanol, (d) water O – water H, (e) methanol O – methanol O, (f) water O – water O, (g) methanol O – water O. The curves were obtained using the TIP4P/2005 water model.

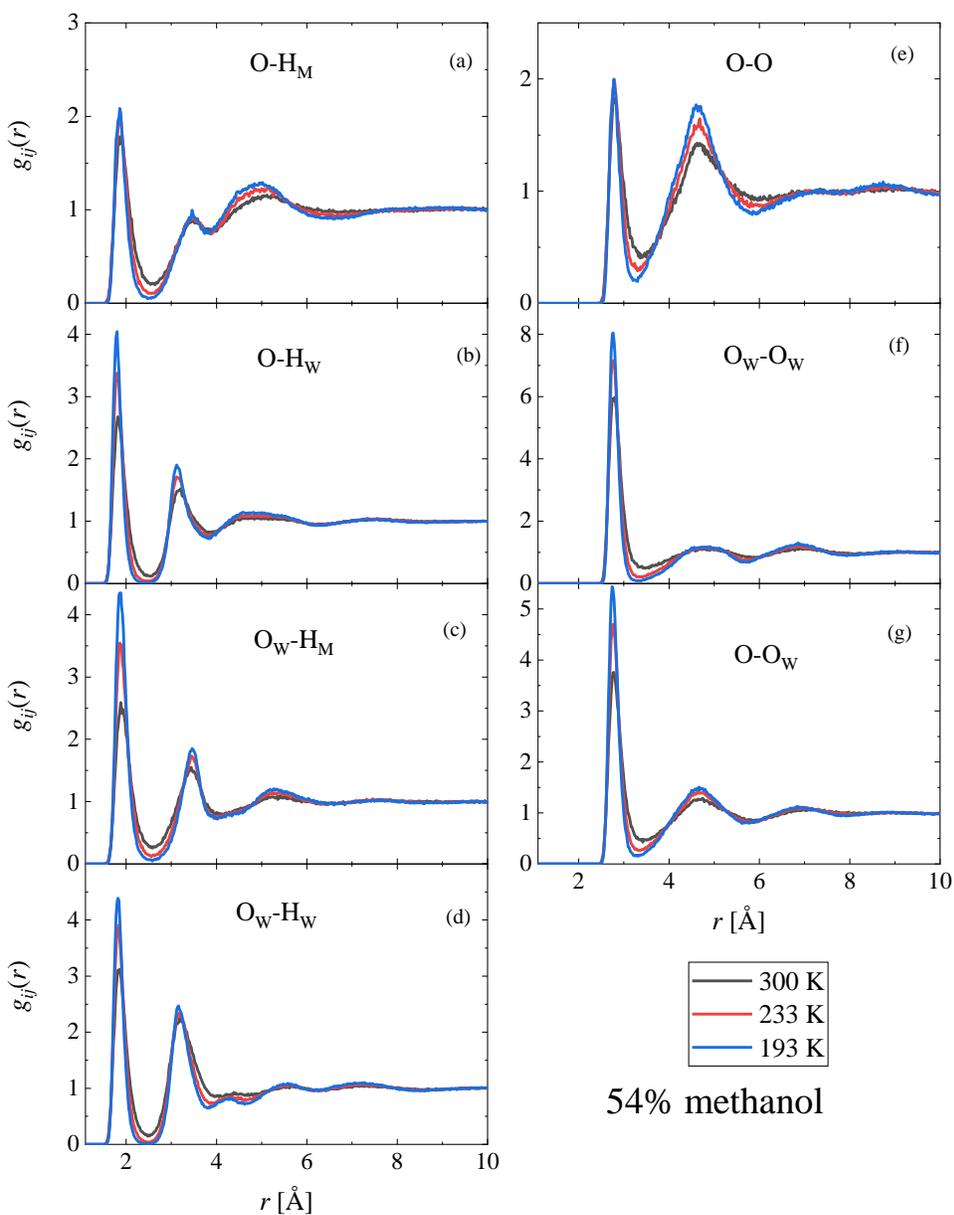

**Figure S54** Temperature dependence of simulated partial radial distribution functions of the methanol-water mixture with 54.42 mol % methanol. The H-bonding related partials are shown: (a) methanol O (denoted as O) – hydroxyl H of methanol (denoted as $H_M$), (b) methanol O – water H (denoted as $H_W$), (c) water O (denoted as $O_W$) – hydroxyl H of methanol, (d) water O – water H, (e) methanol O – methanol O, (f) water O – water O, (g) methanol O – water O. The curves were obtained using the TIP4P/2005 water model.

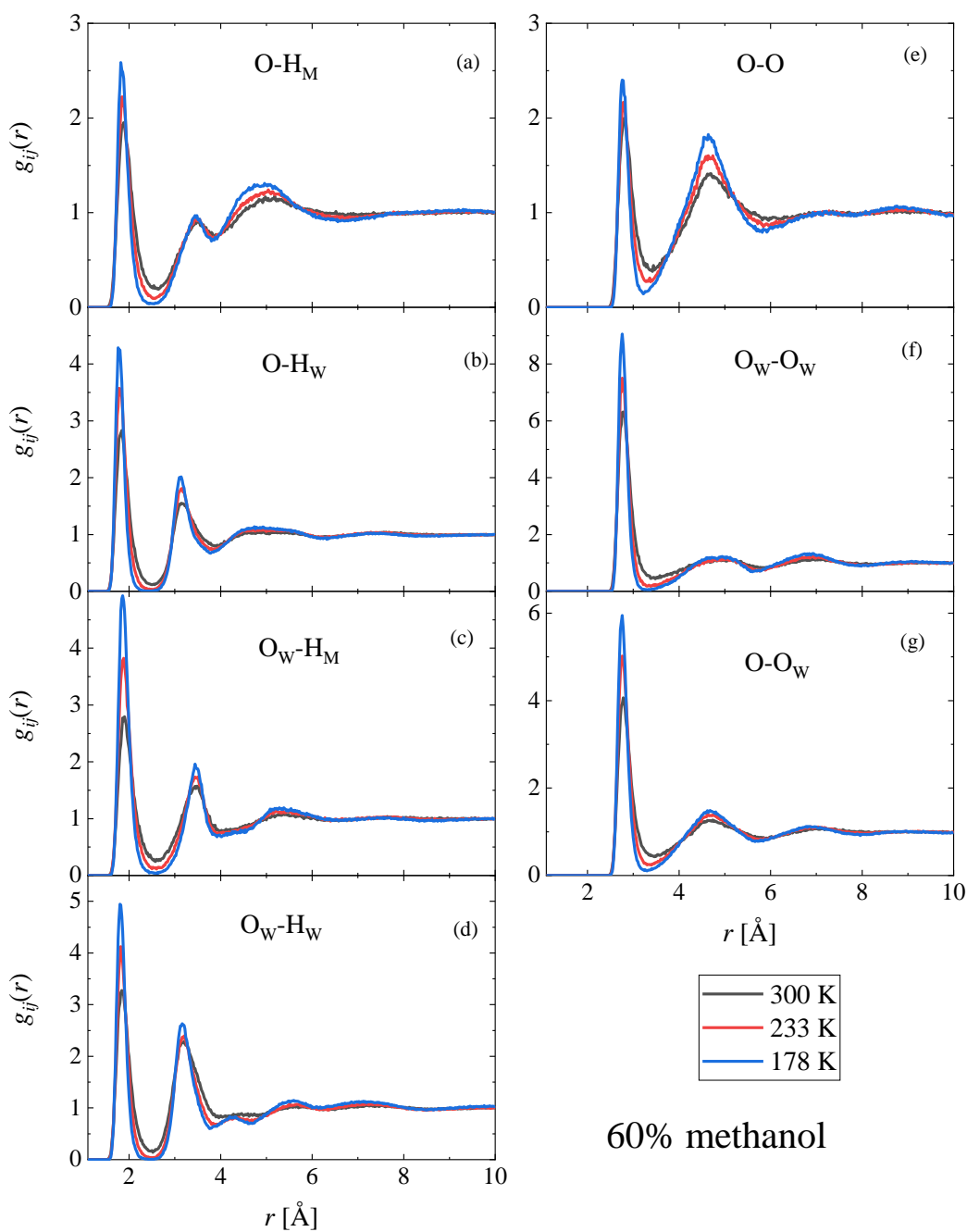

**Figure S55** Temperature dependence of simulated partial radial distribution functions of the methanol-water mixture with 60 mol % methanol. The H-bonding related partials are shown: (a) methanol O (denoted as O) – hydroxyl H of methanol (denoted as $H_M$), (b) methanol O – water H (denoted as $H_W$), (c) water O (denoted as $O_W$) – hydroxyl H of methanol, (d) water O – water H, (e) methanol O – methanol O, (f) water O – water O, (g) methanol O – water O. The curves were obtained using the TIP4P/2005 water model.

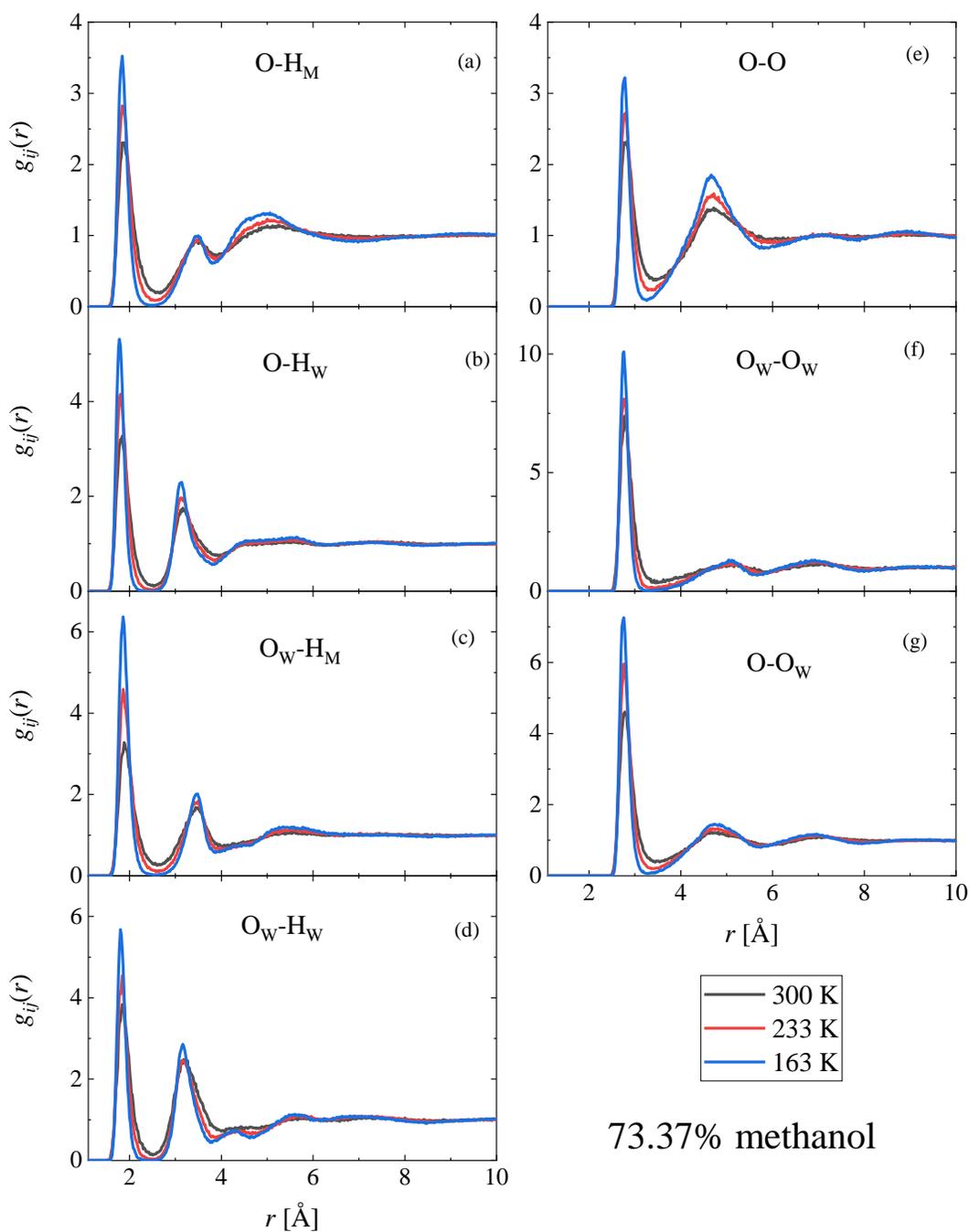

**Figure S56** Temperature dependence of simulated partial radial distribution functions of the methanol-water mixture with 73.37 mol % methanol. The H-bonding related partials are shown: (a) methanol O (denoted as O) – hydroxyl H of methanol (denoted as $H_M$), (b) methanol O – water H (denoted as $H_W$), (c) water O (denoted as $O_W$) – hydroxyl H of methanol, (d) water O – water H, (e) methanol O – methanol O, (f) water O – water O, (g) methanol O – water O. The curves were obtained using the TIP4P/2005 water model.

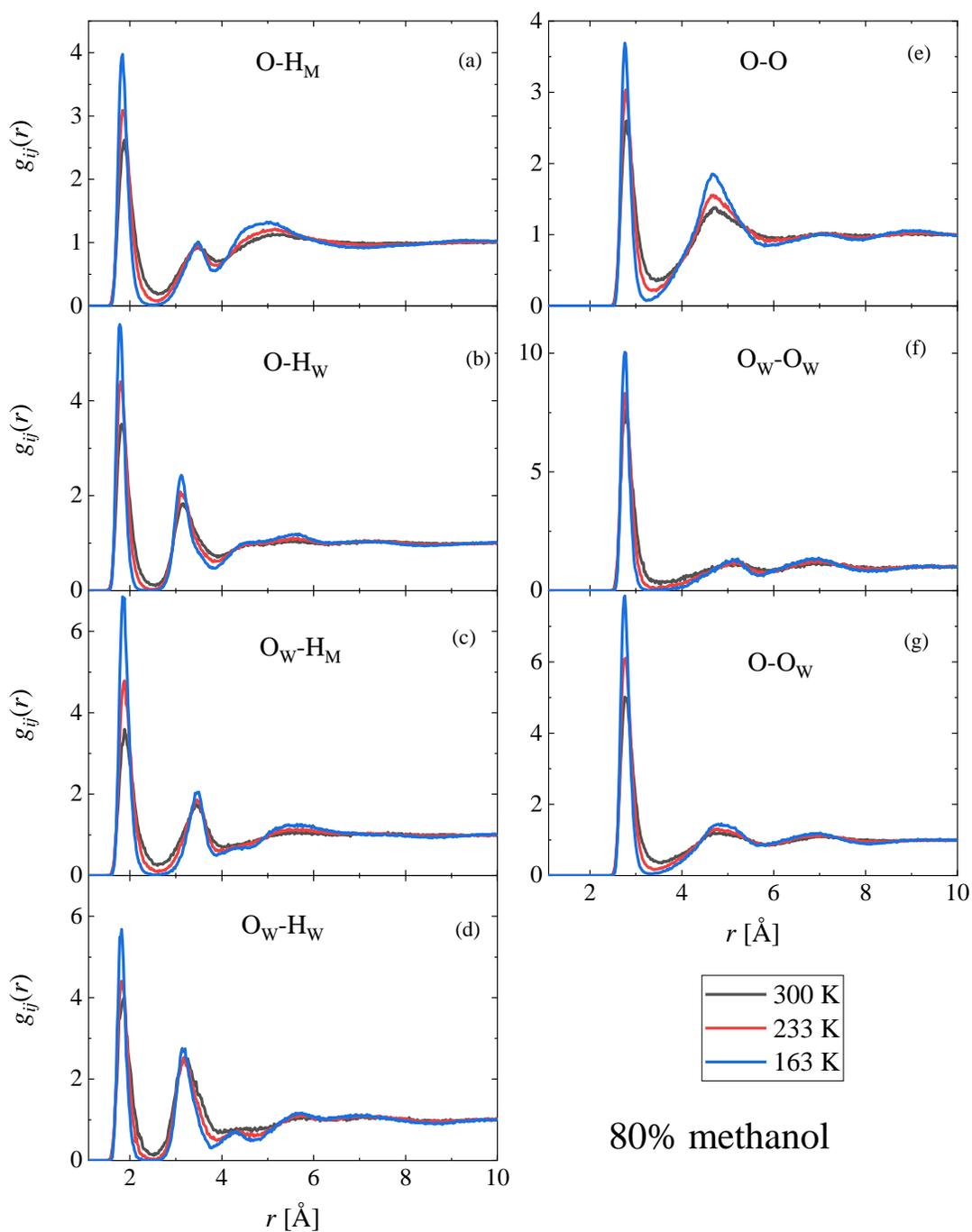

**Figure S57** Temperature dependence of simulated partial radial distribution functions of the methanol-water mixture with 80 mol % methanol. The H-bonding related partials are shown: (a) methanol O (denoted as O) – hydroxyl H of methanol (denoted as $H_M$), (b) methanol O – water H (denoted as $H_W$), (c) water O (denoted as $O_W$) – hydroxyl H of methanol, (d) water O – water H, (e) methanol O – methanol O, (f) water O – water O, (g) methanol O – water O. The curves were obtained using the TIP4P/2005 water model.

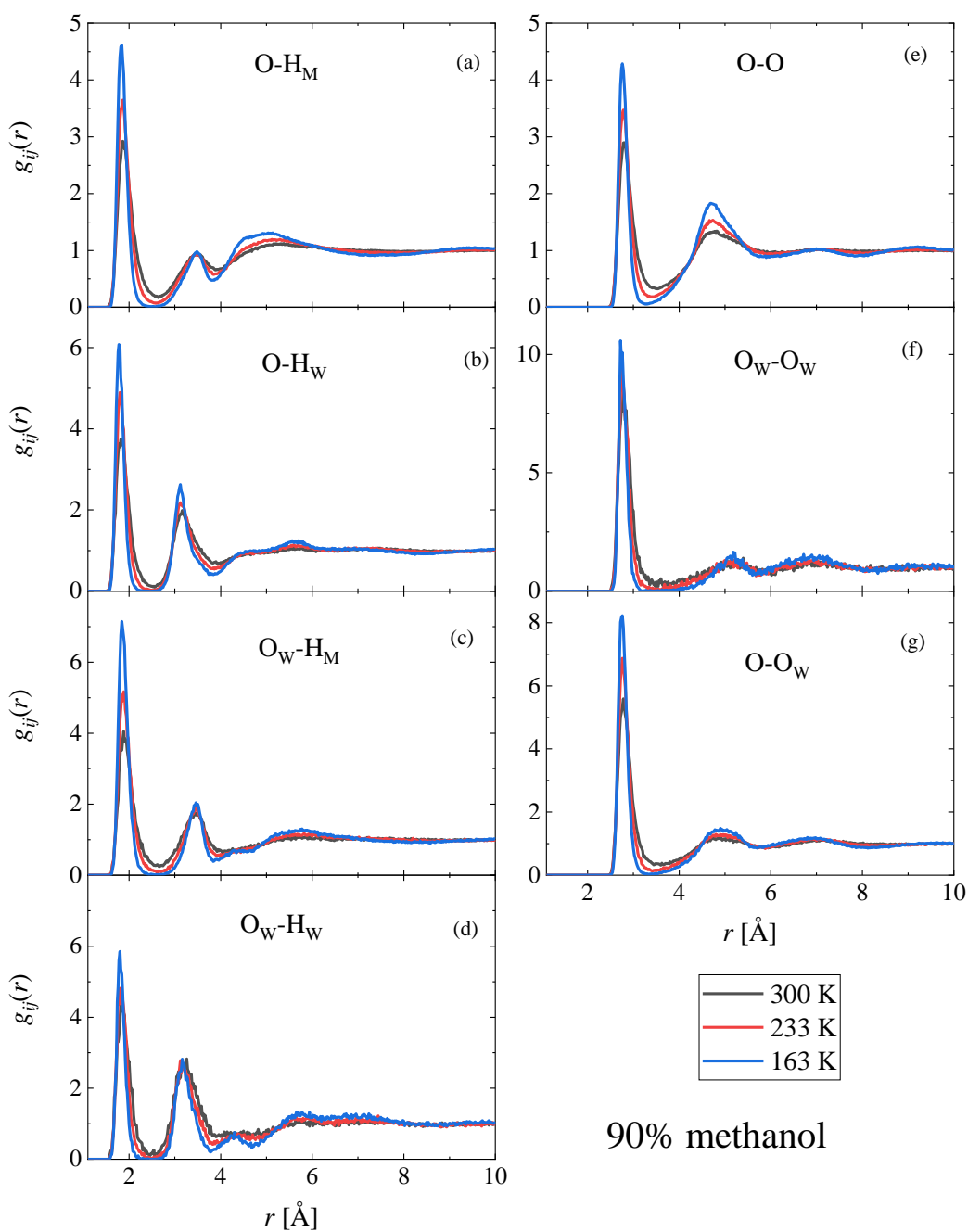

**Figure S58** Temperature dependence of simulated partial radial distribution functions of the methanol-water mixture with 90 mol % methanol. The H-bonding related partials are shown: (a) methanol O (denoted as O) – hydroxyl H of methanol (denoted as $H_M$), (b) methanol O – water H (denoted as $H_W$), (c) water O (denoted as $O_W$) – hydroxyl H of methanol, (d) water O – water H, (e) methanol O – methanol O, (f) water O – water O, (g) methanol O – water O. The curves were obtained using the TIP4P/2005 water model.

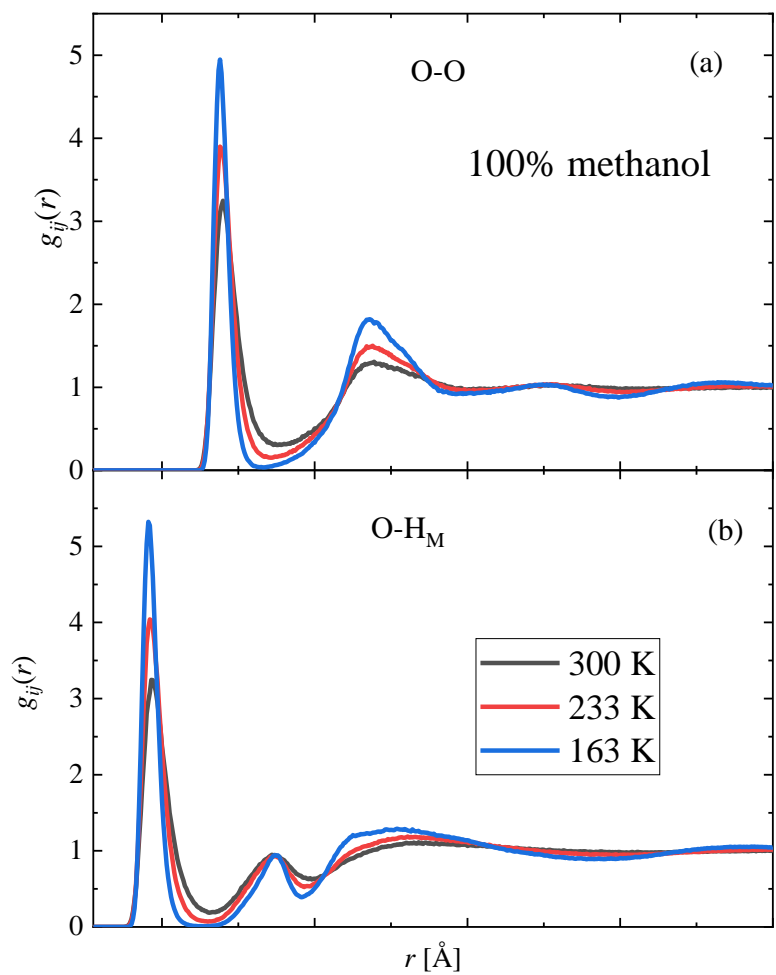

**Figure S59** Temperature dependence of the simulated partial radial distribution functions of pure methanol. The H-bonding related partials are shown: (a) O-O, (b) O – hydroxyl H of methanol (denoted as $H_M$).

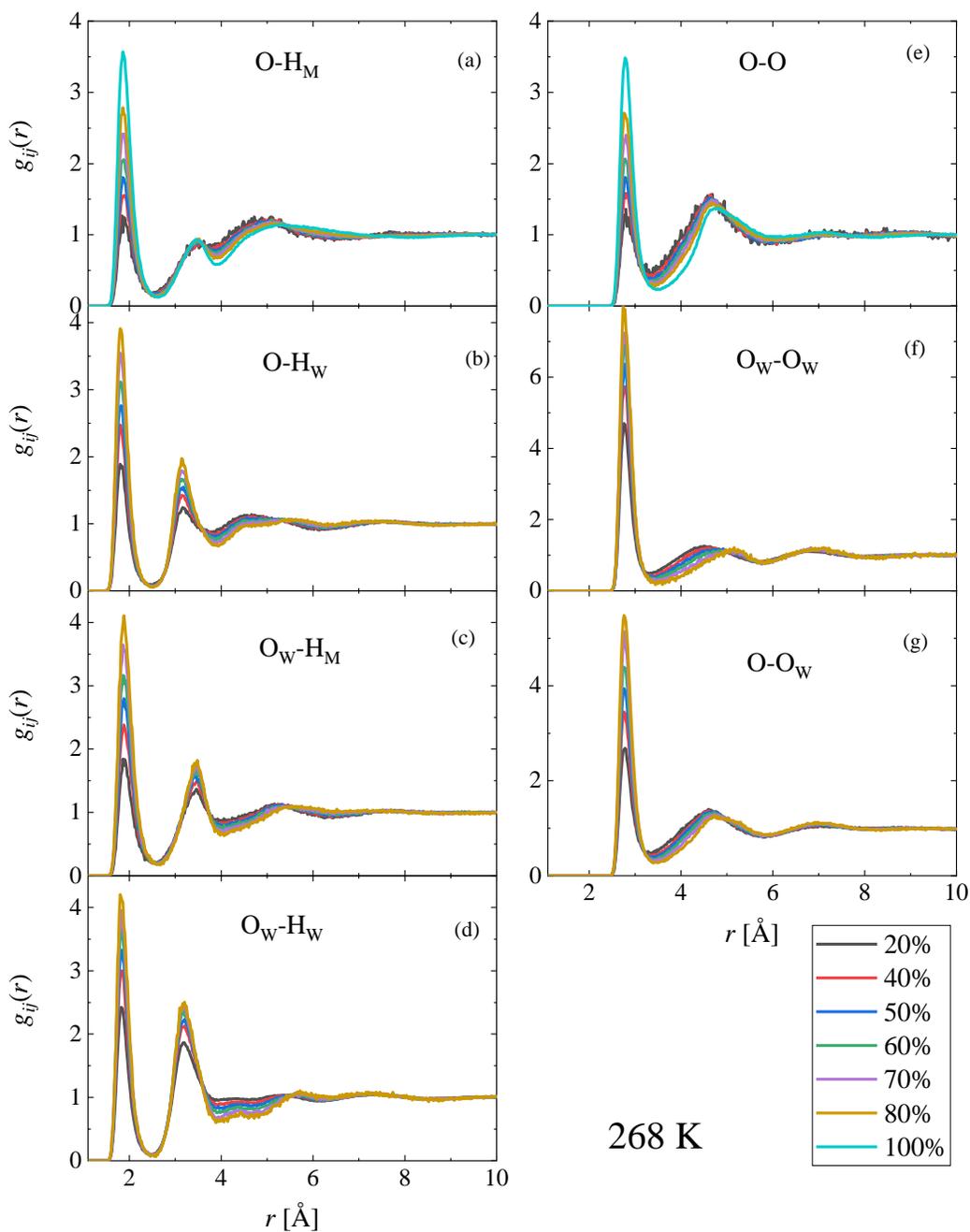

**Figure S60** Concentration dependence of the simulated partial radial distribution functions of the methanol-water mixture at 268 K. The H-bonding related partials are shown: (a) methanol O (denoted as O) – hydroxyl H of methanol (denoted as $H_M$), (b) methanol O – water H (denoted as $H_W$), (c) water O (denoted as $O_W$) – hydroxyl H of methanol, (d) water O – water H, (e) methanol O – methanol O, (f) water O – water O, (g) methanol O – water O. The curves were obtained using the TIP4P/2005 water model. The methanol content of the mixtures are shown in the figure legend.

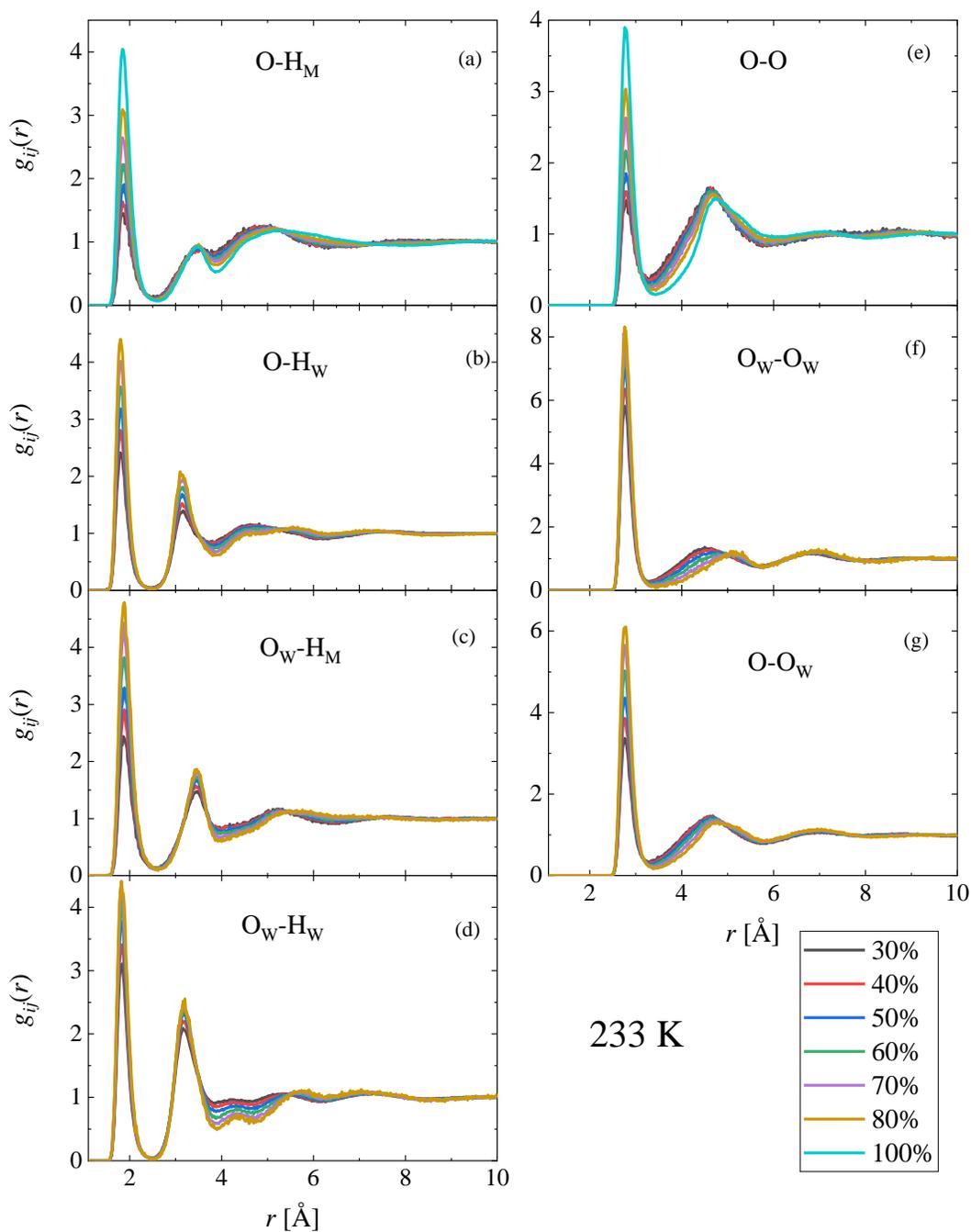

**Figure S61** Concentration dependence of the simulated partial radial distribution functions of the methanol-water mixture at 233 K. The H-bonding related partials are shown: (a) methanol O (denoted as O) – hydroxyl H of methanol (denoted as $H_M$), (b) methanol O – water H (denoted as $H_W$), (c) water O (denoted as $O_W$) – hydroxyl H of methanol, (d) water O – water H, (e) methanol O – methanol O, (f) water O – water O, (g) methanol O – water O. The curves were obtained using the TIP4P/2005 water model. The methanol content of the mixtures are shown in the figure legend.

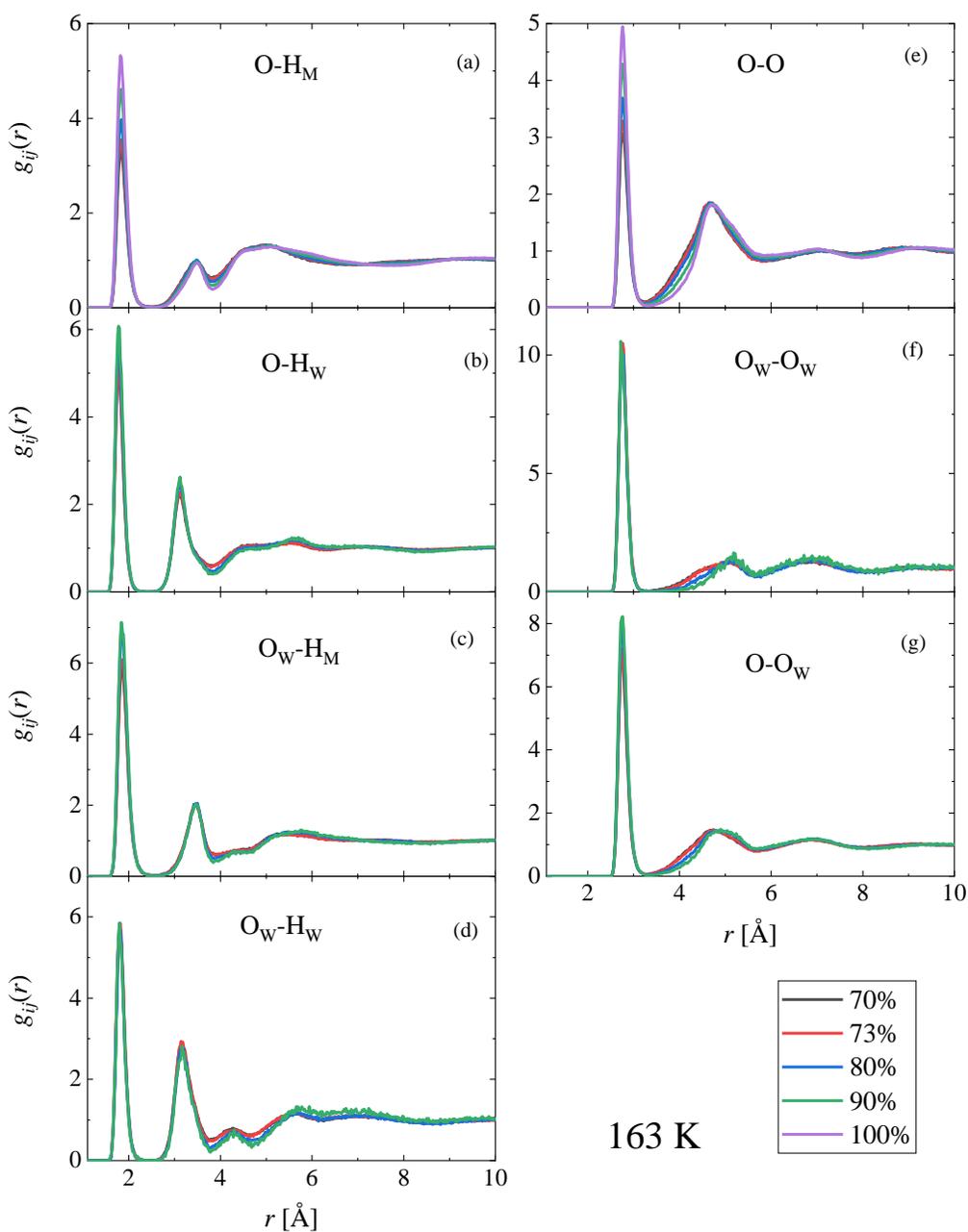

**Figure S62** Concentration dependence of the simulated partial radial distribution functions of the methanol-water mixture at 163 K. The H-bonding related partials are shown: (a) methanol O (denoted as O) – hydroxyl H of methanol (denoted as $H_M$), (b) methanol O – water H (denoted as $H_W$), (c) water O (denoted as $O_W$) – hydroxyl H of methanol, (d) water O – water H, (e) methanol O – methanol O, (f) water O – water O, (g) methanol O – water O. The curves were obtained using the TIP4P/2005 water model. The methanol content of the mixtures are shown in the figure legend.

**H-bond analysis results obtained by using the SPC/E water model**

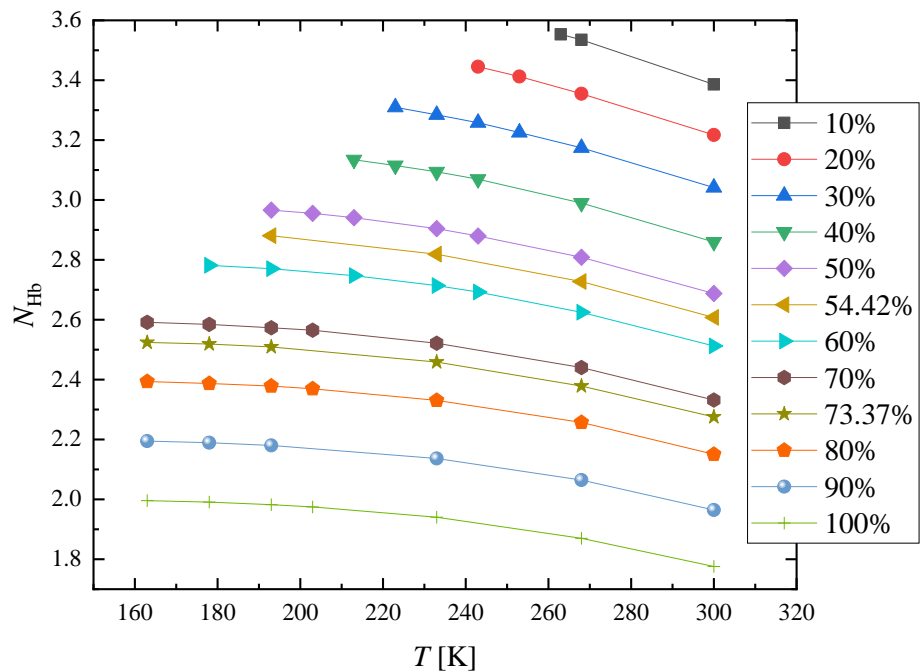

**Figure S63** Average number of hydrogen bonds per molecule in methanol-water mixtures as a function of temperature at different concentrations, obtained from MD simulations using the SPC/E water model.

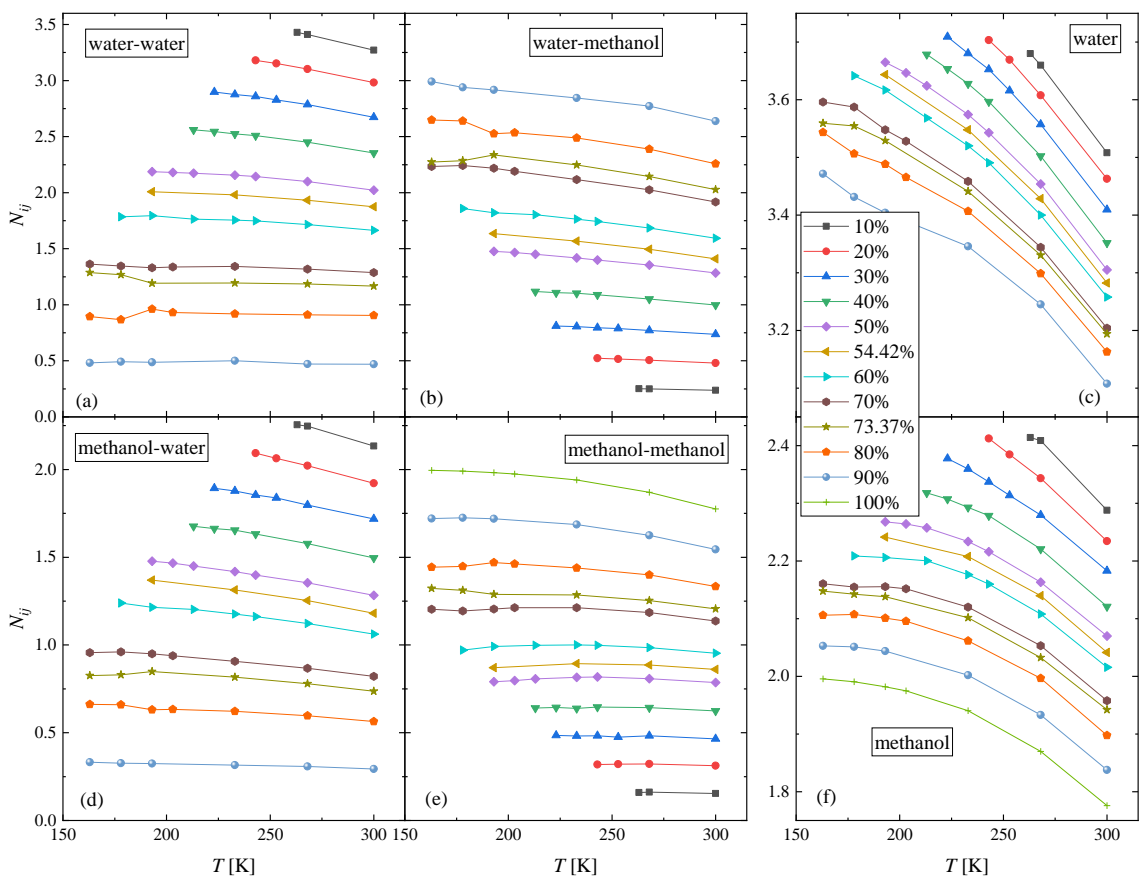

**Figure S64** Temperature dependence of the number of hydrogen bonds at different concentrations, as obtained from MD simulations using the SPC/E water model: (a) average number of H-bonded water molecules around water, (b) average number of H-bonded methanol molecules around water, (c) average number of H-bonded (water and methanol) molecules around water, (d) average number of H-bonded water molecules around methanol, (e) average number of H-bonded methanol molecules around methanol, (f) average number of H-bonded (water and methanol) molecules around methanol.

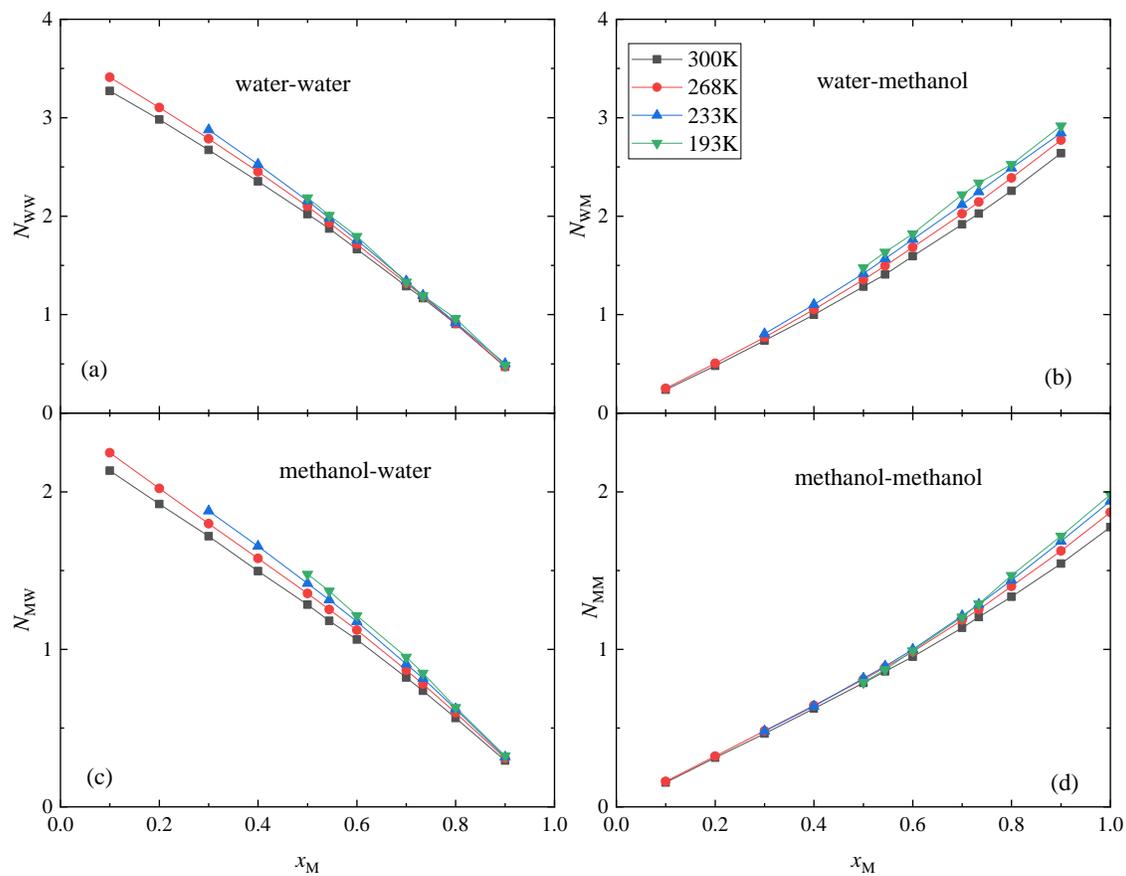

**Figure S65** Concentration dependence of the number of hydrogen bonds at different temperatures, as obtained from MD simulations using the SPC/E water model: (a) average number of H-bonded water molecules around water, (b) average number of H-bonded methanol molecules around water, (c) average number of H-bonded water molecules around methanol, (d) average number of H-bonded methanol molecules around methanol.